\documentclass[12pt,a4paper]{article}
\usepackage[longnamesfirst, authoryear, round]{natbib}

\let\counterwithin\relax
\usepackage[T1]{fontenc}
\usepackage[utf8]{inputenc}
\usepackage[width=1\textwidth, labelfont=bf]{caption}
\usepackage{enumitem}
\usepackage{booktabs}
\usepackage[all]{nowidow} 
\usepackage{amsmath}
\usepackage{amsthm} 
\usepackage{amsfonts}
\usepackage{tikz} 
\usepackage{calc}
\usepackage{siunitx}
\newcommand{\widesim}[2][1.5]{
  \mathrel{\overset{#2}{\scalebox{#1}[1]{$\sim$}}}
}

\def\bigsup#1{^{\vbox{\hbox{$\scriptstyle#1$}\nointerlineskip\hbox{}}}}

\usepackage{tikz-qtree} 
\usepackage{pgfplots}
\pgfplotsset{compat = 1.14}
\usepackage{mathtools,amssymb}
\usepackage{xcolor}
\usepackage{cooltooltips}
\usepackage{MnSymbol}
\usepackage{graphicx}
\usepackage{microtype}
\newcommand{\bigCI}{\mathrel{\text{\scalebox{1.07}{$\perp\mkern-10mu\perp$}}}} 
\usepackage{subcaption}
\usepackage[flushleft]{threeparttable}

\usepackage{placeins} 
\usepackage{indentfirst} 
\usepackage{lscape}
\usepackage{rotating}
\usepackage[pdfborderstyle={/S/U/W 1},breaklinks,colorlinks]{hyperref}
\AtBeginDocument{%
  \hypersetup{
    citecolor=blue,
    urlcolor =blue,
    menucolor = black,
    linkcolor = black}}
    
\usepackage{tocloft}

\usepackage{chngcntr} 
\counterwithin{figure}{section}
\counterwithin{table}{section}

\usepackage{tcolorbox}
\tcbset{colframe=black, colback=white,size=fbox}

\usepackage{algpseudocode}
\usepackage[linesnumbered,ruled]{algorithm2e}

\usepackage[left=2.8cm,right=2.8cm,top=3.5cm,bottom=3.5cm]{geometry}

\usepackage{listings}
\usepackage{xcolor}
\makeatletter
\def\thanks#1{\protected@xdef\@thanks{\@thanks
        \protect\footnotetext{#1}}}
\makeatother

\title{Cross-Fitting and Averaging for Machine Learning Estimation of Heterogeneous Treatment Effects\thanks{\scriptsize{We thank Michael Knaus, Gabriel Okasa, Stefan Lessmann, Wolfgang Karl Härdle and Victor Chernozhukov for valuable comments. Financial support from the Deutsche Forschungsgemeinschaft via the IRTG 1792 “High Dimensional Non Stationary Time Series”, Humboldt-Universität zu Berlin, is gratefully acknowledged.} }\\ }

\date{\today}

\author{Daniel Jacob \\ \href{mailto:daniel.jacob@hu-berlin.de}{daniel.jacob@hu-berlin.de}}

\begin{document}

    \maketitle
\thispagestyle{empty}

\begin{abstract}
\scriptsize
We investigate the finite sample performance of sample splitting, cross-fitting and averaging for the estimation of the conditional average treatment effect. Recently proposed methods, so-called meta-learners, make use of machine learning to estimate different nuisance functions and hence allow for fewer restrictions on the underlying structure of the data. To limit a potential overfitting bias that may result when using machine learning methods, cross-fitting estimators have been proposed. This includes the splitting of the data in different folds to reduce bias and averaging over folds to restore efficiency. To the best of our knowledge, it is not yet clear how exactly the data should be split and averaged. We employ a Monte Carlo study with different data generation processes and consider twelve different estimators that vary in sample-splitting, cross-fitting and averaging procedures. We investigate the performance of each estimator independently on four different meta-learners: the doubly-robust-learner, R-learner, T-learner and X-learner. We find that the performance of all meta-learners heavily depends on the procedure of splitting and averaging. The best performance in terms of mean squared error (MSE) among the sample split estimators can be achieved when applying cross-fitting plus taking the median over multiple different sample-splitting iterations. Some meta-learners exhibit a high variance when the lasso is included in the ML methods. Excluding the lasso decreases the variance and leads to robust and at least competitive results. 

\textbf{JEL classification:} C01, C14, C31, C63\\

\noindent
\textbf{Keywords:}
\textit{causal inference, sample splitting, cross-fitting, sample averaging, machine learning, simulation study} 

\end{abstract}

\newpage
\setcounter{page}{1}

\section{Introduction}

Recent methods to estimate the conditional average treatment effect (CATE) propose using machine learning (ML) methods when the underlying data is high-dimensional or has non-linear dependencies. The simplest approach is to estimate two conditional mean functions, one for the treated observations and one for the non-treated and then take the difference. Even if this method works for observational studies, where the treatment assignment mechanism depends on observed covariates (self-selection into treatment), there are more efficient methods to estimate the CATE. One way is to explicitly control for such selection bias by estimating the probability of treatment and control for it. The building of such a model or function is a classic classification task for which machine learning methods are well suited to find generalizable predictive patterns. Since we are only interested in getting a good prediction of the probability of treatment, we don't need to know the underlying structural form of this function which enables black-box ML methods to be sufficient. Such a function is called a nuisance function. Another nuisance function in this context is, for example, the conditional mean function from the outcome variable. 

ML methods use regularization to decrease the variance of an estimator. However, there is a trade-off between an introduced bias on the parameter of interest through regularization and overfitting. Sample splitting helps to limit overfitting and allows for less restrictive assumptions on the nuisance functions. The idea is to use at least two different samples or folds (say, A and M), one to fit (or train) the nuisance functions (fold A) and one for the estimation of the parameter of interest (fold M). To overcome a potential loss in efficiency, since only a subset of the data is used when estimating the CATE, cross-fitting is an increasingly popular approach to combine ML methods with semi-parametric estimation problems; see, for example, \cite{chernozhukov2018double}, \cite{newey2018cross} and \cite{athey2017efficient}. Cross-fitting estimates the parameter of interest using the subset M and then switches the roles of the sets now using subset M for training and subset A for estimation. The two results are then averaged. If we want to use this procedure to make predictions, we would build two prediction models based on the roles of the samples and average the resulting values for each observation. 

So far there are no clear proposals on how to exactly use sample splitting and cross-fitting. We can split the sample in two or more folds and average among those folds as suggested by \cite{chernozhukov2018double} for the average treatment effect (ATE) or \cite{nie2017quasi} as they do for their R-learner method to estimate the CATE. The former use two- and five-folds, while the latter use five- and ten-folds. In both cases, the estimation of models for the nuisance parameters (we sometimes refer to this as training of a model or function) is done on all folds but the one which is used for the prediction of the nuisance parameters and hence for the estimation of the CATE. This is quite similar to cross-validation. \cite{newey2018cross} and \cite{kennedy2020optimal} suggested for the doubly-robust (DR) estimator to not only use different folds for training and estimation but also to train each nuisance function on a different fold. \cite{zivich2020machine} provide a simulation study on the aforementioned so-called double sample splitting as well as cross-fitting estimators and demonstrate the performance for the ATE. 

While the sample splitting procedure allows for less restrictive assumptions on the ML estimators, sample splitting can introduce a new bias due to a specific sample. This can become more problematic the smaller the whole sample is. If we have a sample with only 500 observations, 5-fold cross-fitting would use 400 observations for training and 100 for estimation. If we furthermore train each nuisance function with a different fold, we could only use 125 observations for building any estimator that includes up to four different functions to estimate. To average any potential bias from sample splitting, we can repeat the estimation multiple times and take the mean or the median over all the estimators. This approach was first proposed for the ATE by \cite{chernozhukov2018double} as well as for the CATE when the nuisance functions might be misspecified due to high-dimensional problems as in \cite{chernozhukov2018generic}. 

For all the different approaches, 2-fold vs. K-fold, double sample splitting, cross-fitting and repeated sample splitting and averaging, there is little to no guidance for practitioners. In this paper, we consider different variations of the approaches and evaluate them independently on a class of recent methods that estimate the CATE. We consider the T-learner, DR-learner, a recent version of the class of orthogonal learners (R-learner) and X-learner in a Monte Carlo study to assess the performance of the CATE on an independent test-set. First, we briefly describe the four different meta-learners that we consider. We then discuss the idea of cross-fitting and double sample splitting as well as the idea of taking the median over multiple iterations. Since our results are based on simulated data, we describe in detail how we generate our data and why. Last, we show and discuss our results for twelve different estimators that are based on variations of splitting, cross-fitting and averaging for each of the meta-learners.







\section{Methods}
\setcounter{equation}{0}

When reviewing recently proposed methods for the estimation of the CATE, we can categorize them into two groups. The first group contains methods that transform the variables to a pseudo-outcome which is used as a proxy for the CATE function (the literature calls these transformed outcome approaches, meta-learners or generic ML algorithms). In a second step, off the shelf machine learning methods can be used to estimate the final CATE. The second group of methods leaves the variables untouched but alters existing machine learning methods in a way that they can be used to estimate the CATE directly (examples are Causal Boosting by \cite{powers2018methods}, Causal Forest by \cite{athey2019generalized} or the Bayesian Regression Tree Models for Causal Inference by \cite{hahn2020}). See \cite{kuenzel2019meta} for a comparison between the S-, T-, and X-learner as well as the Causal Forest in a simulation study. \cite{knaus2018machine} compare the inverse probability weighting (IPW) estimator, doubly-robust (DR), modified covariate method (MCM), R-learner and different versions of the Causal Forest in an empirical Monte Carlo study while \cite{nie2017quasi} compare their R-learner with the S-, T-, X- and U-learner as well as causal boosting. Regarding the base learners (the ML methods), \cite{kuenzel2019meta} use a Random Forest (RF) and a Bayesian Additive Regression Trees (BART) algorithm. \cite{knaus2018machine} use RF and lasso while \cite{nie2017quasi} use boosting and lasso for the estimation of the nuisance functions. 

In this paper, we concentrate only on methods from the first group, hence meta-learners that are flexible in the use of machine learning methods. Since these methods need to estimate different nuisance functions, we have more options to split and average samples and hence to estimate the CATE.

All aforementioned methods are based on the potential outcome framework for which we use the following notations: Each observation has two potential outcomes, $Y^1$ and $Y^0$ of which we only observe one, namely the former if someone was treated or the latter if not. We denote this by the binary treatment indicator $D \in \{0;1\}$ and denote observed covariates $X \in \mathbb{R}^{p}$. To interpret the estimated parameter as a causal relationship, the following assumptions are needed; see, for example, \cite{rubin1980randomization}:  

\vskip 1.0cm

1. Conditional independence ( or conditional ignorability/exogeneity or conditional unconfoundedness):

\begin{align}
\left(Y_{i}^{1}, Y_{i}^{0}\right) \bigCI D_{i}|X_{i}.
\end{align}	

2. Stable Unit Treatment Value Assumption (SUTVA) (or counterfactual consistency):

\begin{align}
Y_i = Y_i^0 + D_i (Y_i^1-Y_i^0).
\end{align}

3. Overlap Assumption (or common support or positivity):

\begin{gather}
\forall x \in supp(X), \quad 0 < P(D=1|X=x) < 1, \\
P(D=1|X=x) \equiv e(x).\label{equ:prop}
\end{gather}

4. Exogeneity of covariates:

\begin{align}
X_i^1 = X_i^0.
\end{align}

Assumption 1 together with Assumption 4 state that the treatment assignment is independent of the two potential outcomes and that the covariates are not affected by the treatment. Assumption 2 ensures that there is no interference, no spillover effects and no hidden variation between treated and non-treated observations. Assumption 3 states that no subpopulation defined by $X = x$ is entirely located in the treatment or control group, hence the treatment probability needs to be bounded away from zero and one. Equation \ref{equ:prop} is the propensity score. 

We define the conditional expectation of the outcome for the treatment or control group as 

\begin{align}
\mu_d(x) = \operatorname{E}[Y_i | X_i = x, D_i = d] \quad with \quad D \in \{0,1\}.
\end{align}
If we don't use any subscript, we refer to this function as the general conditional expectation.

\noindent
\textbf{How we select ML methods:}

The accuracy of the CATE estimation depends on the accuracy of the nuisance functions and hence on the choice of the ML method. For example, \cite{knaus2018machine} finds that when using the lasso, the estimators can have heavy tails in smaller samples. To minimize the dependence of the ML methods on our estimates, we do not assign specific machine learning methods for the estimation but consider a range of different popular methods. To choose which ML method to use for each nuisance function as well as for any additional functions, we use an ensemble and stacking method. In such a setting, not only one ML method may be chosen but an ensemble of methods which are stacked together with different weights. We use the SuperLearner package in R as proposed by \cite{polley2011super}. It also enables us to choose different models for each nuisance function and setting. 
The package offers a general class of prediction methods to be considered by the ensemble. From the 42 different algorithms, we select the following subset of algorithms for our analysis: Linear model (lm), lasso (glmnet), Random Forest (ranger), Gradient boosted trees (xgboost) and simply the mean of the input variable (mean). 

We use 5-fold cross-validation to estimate the performance on all machine learning models and create an weighted average of those models by using a validation sample. Using stacking, we can find the optimal combination of a collection of prediction algorithms or even different settings within one model. 

\subsection{Meta-Learners}

In the following, we briefly describe the considered meta-learners. Except for the T-learner, all other methods generate a pseudo-outcome in the first step which can be seen as an approximation of the conditional average treatment effect. The last step regresses this function on the covariates to get the final estimate. The DR-, R- and X-learner also require to estimate the propensity score as an additional nuisance function if the data does not come from a randomized control trial (RCT).

\noindent
\textbf{Two-model learner (T-learner):}

The T-learner is a two step approach where the conditional mean functions $\mu_1(x) = \operatorname{E}[Y^1 |X_i=x]$ and $\mu_0(x) = \operatorname{E}[Y^0 |X_i=x]$ are estimated separately with any generic machine learning algorithm. The difference between the two functions results in the CATE as shown in Table \ref{tab:learners}. 

One problem with the T-learner is that it aims to minimize the mean squared error for each separate function rather than to minimize the mean squared error of the treatment effect. See, for example, \cite{kuenzel2019meta, kennedy2020optimal} for settings when the T-learner is not the optimal choice.

\noindent
\textbf{Doubly-robust learner (DR-learner):}

A more efficient method than the T-learner might be the DR-learner. It builds on the T-learner and adds a version of inverse probability weighting (IPW) scheme on the residuals of both regression functions ($Y - \mu_D(x)$). We can think of it as combining two different models and hence avoid drawbacks like the minimization goal from the T-learner and a potentially high variance from an IPW model when some propensity scores are small.
The doubly-robust learner takes its name from a double robustness property which states that the estimator remains consistent if either the propensity score model or the conditional outcome model is correctly specified. This is at least true for the average treatment effect \citep{lunceford2004stratification}. Recently, this estimator has gained popularity to estimate the CATE, especially in high-dimensional settings. See, for example, the work by \cite{fan2019estimation} and  \cite{zimmert2019group}. \cite{kennedy2020optimal} which find that for estimating the CATE, the finite-sample error-bound from the DR-learner at most deviates from an oracle error rate by the product of the mean squared error of the propensity score and the conditional mean estimator. 

\noindent
\textbf{Orthogonal-learners (here: R-learner):}

The orthogonal-learner makes use of the idea of orthogonalization to cancel out any selection bias that may arise in observational studies from observed covariates. Here, the residuals from the regression of $Y$ on $X$ are regressed on the residuals from the regression of $D$ on $X$ and weighted by the squared residuals, $(D-\hat{e}(x))^2$. This is similar to the double machine learning approach from \cite{chernozhukov2018double} where their estimator of interest is the ATE.  \cite{nie2017quasi} develop a general class of two-step algorithms for the estimation of the CATE. Their so-called ``R-learner'', as from residualization and a homage to Robinson 1988, makes explicit use of machine learning methods. 

Achieving Neyman orthogonality using a residuals-on-residuals (or debiasing) approach has a long history in econometrics (see the Frisch–Waugh–Lovell theorem from the 1930s for linear regression) and mainly builds on the work by \cite{robinson1988root} who replaces the linear parts by non-parametric kernel regression. \cite{chernozhukov2018generic} adopt the debiasing approach using ML methods in RCT's where the parameter of interest is some feature (like a best linear predictor) of the CATE.


\noindent
\begin{minipage}{\textwidth}
\textbf{X-learner:}

\cite{kuenzel2019meta} propose the X-learner which estimates a treatment effect separately for the control and the treatment group. This might be especially helpful in situations where the proportion of the two groups is highly imbalanced.  
The X-learner has several steps. The first step is identical to the T-learner, namely estimating the two conditional mean functions. In the second step, however, the difference is found in the observed outcome for the treated and control group, respectively. The two imputed treatment effects ($\hat{\psi}_{X}^1:=Y^{1}-\hat{\mu}_{0}\left(x^{1}\right)$ and
$\hat{\psi}_{X}^0:=\hat{\mu}_{1}\left(x^{0}\right)-Y^{0}$) are now used in a third step to regress them individually on the covariates to obtain $\hat{\tau}_0(x)$ (the CATE for the control group) and $\hat{\tau}_1(x)$ (the CATE for the treatment group). The final estimator combines the two estimators plus some weights, $g(x)$:

\begin{align*}
\hat{\tau}(x)=g(x) \hat{\tau}_{0}(x)+(1-g(x)) \hat{\tau}_{1}(x)
\end{align*}

 The weights can, for example, be set to $1 -\hat{e}(x)$ for the treatment group and $\hat{e}(x)$ for the control group estimate, respectively.
\vspace{0.5cm}
\end{minipage}

\noindent
\textbf{Summary of meta-learners:}

We summarise the considered meta-learners in Table \ref{tab:learners} where $\hat{\psi}$ states the pseudo-outcome or estimator for each of the learners. 
The last column counts the number of nuisance functions to train to estimate the pseudo-outcome or estimator and in brackets, we state the total number of models needed to get the final CATE estimate on the whole dataset. Note that the X-learner is regressed only for the treated observations and again only for the observations in the control group. This is why we need two more additional models for the final estimate.


\begin{table}[ht]
\centering
{\renewcommand{\arraystretch}{1.5}%
\resizebox{0.9\textwidth}{!}{%
    \begin{threeparttable}
\caption{Summary of meta-learners}
\label{tab:learners}

\begin{tabular}{lccc}
\hline \hline
Method                 & Estimator/Pseudo-outcome & Weights ($w_i$) & \# of Models   \\  \hline

T-learner              &  $\hat{\psi}_{T} = \hat{\mu}_1(x) - \hat{\mu}_0(x)$               & 1                 &    2 (3)                      \\
DR-learner             &  \begin{tabular}[c]{@{}l@{}}$\hat{\psi}_{DR} = \hat{\psi}_{T}+\dfrac{D\left(Y-\hat{\mu}_{1}\left(x \right)\right)}{\hat{e}\left(x\right)}$\\ \quad \quad $-\dfrac{\left(1-D\right)\left(Y-\hat{\mu}_{0}\left(x \right)\right)}{\left(1-\hat{e}\left(x\right)\right)}$\end{tabular}    & 1  &  3  (4) 					\\ 
R-learner              &  $\hat{\psi}_{R} = \dfrac{(Y-\hat{\mu}\left(x\right))}{(D-\hat{e}\left(x\right))}$      &$\left(D-\hat{e}(x)\right)^2$      &  2 (3)  					 \\
X-learner              &  \begin{tabular}[c]{@{}l@{}} $\hat{\psi}_{X}^1 :=Y^{1}-\hat{\mu}_{0}\left(x^{1}\right)$ \\
$\hat{\psi}_{X}^0:=\hat{\mu}_{1}\left(x^{0}\right)-Y^{0}$ \end{tabular}    & 1   &  3  (5)                     \\ \hline \hline
\end{tabular}

\begin{tablenotes}
      \small
      \item \textit{Notes:} Considered meta-learners that estimate the CATE. \# of Models counts the number of nuisance functions to estimate the pseudo-outcome. Numbers in brackets count the total number of models to train to get the final CATE estimate.
    \end{tablenotes}
  \end{threeparttable}
}
}
\end{table}

The estimators from Table \ref{tab:learners} can be represented as a weighted minimization problem which solves the following:

\begin{align*}
\min _{\tau}\left\{\frac{1}{N} \sum_{i=1}^{N} w_{i}\left[\hat{\psi}_{i}-\tau\left(X_{i}\right)\right]^{2}\right\}.
\end{align*}

\subsection{Sample splitting and cross-fitting}

To aim for a consistent estimator, we need to assume certain complexity conditions on the nuisance functions. Specifically, we want them to be smooth (i.e. differentiable) and the entropy of the candidate nuisance functions to be small enough to fulfil Donsker conditions (e.g. if we assume Lipschitz parametric functions or VC classes). In high-dimensional settings (p>n) or when using ML methods that are complex or adaptive, the Donsker conditions might not hold; see, for example, \cite{robins2013new}, \cite{chernozhukov2016locally} and \cite{rotnitzky2017multiply}. As \cite{chernozhukov2018double} noticed, verification of the entropy condition is so far only available for certain classes of machine learning methods, such as lasso and post-lasso. For classes that employ cross-validation or for hybrid methods (like the SuperLearner), it is likely difficult to verify such conditions. Luckily, there is an easy solution available: sample-splitting. When splitting the sample we can use independent sets for estimating the nuisance functions and constructing the estimating equation. By using different sets, we can treat the nuisance functions as fixed functions which allow avoiding conditions on the complexity. It also allows us to use any ML method such as random forest or boosting or even an ensemble of different methods. The sample splitting approach to avoid smoothness conditions dates back at least to \cite{bickel1982adaptive} and was extended to also use cross-fitting by \cite{schick1986asymptotically}.

Concerning sample splitting and cross-fitting, there are different strategies on how to split the data and average the results. First, consider the 50:50 sample splitting where the data is split into two equal folds, namely an auxiliary sample (A) (which is used to train all nuisance functions) and the main sample (M) for the estimation. Second, the data can be split into that many equal parts as we have functions to estimate and we use each fold to train a different nuisance function. \cite{newey2018cross} use this approach with undersmoothing to reduce bias (they call this approach double cross-fitting). \cite{kennedy2020optimal} adopts this approach and shows that one can achieve faster rates in estimating the CATE using the DR-learner. He restricts the estimation of the two conditional mean functions ($\hat{\mu}_1(x)$ and $\hat{\mu}_0(x)$) to the same fold, the propensity score to another fold and the estimation of the parameter to the remaining fold. Hence we need three folds for the DR-learner. To not get confused about the names of the different approaches, we call the technique where we use different folds on each nuisance function but without cross-fitting ``double sample splitting'' (following \cite{kennedy2020optimal}).

Both approaches can be extended to not only two or three folds but $K$ folds. The idea here is to use more observations to train the nuisance functions. For example, we can assign ($100 - \dfrac{100}{K}$)\% of the observations to sample A and $(\dfrac{100}{K})$\% to sample M. 
For simplicity, we refer to the two- and K-fold estimator as $\hat{\tau}_K$ while $K$ denotes the number of folds. This is independent of whether we use standard or double sample splitting.

The two estimators above might however not be efficient. Sample splitting reduces the available amount of data used for both, the estimation of the nuisance function and the final parameter by construction.
This leads to a loss in efficiency and statistical power in finite samples. If we want to make use of the full sample to restore efficiency, we can switch the roles of the samples thereby using sample M for training and sample A for estimation and again to regress the pseudo-outcome on the covariates. Prediction on the whole data or an independent test set leads to two estimates which are then averaged. If we use $K$ folds, we repeat the procedure until each fold was used for estimation and average over the $K$ estimates. Let $S= \{Y_i,D_i,X_i\}_{i \in \{1...N\}}$ be the set of observations from the whole training sample. Then, we define  $S_k= \{Y_i,D_i,X_i\}_{i \in N_k}$ as the set of observations for each fold using only observations $N_k$ while $\bigcupdot\limits_{k=1}^{K}N_k = \{1...N\}$.  We use data from the set $S_k$ of the training sample to estimate the pseudo-outcome while using independent test data to estimate the final estimator: 

\begin{align}
\hat{\tau}_{k}(x)=\operatorname{E}[\hat{\psi}(S_k) \mid X_i=x],
\end{align}

\begin{align}
\tilde{\tau}_{K}(x) = \frac{1}{K}\sum_{k=1}^{K}\hat{\tau}_{k}(x).
\end{align}

\begin{figure}[ht]
\begin{center}
\includegraphics[width=0.8\textwidth]{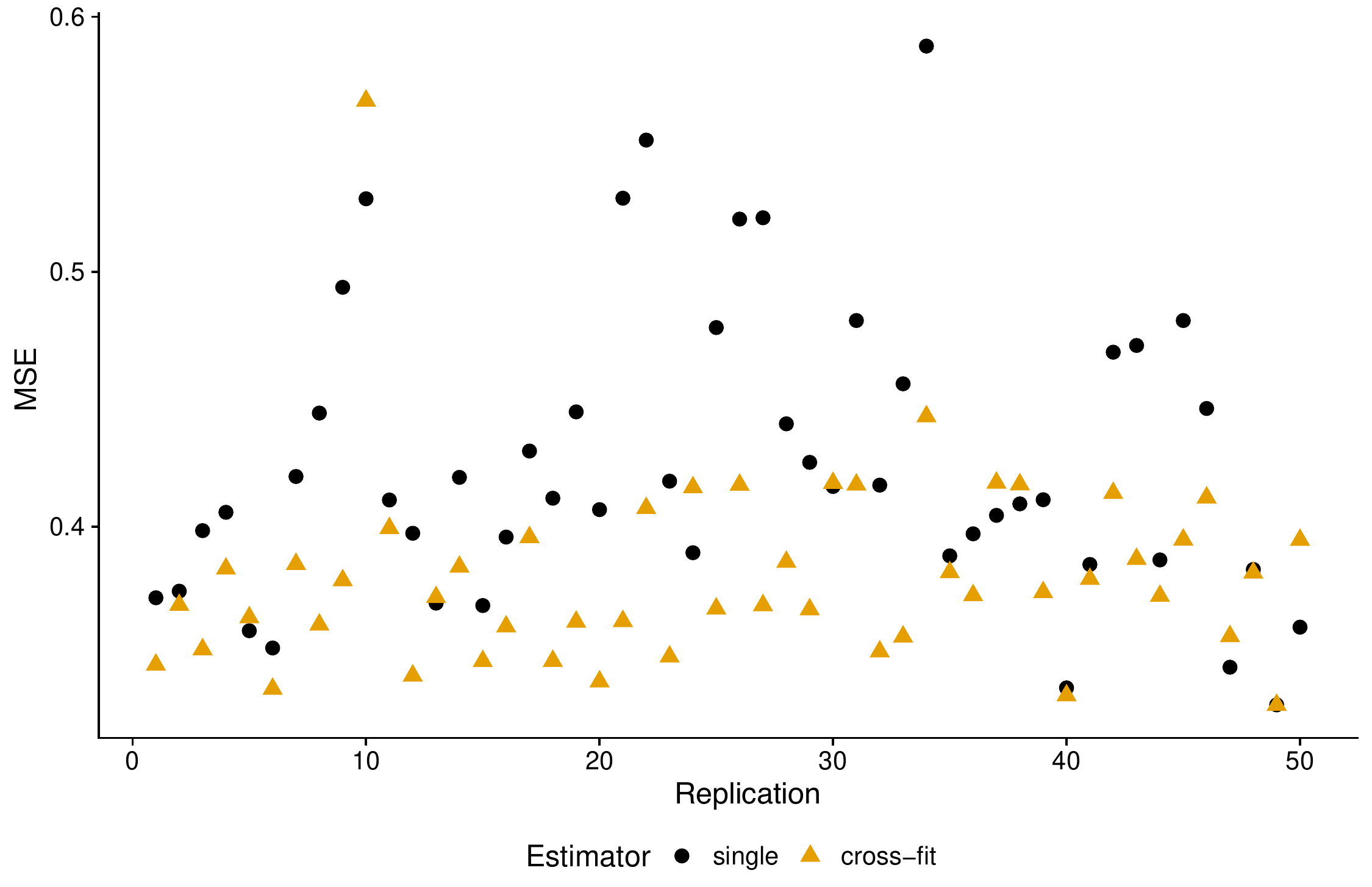}
\caption{Single vs. cross-fit estimation of CATE. }
\label{fig:single_vs_cross-fit}
\end{center}
\end{figure}

We give an example of the benefit from cross-fitting in Figure \ref{fig:single_vs_cross-fit}. We show the MSE from the true treatment effect for a single estimator ($\hat{\tau}_2(x)$) and the cross-fit estimator ($\tilde{\tau}_{2}(x)$) based on a 50:50 sample split. We used the R-learner as the meta-learner and create 50 Monte Carlo replications of the data using the same data generating process (DGP) which simulates a RCT and has the following properties: $N = 2000$, $X = \mathbb{R}^{10}$, $e_{0}(X)$ = 0.5, and $\tau(x) = X_1 + X_2>0 + W \quad \text{with} \quad W \sim  \mathcal{N}(0,0.5)$. Using cross-fitting decreases the MSE compared to the single estimator in about 90\% of the cases. We also find that the variance is smaller compared to the single estimator. 

\begin{figure}[ht]
\begin{center}
\includegraphics[width=0.9\textwidth]{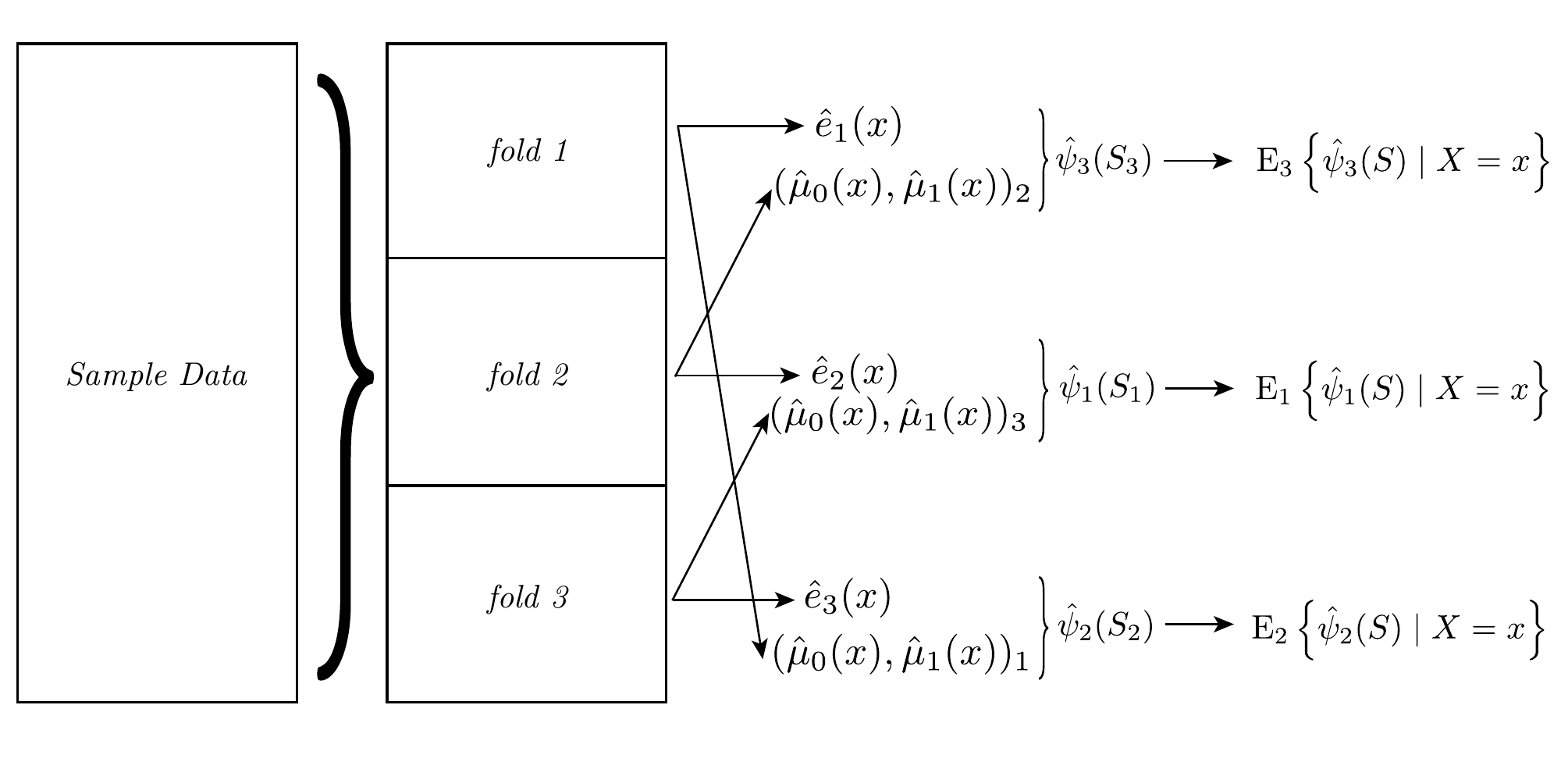}
\caption{Double sample splitting procedure and cross-fitting.}
\label{fig:double sample-split}
\end{center}
\end{figure}

Figure \ref{fig:double sample-split} illustrates the double sample splitting with cross-fitting in detail. Let fold 1 to fold 3 denote three independent samples that include the set of observations $S_k$ as stated above. Additional subscripts indicate on which sample we train the corresponding function. The fold in brackets indicates the fold on which we predict. For example, we use fold 1 to train $\hat{e}_1(x)$ and fold 2 to train $(\hat{\mu}_0,\hat{\mu}_1)_2$. We use fold 3 for the prediction and to calculate the pseudo-outcome $\hat{\psi}_{3}(S_3)$. It has the same subscript as we use the same fold for the regression on the covariates. We then use all the data $S$ to predict the CATE. The resulting estimates from the three pseudo-outcomes are averaged to get the final CATE estimate. 


Splitting the sample in $K$ folds has the consequence that only $\frac{1}{K}$ of the observations is used for estimation. \cite{chernozhukov2018double} find that this approach could lead to an unstable empirical Jacobian in the estimation sample when the sample size is small. To overcome this problem, they suggest to calculate the pseudo-outcome (the residuals in the case of double machine learning) over all folds and then estimate the final parameter using all observations. We call this the combined approach and state the following notation:

\begin{align}
\hat{\hat{\tau}}(x)=\operatorname{E}[\hat{\psi}(S_c) \mid X_i=x].
\end{align}

with $\hat{\psi}(S_c) = \{\hat{\psi}(S_1),...,\hat{\psi}(S_K)\}$ being the combined pseudo-outcomes. Note that this is different from the naive approach where we use the whole training sample to estimate the pseudo-outcome (i.e. $\hat{\psi}(S)$).

While we can use the combined approach for any number of $K$ splits, we only focus on 5-fold splits since in this case we have more observations for estimating the nuisance functions and might see a bigger difference when combining the pseudo-outcomes over 5-folds vs. using $\frac{1}{5}$ for the estimation of the CATE. 

\subsection{Bias reduction due to specific sample splitting}

\begin{figure}[ht]
\begin{center}
\includegraphics[width=1.0\textwidth]{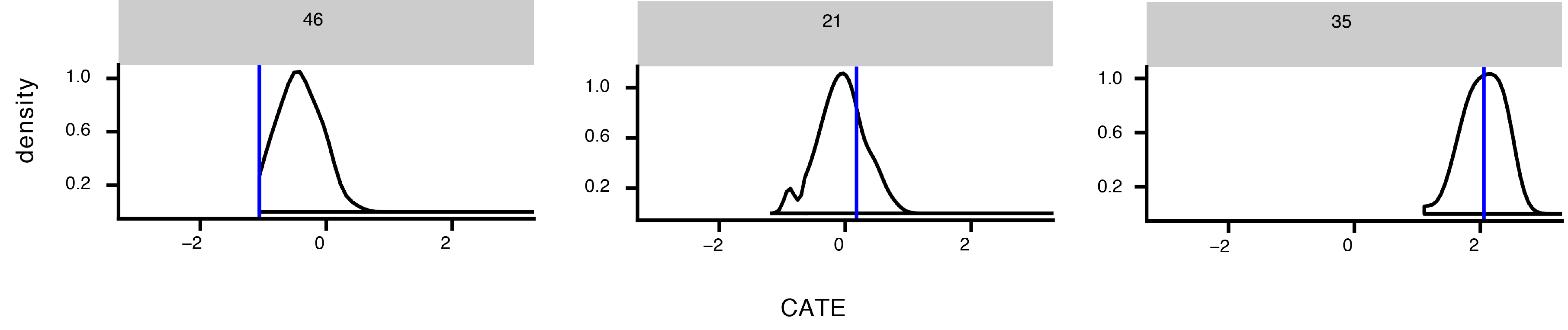}
\caption{Distribution of estimated CATE for 3 selected individuals.}
\label{fig:density}
\end{center}
\end{figure}

Since we only partition our sample once in $K$ folds, we end up with a specific sample used for estimation. Even if the splitting is random and if the specific partition has no impact on the results asymptotically, in finite samples the effect of the specific partition can lead to a bias. To see this, we show the distribution over 50 estimates for the CATE for 3 representative observations from the test set in Figure \ref{fig:density}. Different from the example above, we only create one dataset but repeat the sample splitting in fold A and fold M 50 times. Each time we estimate the CATE for each observation. The simulated data has the same properties as the cross-fit example. The blue lines show the true treatment effect for each observation. We can distinguish three different cases. First consider the left plot: The right tail from the distribution is above zero, leading to the wrong sign of the treatment effect in some cases. Through averaging (either with the mean or the median) we would at least get the right sign. The plot in the middle is similar and again shows a more heavy right tail. Here we would get the right sign and come close to the true value if we take the median instead of the mean. The plot on the right is quite normally distributed and by taking the median we would even get an unbiased estimate. The plot shows that the sample splitting plays a role for the variance in the final estimates, even if the nuisance functions are not complicated (the propensity score is a constant and the treatment effect is linear). Figure \ref{fig:density49} in the Appendix shows 49 randomly selected observations and their distribution of the CATE due to sample splitting. While for most of the observations the distribution is concentrated on either positive or negative values, some observations are centred around zero and hence show positive and negative values given a particular sample split.

\vskip 1.0cm

As suggested by \cite{chernozhukov2018double}, we take the median over all iterations ($B$) for each observation. This may lead to a more stable conditional average treatment effect function:	
\begin{align}
\tilde{\tau}_{median}(x) = median\{\tilde{\tau}_{b}(x)\}^{1:B} 
\end{align}

We neglect the subscript for the number of folds for readability. Here we show the median estimator based on cross-fitting estimators. 


\section{Simulation Study}
\setcounter{equation}{0}

To evaluate the performance of the estimators, we perform a Monte Carlo study in which we repeat the estimation of all parameters 300 times. 
In the following, we describe the data generating process and show the variations that we consider. Since simulations can be biased towards a specific setting, we consider 6 different DGPs and also two different sample sizes. We include two settings for randomized control trials (one with a balanced treatment assignment and one where only 20\% of the observations are treated). In all the other settings we assume selection into treatment and vary the difficulty to estimate the propensity score function as well as the treatment effect function. 
We are interested in finite sample performance and evaluate all estimators and configurations for $N = 2000$ (setting A:F) and $N = 500$ (setting G:L) observations, respectively. In total, we use 12 different settings on which we compare the estimators and meta-learners.

\begin{figure}
\begin{center}
\begin{tikzpicture}[every node/.style = {shape=rectangle, rounded corners,
    draw,top color=white},level 1/.style={sibling distance=-150pt},level 2/.style= {level distance = 100pt, sibling distance=-5pt},level 3/.style= {sibling distance=10pt},font={\fontsize{10pt}{12}\selectfont}  ]

\Tree [.DGP \edge[dashed]; {test data (T) (10,000)} 
[.{sample data (S)}
	[.{naive (no split)}
     	[.{single} 
     		 ]
   	] 
   	[.{50:50 split}
		[.{single} 
     		 ]
		[.cross-fit
			[{median} ] [{single} ] ]     	
   	] 
   	[.{double sample split }
   	[.{single} 
     		 ]
		[.cross-fit
			[{median} ] [{single} ] ]
   	] 
   	[.{5-fold split}
   	[.{single} 
     		 ]
		[.cross-fit
			[{median} ] [{single} ] ]
		[.{combined} 
		[{median} ] [{single} ] ]
   	] 
] 
]

\end{tikzpicture}
\end{center}
\caption{Structure of the different splitting and averaging scenarios.}
\label{Fig:splitting-tree}
\end{figure}
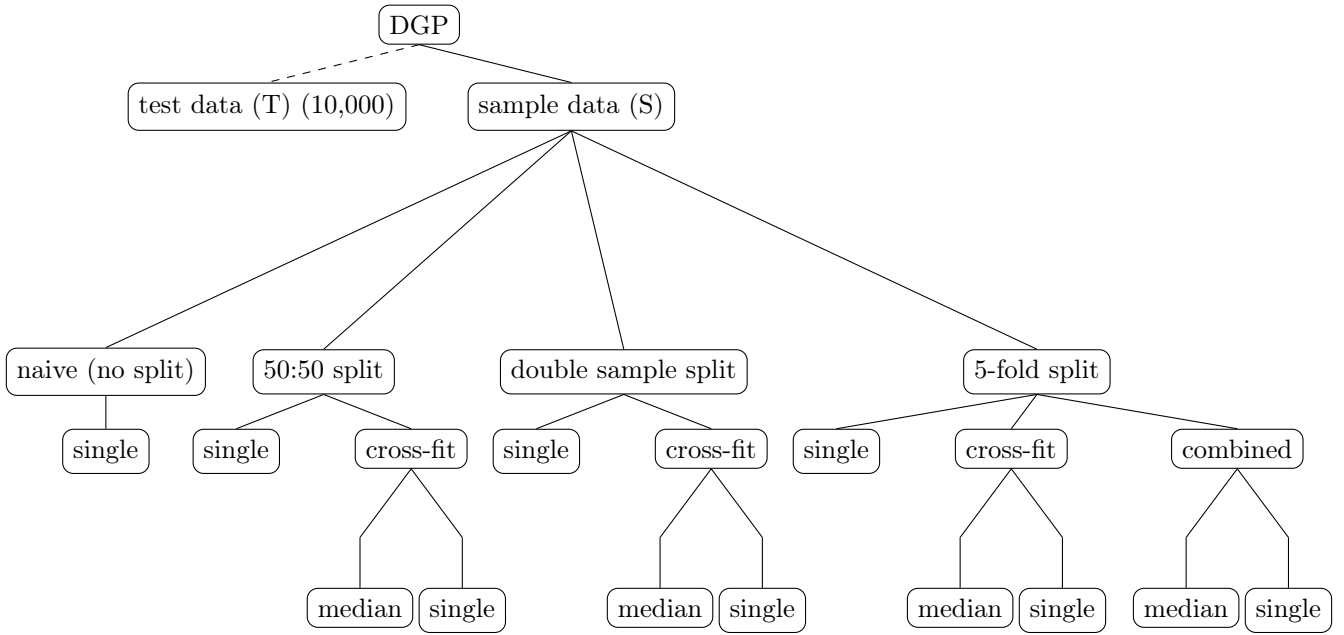

Figure \ref{Fig:splitting-tree} shows the different splitting procedures that we consider in our study. We first set aside a test set ($T$) of 10,000 observations which we use to compare the performance measures on. We choose a higher amount of observations to limit the noise from our Monte Carlo study. We consider 100 replications of the sample data ($S$) for each DGP but keep the test set fixed conditional on the DGP.   

The naive setting uses the whole sample data for training and estimation during all steps. In the 2-fold setting, we split the data into equal parts and use one subset for training and the remaining for estimation. When we use more than two folds, we consider two different strategies. 
First, we consider the setting where we split the sample data into three folds (the double sample splitting setting). Next, we assign folds to train a specific nuisance function. We group the nuisance function in: 1. the propensity score, 2. the conditional mean function (either for both, the treatment and control group or separately) and 3. the regression of the pseudo-outcome on the covariates. For the X-learner, we also assign one specific fold to estimate the two imputed treatment effects. In the cross-fitting setting, we then switch the roles of the three folds and repeat the process until each fold estimated the CATE function.

When we use five-folds, we do not split the training and estimation sample into equal parts but use 4 out of the five folds for training the nuisance functions and the remaining fold for estimation. In the cross-fitting setting, we repeat this process until all of the five folds are used for the estimation. It is also possible to use any K-fold splitting together with the double sample splitting. We would use the same proportion ($1/K$) for the estimation but split the remaining folds for training into two parts to use different folds for each nuisance function. Due to computational reasons, we only consider three folds for the double sample splitting approach in our analysis.

Since cross-fitting uses the whole data and hence increases efficiency, we only apply the median procedure on such cross-fitted estimators. It would also be possible to use the median based on single estimators. Preliminary simulations show that these estimators at most behave equally compared to cross-fit estimators if we take the median over at least 30 iterations. Given these results, we do not consider this class of estimators in our Monte Carlo study. 
Algorithm \ref{pseudo:1} in the Appendix shows the pseudo-code for the procedure of double sample splitting, cross-fitting and taking the median.

\noindent
\textbf{Data generating process:} \label{section:DGP}

The basic model used in this simulation study is a partially linear regression model based on \citet{robinson1988root}:

\begin{align}
Y = \tau(X)D + g(X) + U,  &&\operatorname{E}[U | X,D] = 0, \\
D = e(X) + V,  &&\operatorname{E}[V | X] = 0, \label{prop_score} \\
\tau(X) = t(X) + W &&\operatorname{E}[W| X] = 0, 
\end{align}

with $Y$ being a continuous outcome variable (this can also be a binary variable for all the methods we consider). 
$\tau(X)$ is the true treatment effect or population uplift, while $D$ is the treatment status. The vector $X \in \mathbb{R}^p = ({X_{1},...,X_{p}})$ consists of $p$ different features, covariates or confounders. Let $X_i \widesim{iid} \mathcal{N}(0,\Sigma)$, where $\Sigma$ is a correlation matrix (pseudo code \ref{pseudo:Cov_Generation} in the Appendix describes how we estimate the correlation).  $U$, $V$  and $W$ are error terms which follow a normal distribution with mean zero and variance 1 (if not specified otherwise).

Equation \ref{prop_score} is the propensity score. In the case of completely random treatment assignment, the propensity score $e(X_{i}) = c$ for all units ($i=1,...,N$). The scalar $c$ can take any value within the interval (0,1). In the simulation we consider $ c = 0.5$ (balanced) and $c = 0.2$ (imbalanced). The imbalance in treatment assignment given a RCT can often be observed in practice since treatment is generally costly.

We assume a correlation of the covariates through a uniform distribution of the covariance matrix which is then transformed into a correlation matrix. Correlated characteristics are more common in real datasets and help to investigate the performance of ML algorithms, especially the regularization bias, in a more realistic manner. Figure \ref{Cov_matrix} in the Appendix shows the correlation matrix for 10 randomly selected covariates to give an example of the correlation.

The function $g(X)$ takes the following form: 

\begin{align}
g(X) &=  X_{1} + X_{5} +  X_{4} \times X_{5}.
\end{align}

In the simulation, we focus on different functions of the \textbf{treatment assignment}. We use the cumulative density function (CDF) of the standard normal distribution to create probabilities which are then used in a binomial function to create a binary treatment variable. The dependence of covariates within the CDF is defined as $a(X)$, for which we use a variety of functions, namely random assignment with balanced and imbalanced groups, a linear dependence, interaction terms, and non-linear dependence. 
\begin{align}
&e(X) = \Phi\left(\frac{a(X)- \operatorname{E}[a(X)]}{\sigma[a(X)]} \right),  \\
D \widesim{ind.} \text{Bernoulli}[&e(X)] \quad \text{such that} \quad D \in \{0;1\}.
\end{align}

\begin{align*}
\text{random assigment:} \quad &e_0(X) = c \quad \text{with} \quad c \in (0,1),\\
\text{linear:} \quad &a(X) = X_2 + X_{5} + X_{2} - X_{8},\\
\text{interaction:} \quad &a(X)  = X \times b + X_{5} + X_{2} + X_{3} \times X_{8}, \\
\text{non-linear:} \quad &a(X) = X \times b + \sin(X_{5}) + X_2 + \cos(X_{4} \times X_{8}).
\end{align*}
The vector $b = \frac{1}{l}$ with $l \in \{1,2,...,p\}$ represents weights for every covariate.
The \textbf{treatment effect} takes four different settings where we also vary the degree of heterogeneity. It might not always be the case that there is significant heterogeneity in the treatment effect which is why we also consider the case of a binary effect and even no effect at all. We generate the four different settings as follows:

\begin{align*}
\text{linear:} \quad &\tau(X) = X_1 + X_2>0 + W \quad \text{with} \quad W \widesim{}  \mathcal{N}	(0,0.5), \\
\text{binary:} \quad &\tau(X) = 
\begin{dcases}
    2,& \text{if} \quad X_5>0 \\
    1,              & \text{otherwise},
\end{dcases}  \\
\text{non-linear:} \quad &\tau(X) = \sin(X_{1} + 0.5X_{2} + 0.\overline{33}X_{3}) + \cos( X_{10}),  \\
\text{zero:} \quad &\tau(X) = 0. 
\end{align*}

Similar settings for the treatment assignment and with binary and zero treatment effects are used by \cite{kuenzel2019meta,powers2018methods, nie2017quasi}. 

\section{Simulation Results}
\setcounter{equation}{0}

For each meta-learner, we evaluate the splitting and averaging procedures based on the mean squared error (MSE) from the test data. Following \cite{knaus2018machine}, we also show the absolute bias and the standard deviation. For the absolute bias and the standard deviation, we first average the predicted CATE for each individual over 300 repetitions. In order to summarise the results, we then average all three performance measures over the whole test data observations ($N_T$):

\begin{align}
MSE_{i} &=\frac{1}{R} \sum_{r=1}^{R}\left[\hat{\tau}(X_i)_r-\tau\left(X_{i}\right)\right]^{2} \nonumber \\
\overline{MSE} &= \frac{1}{N_T}\sum_{i=1}^{N_T} MSE_i
\end{align}

\begin{align}
|{Bias}_{i}| &=| \underbrace{\frac{1}{R} \sum_{r=1}^{R} \hat{\tau}\left(X_{i}\right)_{r} }_{\bar{\hat{\tau}}(X_i)} - \tau(X_i)| \nonumber \\ 
\overline{|Bias|} &= \frac{1}{N_T}\sum_{i=1}^{N_T} |{Bias}_{i}| 
\end{align}

\begin{align}
SD_{i} &=\sqrt{\frac{1}{R} \sum_{r=1}^{R}\left[\hat{\tau}\left(X_{i}\right)_{r}-\overline{\hat{\tau}}\left(X_{i}\right)\right]^{2}} \nonumber \\
\overline{SD} &= \frac{1}{N_T}\sum_{i=1}^{N_T} SD_{i}
\end{align}

The estimator ${\hat{\tau}}(X_i)$ refers to either the naive estimator without sample splitting $\hat{\tau}_{naive}(X_i)$, the single estimators without cross-fitting (${\hat{\tau}}_K(X_i)$), the cross-fit estimators which takes the average over $K$ folds ($\tilde{\tau}(X_i)$) or the median estimators ($\tilde{\tau}_{median}(X_i)$) which result only from cross-fit estimators.


Table \ref{tab:DGP} shows the specifications of the DGP for each setting which we refer to in the results.

\begin{table}[ht]
\centering
\resizebox{0.9\textwidth}{!}{%
    \begin{threeparttable}
\caption{DGP settings}
\label{tab:DGP}

\begin{tabular}{l|llllll}
\hline \hline \\
Scenarios & A/G      & B/H    & C/I        & D/J    & E/K   & F/L       \\ \hline
N         & 2000/500      & 2000/500  & 2000/500        & 2000/500     & 2000/500   & 2000/500       \\
$\mathbb{R}\bigsup{p}$       & 20       & 20     & 20         & 20       & 20     & 20         \\
$P(D=1)$ &$0.5 $ &0.2   &linear & interaction & non-linear & linear \\
$\tau(X)$ 	&linear &linear &non-linear &binary &non-linear &zero \\
\hline \hline
\end{tabular}
\begin{tablenotes}
      \small
      \item \textit{Notes:} Simulations A through F use 2000 observations, while G through L only have 500 observations. 
    \end{tablenotes}
  \end{threeparttable}
}
\end{table}

We summarise the results for settings with 2000 observations and 500 observations, respectively. Table \ref{tab:DR-Results_2000} shows the results for the DR-learner. We find that the best performance is among the estimators which use cross-fitting and median averaging. While the absolute bias is competitive among those settings, the standard deviation seems to be the main driver for the different MSE. The same can be observed for the R-learner in Table \ref{tab:R-Results_2000}. To highlight the effect of outliers, we also show the median MSE. In setting C of the R-learner, we find the same mean MSE for the naive and the 5-fold cross-fit with median averaging estimator but the median MSE is slightly lower for the latter. We also see this result in setting I for the DR-learner when comparing the naive and the 5-fold cross-fit with median averaging estimator. In all settings, especially when we have selection into treatment, we find that the MSE can be decreased by applying cross-fitting and again by using median averaging among different sample splits. For the DR-learner, the best performing estimators are the combined method and the naive estimator. There is, however, no clear pattern.  The results for the X-learner, shown in Table \ref{tab:X-Results_2000} differ from the aforementioned estimators. In all settings, except when we have zero treatment effect (setting F), the naive estimator performs best in terms of all performance measures. 

When the sample size is decreased to 500, we find that the variance and hence the MSE increases. The bias, however, stays almost the same. While for the DR-learner the best performing estimator is the combined method, the R-learner seems to perform best using the 5-fold cross-fit with median averaging estimator. We also noticed that with this sample size extreme outliers are present. See, for example, setting J of the DR-learner with a MSE of 37.49 for the double split estimator. The same can be observed for the R-learner in almost all settings with an extreme MSE of 106.63 in setting K. These outliers only occur for estimators without median averaging. All median estimators show comparable MSE values. The Tables for all estimators  and meta-learners with 500 observations  are in the Appendix (see Table \ref{tab:DR-Results_500}, \ref{tab:R-Results_500} and \ref{tab:X-Results_500}). 

In the Appendix (Section \ref{sec:R-learner_w/o_Lasso}) we also show results for experiments where we exclude the lasso and all linear models from the DR- and R-learner. By doing so, the results are more robust in terms of outliers and also at least competitive if not better in terms of MSE. The best performing estimator across all DGPs is the 5-fold cross-fit with median averaging estimator when the lasso and linear models are excluded. This holds at least for the R-learner while using the DR-learner settings E, F, K and L favour the naive estimator. We see the same behaviour for the T-learner, exactly in the aforementioned settings the naive estimator shows the smallest MSE. Otherwise, we find that the cross-fit estimators outperforms the single estimators. When self-selection into treatment is introduced, the median estimators (especially the 5-fold) seems to outperform the cross-fit estimators.

\begin{table}[ht]
\centering
\resizebox{1\textwidth}{!}{%
    \begin{threeparttable}
\caption{Performance measures for the considered estimators based on the DR-learner.}
\label{tab:DR-Results_2000}

\begin{tabular}{c|rrrrrrrrrrrr}
\hline \hline \\

Scenarios & \multicolumn{4}{c}{A} & \multicolumn{4}{c}{B} & \multicolumn{4}{c}{C}   \\    \\ \hline

 & $\overline{MSE}$ & $\overline{|Bias|}$ & $\overline{SD}$  & \textit{Median MSE}  & $\overline{MSE}$ & $\overline{|Bias|}$ & $\overline{SD}$  & \textit{Median MSE} & $\overline{MSE}$ & $\overline{|Bias|}$ & $\overline{SD}$  & \textit{Median MSE}\\
  \hline
naive 							& 0.72 & 0.50 & 0.57 & 0.51 & 0.82 & 0.50 & 0.66 & 0.65 & 0.72 & 0.64 & 0.30 & 0.43 \\ 
  50:50							& 1.40 & 0.47 & 1.02 & 1.01 & 0.90 & 0.47 & 0.75 & 0.75 & 1.03 & 0.46 & 0.82 & 0.69 \\ 
  50:50 cross-fit & 0.59 & 0.47 & 0.43 & 0.37 & 0.69 & 0.48 & 0.58 & 0.54 & 0.91 & 0.49 & 0.72 & 0.64 \\
  double split 						& 2.99 & 0.49 & 1.62 & 1.84 & 1.71 & 0.48 & 1.16 & 1.48 & 1.19 & 0.50 & 0.88 & 0.85 \\ 
  double split cross-fit 				& 0.74 & 0.48 & 0.62 & 0.56 & 0.69 & 0.48 & 0.58 & 0.54 & 0.74 & 0.51 & 0.57 & 0.46 \\ 
  5-fold 						& 0.99 & 0.48 & 0.79 & 0.83 & 2.09 & 0.50 & 1.30 & 1.77 & 1.12 & 0.49 & 0.85 & 0.78 \\ 
  5-fold cross-fit 					& 0.55 & 0.49 & 0.42 & 0.38 & 0.70 & 0.48 & 0.57 & 0.54 & 0.62 & 0.48 & 0.49 & 0.39 \\ 
  5-fold combined &\underline{0.53} & 0.49 & 0.12 & \underline{0.36} & 0.65 & 0.48 & 0.15 & 0.46 & \underline{0.53} & 0.48 & 0.32 & \underline{0.31} \\
  50:50 cross-fit median 				& 0.57 & 0.47 & 0.47 & 0.41 & \underline{0.54} & 0.47 & 0.45 & \underline{0.39} & 0.69 & 0.48 & 0.56 & 0.42 \\ 
  double split cross-fit median 			& 0.59 & 0.48 & 0.48 & 0.42 & 0.56 & 0.48 & 0.46 & 0.41 & 0.68 & 0.50 & 0.50 & 0.40 \\ 
  5-fold cross-fit median 				& 0.57 & 0.49 & 0.45 & 0.40 & 0.61 & 0.48 & 0.49 & 0.44 & 0.60 & 0.49 & 0.45 & 0.36 \\
  5-fold combined median &\underline{0.53} & 0.49 & 0.12 & \underline{0.36} & 0.65 & 0.48 & 0.15 & 0.46 & \underline{0.52} & 0.48 & 0.31 & \underline{0.30} \\

\hline \hline \\
Scenarios & \multicolumn{4}{c}{D} & \multicolumn{4}{c}{E} & \multicolumn{4}{c}{F}   \\    \\ \hline

& $\overline{MSE}$ & $\overline{|Bias|}$ & $\overline{SD}$  & \textit{Median MSE}  & $\overline{MSE}$ & $\overline{|Bias|}$ & $\overline{SD}$  & \textit{Median MSE} & $\overline{MSE}$ & $\overline{|Bias|}$ & $\overline{SD}$  & \textit{Median MSE}\\
  \hline
naive 										& 0.92 & 0.82 & 0.32 & 0.85 & \underline{0.51} & 0.50 & 0.36 & 0.33 & {0.41} & 0.18 & 0.60 & 0.29 \\ 
  50:50 										& 1.21 & 0.50 & 0.89 & 0.83 & 2.58 & 0.67 & 1.35 & 1.79 & 1.95 & 0.34 & 1.33 & 1.38 \\ 
  50:50 cross-fit & 0.58 & 0.38 & 0.58 & 0.41 & 2.26 & 0.66 & 1.13 & 1.32 & 0.82 & 0.33 & 0.81 & 0.62 \\
  double split 							& 1.13 & 0.50 & 0.84 & 0.76 & 2.91 & 0.74 & 1.42 & 2.17 & 3.08 & 0.33 & 1.71 & 2.19 \\ 
  double split cross-fit 				& 0.81 & 0.51 & 0.62 & 0.52 & 1.64 & 0.72 & 0.88 & 1.07 & 0.78 & 0.35 & 0.78 & 0.63 \\ 
  5-fold 										& 1.29 & 0.49 & 0.95 & 1.03 & 2.84 & 0.66 & 1.46 & 2.18 & 3.21 & 0.27 & 1.76 & 2.41 \\ 
  5-fold cross-fit 							& 0.63 & 0.49 & 0.49 & 0.39 & 1.22 & 0.67 & 0.69 & 0.74 & 0.73 & 0.31 & 0.76 & 0.59 \\ 
  5-fold combined & 0.59 & 0.48 & 0.42 & \underline{0.32} & 1.16 & 0.67 & 0.51 & {0.68}& \underline{0.31} & 0.30 & 0.45 & \underline{0.29}\\ 
  50:50 cross-fit median 			& 0.77 & 0.51 & 0.57 & 0.44 & 1.35 & 0.70 & 0.72 & 0.79 & 0.58 & 0.33 & 0.64 & 0.44 \\ 
  double split cross-fit median 	& 0.74 & 0.52 & 0.55 & 0.44 & 1.30 & 0.72 & 0.67 & 0.75 & 0.70 & 0.34 & 0.73 & 0.54 \\ 
  5-fold cross-fit median 			& \underline{0.39} & 0.49 & 0.45 & 0.35 & 1.11 & 0.68 & 0.58 & 0.62 & 0.57 & 0.31 & 0.66 & 0.45 \\
  5-fold combined median & 0.59 & 0.48 & 0.41 & \underline{0.31} & 1.16 & 0.67 & 0.50 & {0.68} & \underline{0.31} & 0.29 & 0.44 &\underline{0.28} \\

\hline \hline \\

\end{tabular}
\begin{tablenotes}
      \small
      \item \textit{Notes:} Number of observations is 2000. Averages over 300 repetitions. Lowest MSE values are underlined. If values are similar we underline the median. For the median estimators we take the median over 20 iterations of sample splitting.  
    \end{tablenotes}
  \end{threeparttable}
}
\end{table}

\begin{table}[ht]
\centering
\resizebox{\textwidth}{!}{%
    \begin{threeparttable}
\caption{Performance measures for the considered estimators based on the R-learner.}
\label{tab:R-Results_2000}

\begin{tabular}{c|rrrrrrrrrrrr}
\hline \hline \\

Scenarios & \multicolumn{4}{c}{A} & \multicolumn{4}{c}{B} & \multicolumn{4}{c}{C}   \\    \\ \hline

 & $\overline{MSE}$ & $\overline{|Bias|}$ & $\overline{SD}$  & \textit{Median MSE}  & $\overline{MSE}$ & $\overline{|Bias|}$ & $\overline{SD}$  & \textit{Median MSE} & $\overline{MSE}$ & $\overline{|Bias|}$ & $\overline{SD}$  & \textit{Median MSE}\\
  \hline
naive & 0.94 & 0.76 & 0.24 & 0.50 & 0.96 & 0.77 & 0.23 & 0.49 & \underline{0.54} & 0.52 & 0.36 & 0.36 \\ 
  50:50 & 1.05 & 0.48 & 0.84 & 0.85 & 1.76 & 0.49 & 1.17 & 1.34 & 1.65 & 0.42 & 1.14 & 1.21 \\ 
  50:50 cross-fit & 0.62 & 0.47 & 0.52 & 0.48 & 0.86 & 0.49 & 0.70 & 0.67 & 0.89 & 0.40 & 0.75 & 0.58 \\ 
  double split & 1.22 & 0.49 & 0.92 & 1.00 & 1.42 & 0.51 & 1.01 & 1.21 & 1.65 & 0.42 & 1.14 & 1.24 \\ 
  double split cross-fit & 0.77 & 0.48 & 0.63 & 0.61 & 0.85 & 0.50 & 0.67 & 0.68 & 0.77 & 0.41 & 0.66 & 0.48 \\ 
  5-fold & 8.39 & 0.50 & 2.83 & 4.41 & 1.75 & 0.51 & 1.16 & 1.53 & 3.28 & 0.43 & 1.71 & 2.45 \\ 
  5-fold cross fit & 0.65 & 0.49 & 0.53 & 0.49 & 0.80 & 0.51 & 0.63 & 0.63 & 0.89 & 0.44 & 0.75 & 0.67 \\ 
  5-fold combined &0.54 & 0.48 & 0.42 & 0.37 & \underline{0.33} & 0.45 & 0.37 & 0.27 & 1.58 & 0.37 & 1.20 & 1.06 \\ 
  50:50 cross-fit median & \underline{0.50} & 0.47 & 0.39 & 0.34 & {0.54} & 0.49 & 0.42 & 0.37 & 0.71 & 0.40 & 0.62 & 0.42 \\ 
  double split cross-fit median & 0.57 & 0.48 & 0.46 & 0.41 & 0.60 & 0.50 & 0.45 & 0.41 & 0.63 & 0.42 & 0.54 & 0.37 \\ 
  5-fold cross-fit median & 0.60 & 0.49 & 0.48 & 0.43 & 0.62 & 0.51 & 0.46 & 0.42 & \underline{0.54} & 0.43 & 0.47 & \underline{0.34} \\ 
  5-fold combined median & 0.52 & 0.48 & 0.39 & 0.34 & 0.35 & 0.45 & 0.39 & 0.29 & 1.46 & 0.37 & 1.14 & 0.97 \\

\hline \hline \\
Scenarios & \multicolumn{4}{c}{D} & \multicolumn{4}{c}{E} & \multicolumn{4}{c}{F}   \\    \\ \hline

 & $\overline{MSE}$ & $\overline{|Bias|}$ & $\overline{SD}$  & \textit{Median MSE}  & $\overline{MSE}$ & $\overline{|Bias|}$ & $\overline{SD}$  & \textit{Median MSE} & $\overline{MSE}$ & $\overline{|Bias|}$ & $\overline{SD}$  & \textit{Median MSE}\\
  \hline
naive & 0.43 & 0.32 & 0.51 & 0.43 & 0.55 & 0.46 & 0.46 & 0.39 & \underline{0.19} & 0.05 & 0.43 & 0.13 \\ 
  50:50 & 0.69 & 0.29 & 0.74 & 0.50 & 1.15 & 0.38 & 0.93 & 0.81 & 1.99 & 0.14 & 1.40 & 1.53 \\ 
  50:50 cross-fit & 0.50 & 0.30 & 0.60 & 0.37 & 0.77 & 0.38 & 0.69 & 0.48 & 0.86 & 0.12 & 0.91 & 0.67 \\ 
  double split & 0.88 & 0.30 & 0.86 & 0.70 & 1.61 & 0.41 & 1.14 & 1.21 & 2.28 & 0.12 & 1.50 & 1.83 \\ 
  double split cross-fit & 0.42 & 0.31 & 0.53 & 0.33 & 0.72 & 0.40 & 0.64 & 0.47 & 0.88 & 0.15 & 0.92 & 0.65 \\ 
  5-fold & 1.19 & 0.30 & 1.03 & 0.94 & 3.46 & 0.43 & 1.78 & 2.68 & 6.20 & 0.17 & 2.48 & 4.72 \\ 
  5-fold cross fit & 0.46 & 0.32 & 0.57 & 0.37 & 0.68 & 0.41 & 0.62 & 0.49 & 1.12 & 0.10 & 1.05 & 0.85 \\ 
  5-fold combined & 0.55 & 0.24 & 0.71 & 0.36 & 1.02 & 0.34 & 0.93 & 0.68 & 1.38 & 0.14 & 1.15 & 0.91 \\
  50:50 cross-fit median & 0.37 & 0.29 & 0.49 & 0.27 & 0.61 & 0.38 & 0.56 & 0.35 & 0.39 & 0.11 & 0.61 & 0.30 \\ 
  double split cross-fit median & \underline{0.32} & 0.30 & 0.44 & 0.25 & 0.56 & 0.40 & 0.51 & 0.34 & 0.32 & 0.12 & 0.54 & 0.24 \\ 
  5-fold cross-fit median & 0.34 & 0.32 & 0.46 & 0.29 & \underline{0.49} & 0.41 & 0.44 & 0.31 & 0.43 & 0.08 & 0.64 & 0.33 \\ 
  5-fold combined median & 0.49 & 0.24 & 0.67 & 0.33 & 0.92 & 0.34 & 0.86 & 0.61 & 1.19 & 0.13 & 1.07 & 0.77 \\

\hline \hline \\

\end{tabular}
\begin{tablenotes}
      \small
      \item \textit{Notes:} Number of observations is 2000. Averages over 300 repetitions. Lowest MSE values are underlined. If values are similar we underline the median. For the median estimators we take the median over 20 iterations of sample splitting.  
    \end{tablenotes}
  \end{threeparttable}
}
\end{table}

\begin{table}[ht]
\centering
\resizebox{\textwidth}{!}{%
    \begin{threeparttable}
\caption{Performance measures for the considered estimators based on the X-learner.}
\label{tab:X-Results_2000}

\begin{tabular}{c|rrrrrrrrrrrr}
\hline \hline \\

Scenarios & \multicolumn{4}{c}{A} & \multicolumn{4}{c}{B} & \multicolumn{4}{c}{C}   \\    \\ \hline

 & $\overline{MSE}$ & $\overline{|Bias|}$ & $\overline{SD}$  & \textit{Median MSE}  & $\overline{MSE}$ & $\overline{|Bias|}$ & $\overline{SD}$  & \textit{Median MSE} & $\overline{MSE}$ & $\overline{|Bias|}$ & $\overline{SD}$  & \textit{Median MSE}\\
  \hline
naive & \underline{0.50} & 0.50 & 0.32 & 0.28 & \underline{0.47} & 0.46 & 0.37 & 0.30 & \underline{0.56} & 0.52 & 0.34 & 0.29 \\ 
  50:50 & 1.58 & 1.01 & 0.06 & 0.75 & 1.22 & 0.87 & 0.11 & 0.55 & 0.85 & 0.72 & 0.32 & 0.58 \\ 
  50:50 cross-fit & 1.58 & 1.01 & 0.06 & 0.75 & 1.21 & 0.87 & 0.09 & 0.55 & 0.85 & 0.72 & 0.32 & 0.58 \\ 
  double split & 1.61 & 1.00 & 0.22 & 0.79 & 1.22 & 0.87 & 0.11 & 0.55 & 1.07 & 0.72 & 0.56 & 0.81 \\ 
  double split cross-fit & 1.60 & 1.00 & 0.18 & 0.77 & 1.21 & 0.87 & 0.10 & 0.55 & 0.92 & 0.72 & 0.41 & 0.65 \\ 
  5-fold & 1.71 & 1.01 & 0.37 & 0.87 & 1.67 & 0.87 & 0.68 & 1.01 & 0.99 & 0.71 & 0.50 & 0.73 \\ 
  5-fold cross fit & 1.63 & 1.00 & 0.24 & 0.80 & 1.30 & 0.86 & 0.36 & 0.65 & 0.88 & 0.72 & 0.37 & 0.62 \\ 
  5-fold combined & 1.58 & 1.01 & 0.07 & 0.75 & 1.20 & 0.86 & 0.10 & 0.55 & 0.75 & 0.72 & 0.24 & 0.48 \\
  50:50 cross-fit median & 1.58 & 1.01 & 0.06 & 0.75 & 1.21 & 0.87 & 0.09 & 0.55 & 0.85 & 0.72 & 0.32 & 0.58 \\ 
  double split cross-fit median & 1.58 & 1.01 & 0.09 & 0.75 & 1.21 & 0.87 & 0.09 & 0.55 & 0.87 & 0.72 & 0.35 & 0.61 \\ 
  5-fold cross-fit median & 1.58 & 1.00 & 0.10 & 0.75 & 1.21 & 0.87 & 0.12 & 0.55 & 0.85 & 0.72 & 0.32 & 0.58 \\ 
  5-fold combined median & 1.58 & 1.01 & 0.07 & 0.75 & 1.20 & 0.86 & 0.10 & 0.55 & 0.75 & 0.72 & 0.23 & 0.48 \\ 

\hline \hline \\
Scenarios & \multicolumn{4}{c}{D} & \multicolumn{4}{c}{E} & \multicolumn{4}{c}{F}   \\    \\ \hline

 & $\overline{MSE}$ & $\overline{|Bias|}$ & $\overline{SD}$  & \textit{Median MSE}  & $\overline{MSE}$ & $\overline{|Bias|}$ & $\overline{SD}$  & \textit{Median MSE} & $\overline{MSE}$ & $\overline{|Bias|}$ & $\overline{SD}$  & \textit{Median MSE}\\
  \hline
naive & \underline{0.23} & 0.29 & 0.32 & 0.16 & \underline{0.54} & 0.51 & 0.34 & 0.28 & 0.27 & 0.31 & 0.30 & 0.14 \\ 
  50:50 & 0.61 & 0.60 & 0.28 & 0.43 & 1.36 & 0.94 & 0.29 & 0.86 & 0.15 & 0.21 & 0.33 & 0.16 \\ 
  50:50 cross-fit & 0.60 & 0.60 & 0.27 & 0.42 & 1.34 & 0.94 & 0.27 & 0.85 & {0.14} & 0.21 & 0.32 & 0.15 \\ 
  double split & 0.70 & 0.62 & 0.38 & 0.46 & 1.50 & 0.95 & 0.45 & 0.99 & 0.31 & 0.20 & 0.52 & 0.30 \\ 
  double split cross-fit & 0.69 & 0.62 & 0.38 & 0.44 & 1.43 & 0.95 & 0.35 & 0.91 & 0.24 & 0.21 & 0.45 & 0.24 \\ 
  5-fold & 0.86 & 0.60 & 0.58 & 0.75 & 1.55 & 0.93 & 0.56 & 1.09 & 0.37 & 0.21 & 0.57 & 0.28 \\ 
  5-fold cross fit & 0.62 & 0.59 & 0.32 & 0.45 & 1.36 & 0.93 & 0.34 & 0.87 & 0.22 & 0.20 & 0.43 & 0.20 \\ 
  5-fold combined & 0.58 & 0.58 & 0.27 & 0.37 & 1.30 & 0.93 & 0.25 & 0.81 & 0.10 & 0.20 & 0.25 & 0.10 \\
  50:50 cross-fit median & 0.60 & 0.60 & 0.27 & 0.42 & 1.34 & 0.94 & 0.27 & 0.85 & {0.14} & 0.21 & 0.31 & 0.15 \\ 
  double split cross-fit median & 0.62 & 0.61 & 0.29 & 0.39 & 1.39 & 0.95 & 0.30 & 0.88 & 0.17 & 0.21 & 0.35 & 0.18 \\ 
  5-fold cross-fit median & 0.58 & 0.59 & 0.26 & 0.45 & 1.32 & 0.93 & 0.27 & 0.83 & 0.15 & 0.21 & 0.32 & 0.16 \\ 
  5-fold combined median & 0.58 & 0.58 & 0.27 & 0.36 & 1.30 & 0.93 & 0.24 & 0.81 & \underline{0.09} & 0.19 & 0.25 & 0.10 \\

\hline \hline \\

\end{tabular}
\begin{tablenotes}
      \small
      \item \textit{Notes:} Number of observations is 2000. Averages over 300 repetitions. Lowest MSE values are underlined. If values are similar we underline the median. For the median estimators we take the median over 20 iterations of sample splitting.  
    \end{tablenotes}
  \end{threeparttable}
}
\end{table}

\clearpage

\section{Discussion}

This paper studies the finite sample performance of estimators based on meta-learners for the estimation of heterogeneous treatment effects. Such meta-learners rely on the estimation of different nuisance functions which facilitates different ways of sample splitting, cross-fitting and averaging. Samples can be equally or unequally split into folds. Each fold can be used to train all nuisance functions or we can select different folds for each function. Given specific folds, we can use cross-fitting to average the results. Furthermore, we can repeat this process to generate new folds and take the median over a sufficiently large number of iterations. 

To study the aforementioned estimators, we generate twelve different artificial datasets that vary in their complexity of the propensity score and the treatment effect as well as the number of observations. We perform a Monte Carlo study with 300 repetitions of the DGP and evaluate each estimator on different performance measures. We further use four popular meta-learners: the DR-learner, R-learner, T-learner and X-learner and apply all estimators based on the splitting and averaging procedures on each of them independently. 

For the T, DR, and R-learner, we can group our findings in three categories. First, if we have an RCT, the median estimators (50:50, double split and 5-fold) perform similarly in terms of mean MSE, mean absolute bias and mean standard deviation and the difference in MSE is not that high compared to the single estimators. Second, if we deviate from RCT towards an observational study, the performance measures differ between the estimators and the learners. Using the DR-learner we find that the combined method overall performs best while for the R-learner it is mainly the 5-fold cross-fit with median averaging estimator. This result does not hold for the T-learner in general. The third group is if we have zero treatment effect. Then we find that a naive estimator (without sample splitting) performs best in all performance measures. An exception of the previous findings holds for the X-learner. Here we find in almost all DGPs that the naive estimator performs best.

Taking the median over at least 20 iterations decreases the MSE further compared to other estimators. Especially for the R-learner, we find significant differences in the performance when taking the median on top of a five-fold cross-fit estimator (e.g. for setting E, the MSE is 3.46 for the 5-fold cross-fit and only 0.49 for the median estimator). The more complicated the functions are and the more dependence we introduce, the harder it is for the ML methods to learn the functions. This is why assigning more observations to train the functions is especially helpful in such complicated settings. Using 80\% of the observations for training also decreases the bias from a particular sample. We see this since the MSE decreases less when we take the median over the 5-fold sample instead of the 3-fold sample (where we only use 33\% for training).

In RCTs we find that the performance between the T-, DR- and R-learner is competitive. This finding vanishes as soon as selection into treatment is present, leading to an up to two times higher mean squared errors for the T-learner for different estimators compared to the DR-learner (e.g. in setting D, the lowest MSE for the T-learner is 1.21 whereas the lowest MSE for the DR-learner is 0.59). Comparing all meta-learners, we find the lowest MSE (both in terms of bias and variance) for the R-learner while the DR-learner sometimes has slightly higher values. The results are quite similar to findings from \cite{kennedy2020optimal} for the comparison of the T and DR-learner and findings from \cite{knaus2018machine} who compares the DR- and R-learner. 

It is interesting to note that we can confirm findings from previous research with respect to a higher variance when using the lasso. Not only are our results more robust and show no outliers when we exclude the lasso and linear models but they also show a smaller MSE for most of the settings while being competitive in the remaining settings. We also noticed that the combined method can exhibit a higher variance than using cross-fitting. The reason might be that if the pseudo-outcome already has some heavy tails, the combined method incorporates them in the regression. Using a cross-fit method might benefit from averaging these outliers out in the last step. This is interesting since we especially observe this higher variance the smaller the sample size is and if we use the lasso or linear models, that might extrapolate. This is in contrast to the suggestions from \cite{chernozhukov2018double} to use this method in finite samples. The reason might be that when estimating the ATE, the outliers cancel out.

Based on our findings, we recommend to not only rely on a specific sample split but use cross-fitting and multiple iterations over which one takes the median. In our simulation, we use a maximum of 50 iterations and find that after around 20, the result stabilises if using the 5-fold cross-fit estimator. For other estimators, one might need even fewer iterations (e.g. for the combined method) as we show in the supplement in section \ref{sub:plots_50} where we plot the MSE for each estimator over 50 iterations. We also experimented with 100 iterations but found no deviations compared to 50 especially since we take the median. If we have prior knowledge about the structural form ( i.e. whether the data is based on an RCT or an observational study), we should deviate from the 50:50 splitting and instead use more observations for the training of the nuisance functions. The number of observations we should use for training might also depend on how many parameters we have to tune while training the nuisance functions.


\FloatBarrier 
\newpage

\phantomsection \addcontentsline{toc}{section}{References}
\bibliographystyle{plainnat}
\bibliography{references}


\clearpage

\appendix

\section{Additional Plots}

\begin{figure}[ht]
\begin{center}
\includegraphics[width=0.7\textwidth]{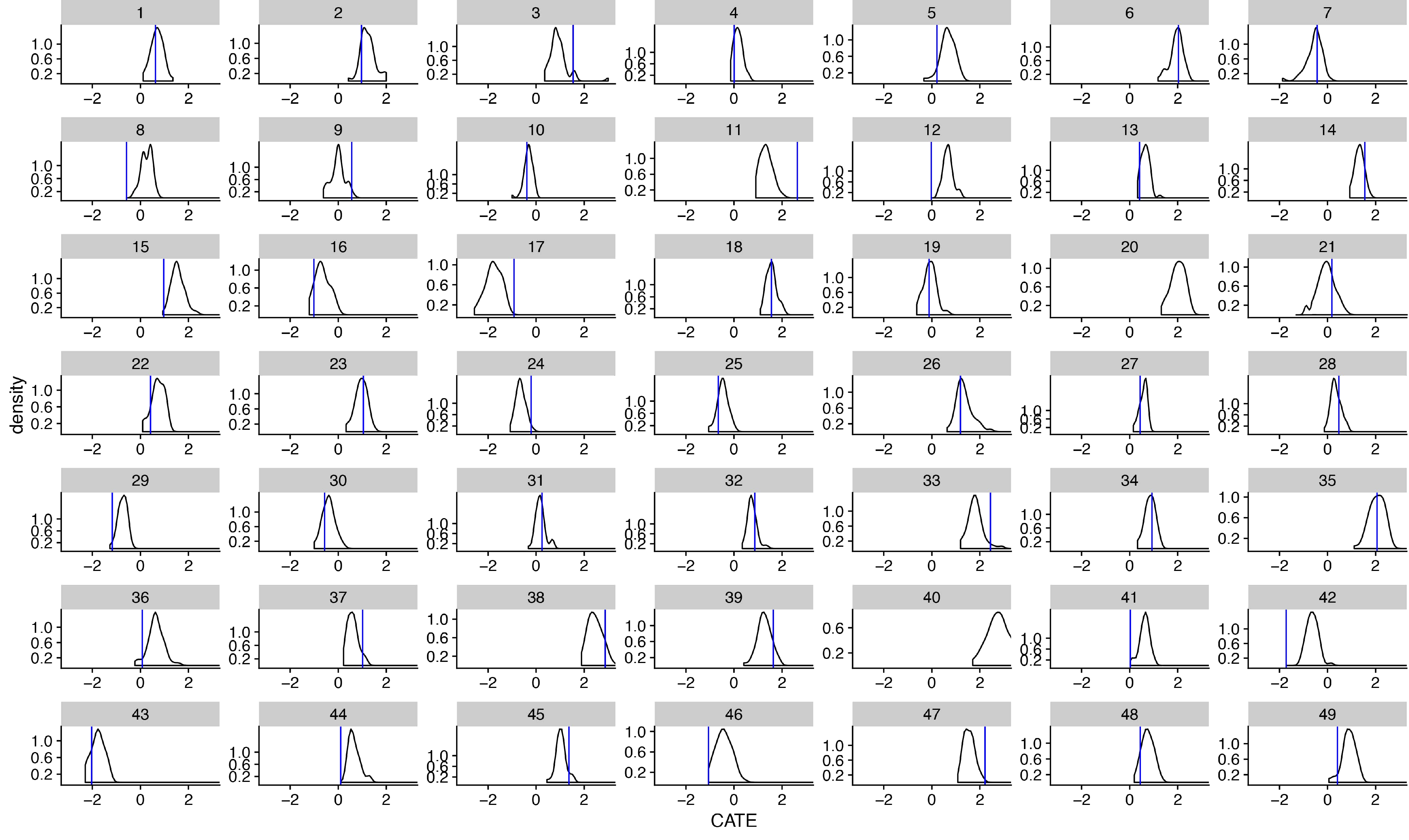}
\caption{Distribution of estimated CATE for randomly selected individuals. }
\label{fig:density49}
\end{center}
\end{figure}

\begin{figure}[ht]
\begin{center}
\includegraphics[scale=0.3]{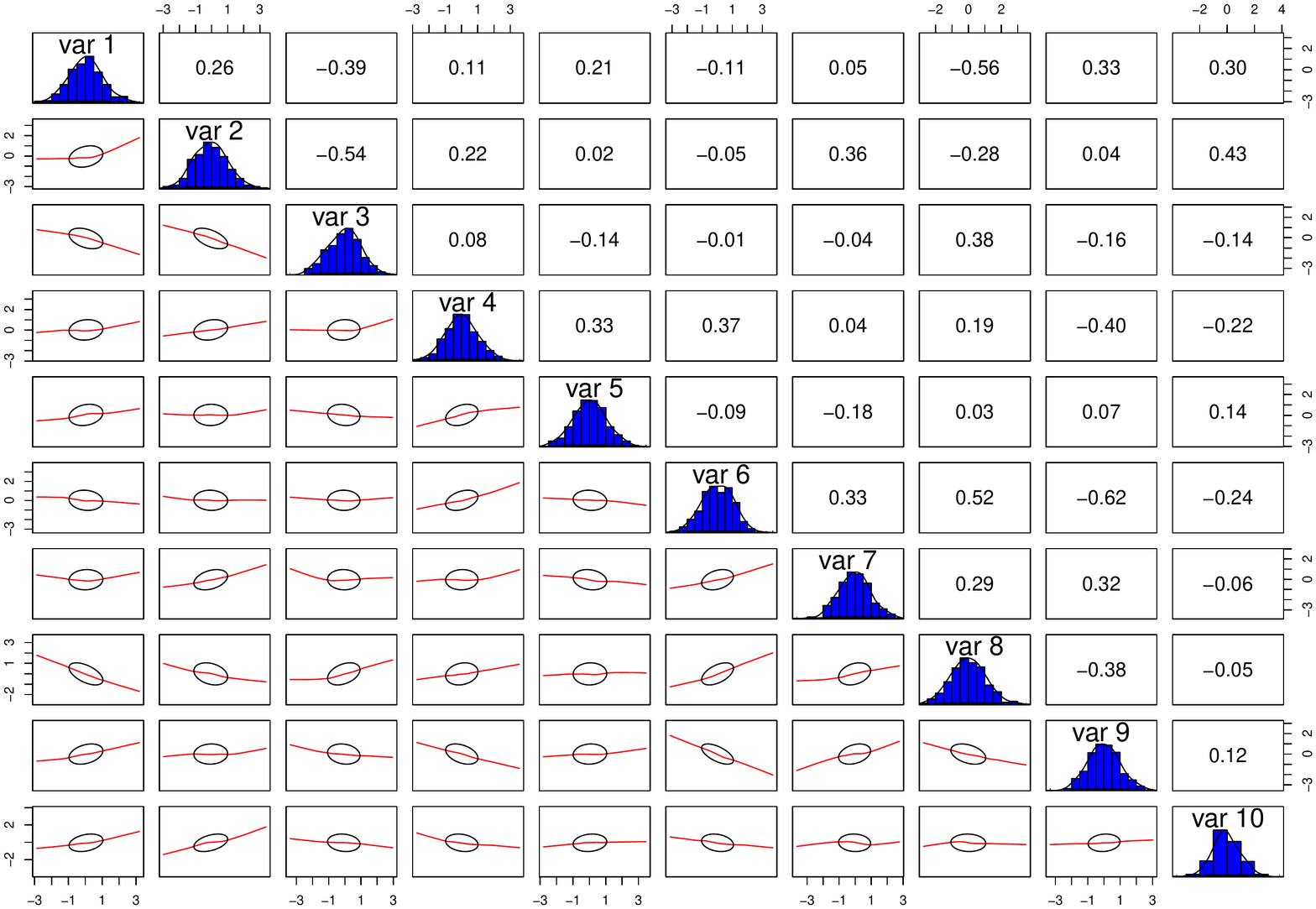}
\caption[Correlation Matrix of Covariates.]{Correlation Matrix of Covariates.}
\label{Cov_matrix}
\end{center}
\end{figure}

\clearpage

\clearpage

\section{Tables}

\subsection{Results with 500 observations and T-learner}

\begin{table}[ht]
\centering
\resizebox{\textwidth}{!}{%
    \begin{threeparttable}
\caption{Performance measures for the considered estimators based on the DR-learner.}
\label{tab:DR-Results_500}

\begin{tabular}{c|rrrrrrrrrrrr}
\hline \hline \\

Scenarios & \multicolumn{4}{c}{G} & \multicolumn{4}{c}{H} & \multicolumn{4}{c}{I}   \\    \\ \hline

 & $\overline{MSE}$ & $\overline{|Bias|}$ & $\overline{SD}$  & \textit{Median MSE}  & $\overline{MSE}$ & $\overline{|Bias|}$ & $\overline{SD}$  & \textit{Median MSE} & $\overline{MSE}$ & $\overline{|Bias|}$ & $\overline{SD}$  & \textit{Median MSE}\\
  \hline
naive & 1.21 & 0.53 & 0.88 & 1.00 & 2.06 & 0.57 & 1.24 & 1.71 & 0.69 & 0.60 & 0.41 & 0.47 \\ 
  50:50 & 2.33 & 0.51 & 1.39 & 2.03 & 3.36 & 0.53 & 1.71 & 2.88 & 1.45 & 0.53 & 0.98 & 1.03 \\ 
  50:50 cross-fit & 2.96 & 0.53 & 1.59 & 2.40 & 2.58 & 0.56 & 1.45 & 2.16 & 3.36 & 0.55 & 1.68 & 2.72 \\ 
  double split & 2.95 & 0.55 & 1.58 & 2.53 & 5.87 & 0.55 & 2.32 & 5.09 & 1.48 & 0.53 & 0.99 & 1.09 \\ 
  double split cross-fit & 1.54 & 0.55 & 1.04 & 1.30 & 2.67 & 0.55 & 1.48 & 2.28 & 0.90 & 0.53 & 0.65 & 0.58 \\ 
  5-fold & 5.87 & 0.56 & 2.32 & 5.14 & 9.94 & 0.57 & 3.07 & 8.36 & 1.33 & 0.50 & 0.95 & 1.03 \\ 
  5-fold cross fit & 1.49 & 0.58 & 0.99 & 1.28 & 2.33 & 0.57 & 1.35 & 2.02 & 0.75 & 0.51 & 0.54 & 0.46 \\ 
  5-fold combined & 	\underline{1.00} & 0.58 & 0.16 & \underline{0.64} & \underline{1.17} & 0.59 & 0.19 & 0.66 & \underline{0.53} & 0.50 & 0.31 &\underline{0.29} \\ 
  50:50 cross-fit median & {0.96} & 0.53 & 0.73 & 0.78 & {1.37} & 0.52 & 0.97 & 1.17 & 0.87 & 0.51 & 0.65 & 0.55 \\ 
  double split cross-fit median & 0.97 & 0.56 & 0.71 & 0.77 & 1.43 & 0.55 & 0.98 & 1.20 & 0.81 & 0.53 & 0.57 & 0.48 \\ 
  5-fold cross-fit median & 1.13 & 0.58 & 0.78 & 0.90 & 1.39 & 0.57 & 0.94 & 1.15 & 0.71 & 0.51 & 0.50 & 0.41 \\ 
  5-fold combined & \underline{1.00} & 0.58 & 0.16 & \underline{0.64} & \underline{1.17} & 0.59 & 0.19 & 0.66 & \underline{0.52} & 0.50 & 0.31 & \underline{0.29} \\

\hline \hline \\
Scenarios & \multicolumn{4}{c}{J} & \multicolumn{4}{c}{K} & \multicolumn{4}{c}{L}   \\    \\ \hline

 & $\overline{MSE}$ & $\overline{|Bias|}$ & $\overline{SD}$  & \textit{Median MSE}  & $\overline{MSE}$ & $\overline{|Bias|}$ & $\overline{SD}$  & \textit{Median MSE} & $\overline{MSE}$ & $\overline{|Bias|}$ & $\overline{SD}$  & \textit{Median MSE}\\
  \hline
naive & 0.99 & 0.71 & 0.60 & 0.95 & \underline{0.70} & 0.59 & 0.42 & 0.43 & 1.35 & 0.23 & 1.13 & 1.06 \\ 
  50:50 & 3.35 & 0.35 & 1.78 & 2.65 & 1.99 & 0.63 & 1.17 & 1.55 & 3.89 & 0.36 & 1.92 & 3.19 \\ 
  50:50 cross-fit & 1.44 & 0.43 & 1.07 & 1.15 & 1.31 & 0.74 & 1.03 & 1.47 & 1.88 & 0.36 & 1.30 & 1.57 \\ 
  double split & 37.49 & 0.41 & 6.10 & 20.61 & 2.20 & 0.68 & 1.22 & 1.78 & 4.74 & 0.51 & 2.09 & 3.93 \\ 
  double split cross-fit & 5.66 & 0.35 & 2.34 & 3.58 & 1.24 & 0.67 & 0.73 & 0.89 & 2.80 & 0.46 & 1.58 & 2.31 \\ 
  5-fold & 4.42 & 0.37 & 2.05 & 3.66 & 1.97 & 0.65 & 1.14 & 1.57 & 7.06 & 0.44 & 2.60 & 5.67 \\ 
  5-fold cross fit & 1.41 & 0.36 & 1.10 & 1.19 & 1.02 & 0.66 & 0.58 & 0.66 & 2.35 & 0.42 & 1.45 & 2.00 \\ 
  5-fold combined & 1.13 & 0.36 & 0.78 & 0.84 & 0.82 & 0.65 & 0.33 & 0.47 & 0.58 & 0.45 & 0.67 & 0.66 \\ 
  50:50 cross-fit median & 1.51 & 0.35 & 1.15 & 1.23 & 1.16 & 0.65 & 0.71 & 0.80 & {1.30} & 0.43 & 1.01 & 1.10 \\ 
  double split cross-fit median & 0.97 & 0.36 & 0.87 & 0.81 & 1.11 & 0.70 & 0.61 & 0.73 & 2.14 & 0.46 & 1.35 & 1.59 \\ 
  5-fold cross-fit median & \underline{0.83} & 0.37 & 0.79 & 0.71 & 0.97 & 0.66 & 0.54 & 0.60 & 1.66 & 0.41 & 1.18 & 1.34 \\ 
  5-fold combined & 1.11 & 0.36 & 0.75 & 0.82 & 0.82 & 0.65 & 0.32 & 0.47 & \underline{0.55} & 0.45 & 0.65 & 0.61 \\ 

\hline \hline \\

\end{tabular}
\begin{tablenotes}
      \small
      \item \textit{Notes:} Number of observations is 500. Averages over 300 repetitions. Lowest MSE values are underlined. If values are similar we underline the median. For the median estimators we take the median over 20 iterations of sample splitting.  
    \end{tablenotes}
  \end{threeparttable}
}
\end{table}

\begin{table}[ht]
\centering
\resizebox{\textwidth}{!}{%
    \begin{threeparttable}
\caption{Performance measures for estimators based on the R-learner.}
\label{tab:R-Results_500}

\begin{tabular}{c|rrrrrrrrrrrr}
\hline \hline \\

Scenarios & \multicolumn{4}{c}{G} & \multicolumn{4}{c}{H} & \multicolumn{4}{c}{I}   \\    \\ \hline

 & $\overline{MSE}$ & $\overline{|Bias|}$ & $\overline{SD}$  & \textit{Median MSE}  & $\overline{MSE}$ & $\overline{|Bias|}$ & $\overline{SD}$  & \textit{Median MSE} & $\overline{MSE}$ & $\overline{|Bias|}$ & $\overline{SD}$  & \textit{Median MSE}\\
  \hline
naive & 1.07 & 0.68 & 0.60 & 0.72 & \underline{1.70} & 0.73 & 0.93 & 1.31 & \underline{0.92} & 0.56 & 0.66 & 0.73 \\ 
  50:50 & 1.77 & 0.52 & 1.16 & 1.53 & 4.94 & 0.54 & 2.12 & 4.06 & 4.41 & 0.54 & 1.97 & 3.64 \\ 
  50:50 cross-fit & 1.09 & 0.51 & 0.83 & 0.92 & 5.73 & 0.56 & 2.29 & 4.28 & 2.50 & 0.51 & 1.41 & 1.95 \\ 
  double split & 3.76 & 0.52 & 1.83 & 3.21 & 29.10 & 0.64 & 5.33 & 15.77 & 8.80 & 0.56 & 2.88 & 6.78 \\ 
  double split cross-fit & 1.40 & 0.51 & 1.00 & 1.20 & 5.43 & 0.64 & 2.19 & 4.47 & 4.63 & 0.56 & 2.02 & 3.58 \\ 
  5-fold & 19.71 & 0.56 & 4.38 & 13.55 & 11.90 & 0.74 & 3.32 & 9.88 & 19.11 & 0.53 & 4.31 & 16.02 \\ 
  5-fold cross fit & 1.83 & 0.54 & 1.17 & 1.57 & 4.02 & 0.65 & 1.83 & 3.33 & 5.00 & 0.54 & 2.12 & 4.05 \\ 
  5-fold combined & 1.26 & 0.49 & 0.94 & 1.06 & 4.36 & 0.55 & 2.00 & 3.58 & 2.52 & 0.46 & 1.46 & 1.97 \\ 
  50:50 cross-fit median & 0.95 & 0.51 & 0.74 & 0.78 & 2.56 & 0.57 & 1.43 & 2.23 & 1.43 & 0.52 & 0.97 & 1.02 \\ 
  double split cross-fit median & 0.95 & 0.53 & 0.72 & 0.76 & 2.28 & 0.62 & 1.30 & 1.96 & 1.62 & 0.54 & 1.05 & 1.18 \\ 
  5-fold cross-fit median & \underline{0.92} & 0.55 & 0.68 & 0.72 & 1.97 & 0.67 & 1.13 & 1.65 & 1.66 & 0.55 & 1.07 & 1.23 \\
  5-fold combined median &   1.26 & 0.49 & 0.94 & 1.06 & 4.50 & 0.55 & 2.04 & 3.70 & 2.29 & 0.45 & 1.38 & 1.76 \\ 

\hline \hline \\
Scenarios & \multicolumn{4}{c}{J} & \multicolumn{4}{c}{K} & \multicolumn{4}{c}{L}   \\    \\ \hline

 & $\overline{MSE}$ & $\overline{|Bias|}$ & $\overline{SD}$  & \textit{Median MSE}  & $\overline{MSE}$ & $\overline{|Bias|}$ & $\overline{SD}$  & \textit{Median MSE} & $\overline{MSE}$ & $\overline{|Bias|}$ & $\overline{SD}$  & \textit{Median MSE}\\
  \hline
naive & 0.78 & 0.39 & 0.72 & 0.79 & \underline{1.44} & 0.52 & 1.01 & 1.16 & \underline{0.47} & 0.07 & 0.68 & 0.39 \\ 
  50:50 & 1.64 & 0.38 & 1.20 & 1.37 & 6.66 & 0.50 & 2.50 & 5.37 & 4.08 & 0.17 & 2.01 & 3.44 \\ 
  50:50 cross-fit & 0.91 & 0.37 & 0.85 & 0.79 & 3.20 & 0.48 & 1.68 & 2.53 & 1.72 & 0.16 & 1.29 & 1.46 \\ 
  double split & 8.52 & 0.35 & 2.89 & 5.47 & 10.79 & 0.40 & 3.24 & 8.89 & 22.31 & 0.38 & 4.70 & 15.79 \\ 
  double split cross-fit & 1.71 & 0.35 & 1.24 & 1.30 & 5.47 & 0.48 & 2.26 & 4.37 & 2.96 & 0.22 & 1.70 & 2.52 \\ 
  5-fold & 3.68 & 0.37 & 1.86 & 3.01 & 106.63 & 0.76 & 10.27 & 59.41 & 19.22 & 0.34 & 4.36 & 15.71 \\ 
  5-fold cross fit & 1.03 & 0.38 & 0.91 & 0.91 & 4.60 & 0.50 & 2.04 & 3.65 & 8.27 & 0.24 & 2.86 & 6.75 \\ 
  5-fold combined & 1.24 & 0.34 & 1.05 & 0.99 & 4.05 & 0.49 & 1.91 & 3.16 & 6.80 & 0.22 & 2.59 & 5.01 \\ 
  50:50 cross-fit median & 0.50 & 0.37 & 0.55 & 0.44 & 1.73 & 0.48 & 1.16 & 1.32 & 1.15 & 0.16 & 1.05 & 0.97 \\ 
  double split cross-fit median & 0.52 & 0.37 & 0.58 & 0.45 & 2.11 & 0.49 & 1.31 & 1.56 & 1.28 & 0.19 & 1.10 & 0.97 \\ 
  5-fold cross-fit median & \underline{0.45} & 0.38 & 0.50 & 0.41 & 1.74 & 0.49 & 1.16 & 1.37 & 1.73 & 0.13 & 1.30 & 1.35 \\ 
  5-fold combined median & 1.12 & 0.34 & 0.99 & 0.91 & 3.91 & 0.49 & 1.88 & 3.06 & 6.17 & 0.23 & 2.47 & 4.46 \\ 

\hline \hline \\

\end{tabular}
\begin{tablenotes}
      \small
      \item \textit{Notes:} Number of observations is 500. Averages over 300 repetitions. Lowest MSE values are underlined. If values are similar we underline the median. For the median estimators we take the median over 20 iterations of sample splitting.  
    \end{tablenotes}
  \end{threeparttable}
}
\end{table}

\begin{table}[ht]
\centering
\resizebox{\textwidth}{!}{%
    \begin{threeparttable}
\caption{Performance measures for the considered estimators based on the X-learner.}
\label{tab:X-Results_500}

\begin{tabular}{c|rrrrrrrrrrrr}
\hline \hline \\

Scenarios & \multicolumn{4}{c}{G} & \multicolumn{4}{c}{H} & \multicolumn{4}{c}{I}   \\    \\ \hline

 & $\overline{MSE}$ & $\overline{|Bias|}$ & $\overline{SD}$  & \textit{Median MSE}  & $\overline{MSE}$ & $\overline{|Bias|}$ & $\overline{SD}$  & \textit{Median MSE} & $\overline{MSE}$ & $\overline{|Bias|}$ & $\overline{SD}$  & \textit{Median MSE}\\
  \hline
naive & \underline{0.70} & 0.54 & 0.49 & 0.45 & 2.35 & 0.48 & 1.41 & 1.99 & \underline{1.28} & 0.64 & 0.74 & 0.80 \\ 
  50:50 & 1.45 & 0.95 & 0.13 & 0.66 & 1.32 & 0.89 & 0.25 & 0.63 & 2.05 & 1.08 & 0.54 & 1.16 \\ 
  50:50 cross-fit & 1.45 & 0.95 & 0.12 & 0.66 & \underline{1.30} & 0.89 & 0.21 & \underline{0.61} & 2.01 & 1.08 & 0.50 & 1.12 \\ 
  double split & 2.44 & 0.97 & 0.98 & 1.72 & 1.33 & 0.89 & 0.28 & 0.64 & 7.58 & 1.13 & 2.37 & 5.84 \\ 
  double split cross-fit & 1.70 & 0.97 & 0.48 & 0.88 &\underline{1.31} & 0.89 & 0.23 &\underline{0.62} & 3.06 & 1.10 & 1.10 & 2.08 \\ 
  5-fold & 3.34 & 0.94 & 1.39 & 2.58 & 4.23 & 0.89 & 1.73 & 3.52 & 5.48 & 1.09 & 1.91 & 4.04 \\ 
  5-fold cross fit & 1.83 & 0.96 & 0.62 & 1.03 & 4.69 & 0.91 & 1.84 & 3.77 & 3.07 & 1.08 & 1.13 & 2.20 \\ 
  5-fold combined & 1.23 & 0.96 & 0.10 & 0.52 & 2.35 & 0.94 & 0.26 & 1.24 & 2.16 & 1.05 & 0.66 & 1.39 \\ 
  50:50 cross-fit median & 1.45 & 0.95 & 0.12 & 0.66 & \underline{1.30} & 0.89 & 0.21 & \underline{0.61} & 2.01 & 1.08 & 0.50 & 1.12 \\ 
  double split cross-fit median & 1.46 & 0.95 & 0.19 & 0.68 & \underline{1.30} & 0.89 & 0.21 & \underline{0.61} & 2.03 & 1.07 & 0.54 & 1.15 \\ 
  5-fold cross-fit median & 1.56 & 0.95 & 0.37 & 0.80 & 1.57 & 0.89 & 0.57 & 0.91 & 2.11 & 1.08 & 0.59 & 1.22 \\
  5-fold combined median & 	1.23 & 0.96 & 0.10 & 0.52 & 2.35 & 0.94 & 0.25 & 1.24 & 2.15 & 1.05 & 0.64 & 1.37 \\ 

\hline \hline \\
Scenarios & \multicolumn{4}{c}{J} & \multicolumn{4}{c}{K} & \multicolumn{4}{c}{L}   \\    \\ \hline

 & $\overline{MSE}$ & $\overline{|Bias|}$ & $\overline{SD}$  & \textit{Median MSE}  & $\overline{MSE}$ & $\overline{|Bias|}$ & $\overline{SD}$  & \textit{Median MSE} & $\overline{MSE}$ & $\overline{|Bias|}$ & $\overline{SD}$  & \textit{Median MSE}\\
  \hline
naive & 0.81 & 0.55 & 0.60 & 0.57 & \underline{1.34} & 0.58 & 0.89 & 0.98 & 0.69 & 0.48 & 0.54 & 0.42 \\ 
  50:50 & 0.62 & 0.55 & 0.43 & 0.42 & 2.78 & 1.28 & 0.57 & 1.59 & 0.24 & 0.19 & 0.45 & 0.25 \\ 
  50:50 cross-fit & {0.58} & 0.54 & 0.39 & 0.39 & 2.71 & 1.28 & 0.51 & 1.52 & 0.20 & 0.17 & 0.41 & 0.21 \\ 
  double split & 1.07 & 0.54 & 0.80 & 0.90 & 3.36 & 1.24 & 1.02 & 2.24 & 1.49 & 0.19 & 1.20 & 1.17 \\ 
  double split cross-fit & 0.70 & 0.54 & 0.52 & 0.53 & 3.35 & 1.25 & 1.01 & 2.27 & 0.55 & 0.17 & 0.72 & 0.48 \\ 
  5-fold & 1.15 & 0.56 & 0.84 & 0.98 & 4.86 & 1.31 & 1.51 & 3.66 & 2.67 & 0.23 & 1.62 & 2.08 \\ 
  5-fold cross fit & 0.68 & 0.54 & 0.50 & 0.50 & 5.76 & 1.28 & 1.81 & 4.09 & 0.76 & 0.17 & 0.86 & 0.58 \\ 
  5-fold combined & \underline{0.54} & 0.54 & 0.36 & \underline{0.35} & 4.63 & 1.26 & 1.03 & 2.94 & \underline{0.19} & 0.18 & 0.39 & \underline{0.19} \\ 
  50:50 cross-fit median & {0.58} & 0.54 & 0.38 & 0.39 & 2.70 & 1.28 & 0.50 & 1.51 & \underline{0.19} & 0.17 & 0.40 & 0.20 \\ 
  double split cross-fit median & 0.59 & 0.54 & 0.40 & 0.40 & 2.67 & 1.25 & 0.55 & 1.52 & 0.23 & 0.18 & 0.45 & 0.23 \\ 
  5-fold cross-fit median & 0.61 & 0.55 & 0.41 & 0.44 & 2.87 & 1.30 & 0.58 & 1.66 & 0.30 & 0.18 & 0.51 & 0.27 \\ 
  5-fold combined median & \underline{0.53} & 0.54 & 0.34 & \underline{0.34} & 4.62 & 1.26 & 1.00 & 2.92 & \underline{0.18} & 0.18 & 0.38 & \underline{0.18} \\ 

\hline \hline \\

\end{tabular}
\begin{tablenotes}
      \small
      \item \textit{Notes:} Number of observations is 500. Averages over 300 repetitions. Lowest MSE values are underlined. If values are similar we underline the median. For the median estimators we take the median over 20 iterations of sample splitting.  
    \end{tablenotes}
  \end{threeparttable}
}
\end{table}

\begin{table}[ht]
\centering
\resizebox{\textwidth}{!}{%
    \begin{threeparttable}
\caption{Performance measures for estimators based on the T-learner.}
\label{tab:T-Results}

\begin{tabular}{c|rrrrrrrrrrrr}
\hline \hline \\

Scenarios & \multicolumn{4}{c}{A} & \multicolumn{4}{c}{B} & \multicolumn{4}{c}{C}   \\    \\ \hline

 & $\overline{MSE}$ & $\overline{|Bias|}$ & $\overline{SD}$  & \textit{Median MSE}  & $\overline{MSE}$ & $\overline{|Bias|}$ & $\overline{SD}$  & \textit{Median MSE} & $\overline{MSE}$ & $\overline{|Bias|}$ & $\overline{SD}$  & \textit{Median MSE}\\
  \hline
naive & 2.01 & 1.04 & 0.57 & 1.11 & 1.83 & 0.97 & 0.60 & 1.00 & 1.26 & 0.83 & 0.46 & 0.80 \\ 
  50:50 & 0.67 & 0.49 & 0.55 & 0.49 & 0.84 & 0.49 & 0.68 & 0.66 & 1.27 & 0.72 & 0.61 & 0.70 \\ 
  50:50 cross-fit & 0.55 & 0.49 & 0.42 & 0.36 & \underline{0.59} & 0.49 & 0.46 & 0.40 & 1.16 & 0.72 & 0.52 & 0.60 \\ 
  5-fold & 0.59 & 0.48 & 0.48 & 0.40 & 0.80 & 0.49 & 0.64 & 0.61 & 1.18 & 0.70 & 0.58 & 0.63 \\ 
  5-fold cross fit & \underline{0.54} & 0.48 & 0.42 & 0.35 & 0.63 & 0.49 & 0.50 & 0.44 & \underline{1.14} & 0.70 & 0.54 & 0.59 \\ 
  50:50 cross-fit median & \underline{0.53} & 0.49 & 0.40 & \underline{0.33} & 0.63 & 0.49 & 0.50 & 0.45 & \underline{1.15} & 0.72 & 0.51 & 0.59 \\ 
  5-fold cross-fit median & \underline{0.54} & 0.48 & 0.42 & 0.35 & 0.62 & 0.49 & 0.48 & 0.41 & \underline{1.14} & 0.70 & 0.54 & \underline{0.57} \\ 

\hline \hline \\
Scenarios & \multicolumn{4}{c}{D} & \multicolumn{4}{c}{E} & \multicolumn{4}{c}{F}   \\    \\ \hline

 & $\overline{MSE}$ & $\overline{|Bias|}$ & $\overline{SD}$  & \textit{Median MSE}  & $\overline{MSE}$ & $\overline{|Bias|}$ & $\overline{SD}$  & \textit{Median MSE} & $\overline{MSE}$ & $\overline{|Bias|}$ & $\overline{SD}$  & \textit{Median MSE}\\
  \hline
naive & 2.78 & 1.50 & 0.52 & 4.02 & \underline{1.30} & 0.80 & 0.59 & 0.88 & \underline{0.19} & 0.03 & 0.44 & 0.16 \\ 
  50:50 & 1.46 & 0.78 & 0.66 & 0.77 & 1.96 & 0.97 & 0.65 & 0.99 & 0.98 & 0.66 & 0.55 & 0.59 \\ 
  50:50 cross-fit & 1.32 & 0.78 & 0.55 & 0.62 & 1.84 & 0.97 & 0.55 & 0.87 & 0.90 & 0.67 & 0.45 & 0.50 \\ 
  5-fold & 1.27 & 0.73 & 0.58 & 0.60 & 1.78 & 0.93 & 0.58 & 0.84 & 0.85 & 0.62 & 0.49 & 0.48 \\ 
  5-fold cross fit & \underline{1.21} & 0.73 & \underline{0.53} & 0.53 & 1.73 & 0.93 & 0.54 & 0.80 & 0.81 & 0.62 & 0.45 & 0.44 \\ 
  50:50 cross-fit median & 1.31 & 0.78 & 0.54 & 0.61 & 1.83 & 0.97 & 0.53 & 0.85 & 0.89 & 0.67 & 0.44 & 0.49 \\ 
  5-fold cross-fit median & \underline{1.21} & 0.73 & \underline{0.53} & 0.53 & 1.73 & 0.93 & 0.54 & 0.80 & 0.81 & 0.62 & 0.45 & 0.45 \\ 
  
  \hline \hline \\
Scenarios & \multicolumn{4}{c}{G} & \multicolumn{4}{c}{H} & \multicolumn{4}{c}{I}   \\    \\ \hline

 & $\overline{MSE}$ & $\overline{|Bias|}$ & $\overline{SD}$  & \textit{Median MSE}  & $\overline{MSE}$ & $\overline{|Bias|}$ & $\overline{SD}$  & \textit{Median MSE} & $\overline{MSE}$ & $\overline{|Bias|}$ & $\overline{SD}$  & \textit{Median MSE}\\
 \hline
naive & 3.50 & 1.11 & 1.27 & 2.59 & 4.57 & 1.05 & 1.68 & 3.67 & 1.36 & 0.82 & 0.63 & 0.99 \\ 
  50:50 & 1.63 & 0.51 & 1.11 & 1.42 & 3.20 & 0.49 & 1.68 & 2.80 & 1.32 & 0.68 & 0.73 & 0.80 \\ 
  50:50 cross-fit & 1.09 & 0.50 & 0.84 & 0.91 & 1.87 & 0.49 & 1.22 & 1.66 & 1.16 & 0.68 & 0.61 & 0.63 \\ 
  5-fold & 0.94 & 0.50 & 0.74 & 0.76 & 1.40 & 0.50 & 1.00 & 1.21 & 1.27 & 0.68 & 0.70 & 0.74 \\ 
  5-fold cross fit & \underline{0.84} & 0.50 & 0.67 & 0.66 & 1.19 & 0.49 & 0.90 & 1.01 & 1.18 & 0.67 & 0.64 & 0.66 \\ 
  50:50 cross-fit median & 0.88 & 0.50 & 0.71 & 0.71 & \underline{1.16} & 0.49 & 0.88 & 0.99 & \underline{1.13} & 0.68 & 0.59 & 0.60 \\ 
  5-fold cross-fit median & 0.86 & 0.50 & 0.69 & 0.68 & 1.18 & 0.49 & 0.90 & 1.01 & 1.18 & 0.67 & 0.63 & 0.66 \\ 
 
 \hline \hline \\
Scenarios & \multicolumn{4}{c}{J} & \multicolumn{4}{c}{K} & \multicolumn{4}{c}{L}   \\    \\ \hline

 & $\overline{MSE}$ & $\overline{|Bias|}$ & $\overline{SD}$  & \textit{Median MSE}  & $\overline{MSE}$ & $\overline{|Bias|}$ & $\overline{SD}$  & \textit{Median MSE} & $\overline{MSE}$ & $\overline{|Bias|}$ & $\overline{SD}$  & \textit{Median MSE}\\
 \hline
naive & 3.88 & 1.50 & 1.17 & 4.33 & \underline{1.37} & 0.80 & 0.63 & 0.91 & \underline{0.98} & 0.03 & 0.99 & 0.78 \\ 
  50:50 & 2.97 & 0.70 & 1.48 & 2.48 & 1.84 & 0.94 & 0.71 & 1.23 & 2.29 & 0.84 & 1.10 & 1.70 \\ 
  50:50 cross-fit & 2.14 & 0.69 & 1.17 & 1.71 & 1.63 & 0.93 & 0.57 & 1.02 & 1.63 & 0.83 & 0.77 & 1.07 \\ 
  5-fold & 1.75 & 0.66 & 1.02 & 1.35 & 1.67 & 0.90 & 0.67 & 1.08 & 1.83 & 0.73 & 1.01 & 1.35 \\ 
  5-fold cross fit & \underline{1.56} & 0.67 & 0.92 &\underline{1.16} & 1.60 & 0.90 & 0.61 & 0.99 & 1.38 & 0.75 & 0.72 & 0.93 \\ 
  50:50 cross-fit median & 1.76 & 0.69 & 0.99 & 1.34 & 1.61 & 0.93 & 0.56 & 0.99 & 1.44 & 0.81 & 0.66 & 0.90 \\ 
  5-fold cross-fit median & \underline{1.56} & 0.67 & 0.91 & \underline{1.16} & 1.60 & 0.90 & 0.61 & 0.99 & 1.38 & 0.76 & 0.71 & 0.92 \\  

\hline \hline \\

\end{tabular}
\begin{tablenotes}
      \small
      \item \textit{Notes:} Setting A to F has 2000 observations and G to F has 500 observations. Averages over 300 repetitions. Lowest MSE values are underlined. If values are similar we underline the median. For the median estimators we take the median over 20 iterations of sample splitting.  
    \end{tablenotes}
  \end{threeparttable}
}
\end{table}

\clearpage

\subsection{Excluding lasso and linear models} \label{sec:R-learner_w/o_Lasso}

The following tables show results for the DR- and R-learner when the lasso algorithm and all linear models are excluded from the stacking method for all functions. This is motivated by the appearance of extreme outliers, especially for settings with 500 observations. We find that when we exclude the aforementioned models the results are more robust in terms of outliers. In RCTs with 2000 observations, the results are competitive or slightly worse. When we only take 500 observations or when we deviate from an RCT, the results are either competitive or better without the lasso and other linear models.

\begin{table}[ht]
\centering
\resizebox{\textwidth}{!}{%
    \begin{threeparttable}
\caption{Performance measures for estimators based on the DR-learner without lasso and linear models.}
\label{tab:DR-Results_without_LASSO_2000}

\begin{tabular}{c|rrrrrrrrrrrr}
\hline \hline \\

Scenarios & \multicolumn{4}{c}{A} & \multicolumn{4}{c}{B} & \multicolumn{4}{c}{C}   \\    \\ \hline

 & $\overline{MSE}$ & $\overline{|Bias|}$ & $\overline{SD}$  & \textit{Median MSE}  & $\overline{MSE}$ & $\overline{|Bias|}$ & $\overline{SD}$  & \textit{Median MSE} & $\overline{MSE}$ & $\overline{|Bias|}$ & $\overline{SD}$  & \textit{Median MSE}\\
  \hline
naive & 0.85 & 0.69 & 0.32 & 0.43 & 1.66 & 0.97 & 0.37 & 0.77 & 0.67 & 0.62 & 0.30 & 0.40 \\ 
  50:50 & 0.82 & 0.49 & 0.66 & 0.58 & 1.61 & 0.46 & 1.13 & 1.27 & 1.10 & 0.48 & 0.85 & 0.74 \\ 
 50:50 cross-fit & 0.61 & 0.48 & 0.49 & 0.39 & 0.94 & 0.44 & 0.79 & 0.75 & 0.63 & 0.37 & 0.63 & 0.44 \\ 
  double split & 0.94 & 0.52 & 0.72 & 0.68 & 2.00 & 0.48 & 1.28 & 1.56 & 1.16 & 0.52 & 0.85 & 0.80 \\ 
  double split cross-fit & 0.63 & 0.51 & 0.46 & 0.40 & 0.92 & 0.48 & 0.75 & 0.73 & 0.76 & 0.53 & 0.57 & 0.48 \\ 
  5-fold & 0.95 & 0.54 & 0.70 & 0.68 & 1.72 & 0.54 & 1.12 & 1.37 & 1.13 & 0.50 & 0.86 & 0.82 \\ 
  5-fold cross fit & 0.62 & 0.54 & 0.39 & 0.36 & 0.77 & 0.54 & 0.56 & 0.54 & 0.64 & 0.50 & 0.50 & 0.42 \\ 
  5-fold combined & 0.65 & 0.47 & 0.42 & 0.45 & 1.04 & 0.44 & 0.63 & 0.85 & 0.92 & 0.46 & 0.52 & 0.60 \\ 
  50:50 cross-fit median & \underline{0.58} & 0.49 & 0.44 & 0.37 & 0.75 & 0.46 & 0.64 & 0.59 & 0.72 & 0.50 & 0.57 & 0.44 \\ 
  double split cross-fit median & \underline{0.59} & 0.51 & 0.42 & 0.37 & 0.72 & 0.48 & 0.60 & 0.55 & 0.70 & 0.53 & 0.52 & 0.43 \\ 
  5-fold cross-fit median & 0.60 & 0.54 & 0.37 &\underline{ 0.34} & \underline{0.70} & 0.54 & 0.50 & 0.47 & \underline{0.48} & 0.50 & 0.47 & 0.39 \\ 
  5-fold combined median & 0.62 & 0.47 & 0.53 & 0.43 & 0.92 & 0.44 & 0.79 & 0.75 & 0.85 & 0.46 & 0.70 & 0.54 \\ 

\hline \hline \\
Scenarios & \multicolumn{4}{c}{D} & \multicolumn{4}{c}{E} & \multicolumn{4}{c}{F}   \\    \\ \hline

 & $\overline{MSE}$ & $\overline{|Bias|}$ & $\overline{SD}$  & \textit{Median MSE}  & $\overline{MSE}$ & $\overline{|Bias|}$ & $\overline{SD}$  & \textit{Median MSE} & $\overline{MSE}$ & $\overline{|Bias|}$ & $\overline{SD}$  & \textit{Median MSE}\\
  \hline
naive & 1.06 & 0.89 & 0.3 & 0.91 & \underline{0.57} & 0.56 & 0.32 & 0.36 & \underline{0.14} & 0.17 & 0.27 & 0.07 \\ 
  50:50 & 1.08 & 0.42 & 0.84 & 0.61 & 1.83 & 0.69 & 0.95 & 0.96 & 0.70 & 0.31 & 0.74 & 0.50 \\ 
  50:50 cross-fit & 0.65 & 0.40 & 0.59 & 0.39 & 1.37 & 0.66 & 0.76 & 0.72 & 0.49 & 0.33 & 0.57 & 0.34 \\
  double split & 1.07 & 0.47 & 0.80 & 0.66 & 1.84 & 0.70 & 0.98 & 1.08 & 0.82 & 0.33 & 0.80 & 0.60 \\ 
  double split cross-fit & 0.71 & 0.46 & 0.55 & 0.36 & 1.40 & 0.71 & 0.69 & 0.64 & 0.41 & 0.32 & 0.49 & 0.28 \\ 
  5-fold & 1.09 & 0.42 & 0.88 & 0.71 & 1.71 & 0.65 & 0.98 & 1.09 & 0.74 & 0.28 & 0.78 & 0.59 \\ 
  5-fold cross fit & 0.54 & 0.41 & 0.48 & 0.29 & 1.11 & 0.65 & 0.60 & 0.53 & 0.29 & 0.28 & 0.41 & 0.21 \\ 
  5-fold combined & 0.96 & 0.40 & 0.51 & 0.52 & 1.75 & 0.68 & 0.65 & 0.84 & 0.61 & 0.28 & 0.43 & 0.41 \\ 
  50:50 cross-fit median & 0.69 & 0.43 & 0.55 & 0.32 & 1.40 & 0.70 & 0.68 & 0.59 & 0.39 & 0.31 & 0.48 & 0.25 \\ 
  double split cross-fit median & 0.67 & 0.46 & 0.50 & 0.30 & 1.33 & 0.72 & 0.62 & 0.55 & 0.36 & 0.33 & 0.43 & 0.22 \\ 
  5-fold cross-fit median & \underline{0.50}& 0.42 & 0.44 & 0.26 & 1.08 & 0.66 & 0.56 & 0.49 & 0.26 & 0.28 & 0.37 & 0.18 \\ 
  5-fold combined median & 0.87 & 0.41 & 0.70 & 0.44 & 1.67 & 0.68 & 0.84 & 0.75 & 0.55 & 0.29 & 0.63 & 0.36 \\  

\hline \hline \\

\end{tabular}
\begin{tablenotes}
      \small
      \item \textit{Notes:} Number of observations is 2000. Averages over 300 repetitions. Lowest MSE values are underlined. If values are similar we underline the median. For the median estimators we take the median over 20 iterations of sample splitting.  
    \end{tablenotes}
  \end{threeparttable}
}
\end{table}

\begin{table}[ht]
\centering
\resizebox{\textwidth}{!}{%
    \begin{threeparttable}
\caption{Performance measures for estimators based on the DR-learner without lasso and linear models.}
\label{tab:DR-Results_without_LASSO_500}

\begin{tabular}{c|rrrrrrrrrrrr}
\hline \hline \\

Scenarios & \multicolumn{4}{c}{G} & \multicolumn{4}{c}{H} & \multicolumn{4}{c}{I}   \\    \\ \hline

 & $\overline{MSE}$ & $\overline{|Bias|}$ & $\overline{SD}$  & \textit{Median MSE}  & $\overline{MSE}$ & $\overline{|Bias|}$ & $\overline{SD}$  & \textit{Median MSE} & $\overline{MSE}$ & $\overline{|Bias|}$ & $\overline{SD}$  & \textit{Median MSE}\\
  \hline
naive & 1.14 & 0.78 & 0.45 & 0.64 & 1.85 & 1.01 & 0.50 & 0.93 & 0.79 & 0.65 & 0.41 & 0.51 \\ 
  50:50 & 1.53 & 0.61 & 0.98 & 1.11 & 4.34 & 0.57 & 1.96 & 3.32 & 1.50 & 0.56 & 1.01 & 1.11 \\ 
  50:50 cross-fit & 0.95 & 0.54 & 0.69 & 0.70 & 2.54 & 0.52 & 1.46 & 1.99 & 0.97 & 0.52 & 0.73 & 0.68 \\ 
  double split & 1.77 & 0.64 & 1.07 & 1.36 & 5.84 & 0.61 & 2.30 & 4.47 & 1.64 & 0.60 & 1.03 & 1.26 \\ 
  double split cross-fit & 1.04 & 0.65 & 0.63 & 0.70 & 2.13 & 0.64 & 1.23 & 1.76 & 1.00 & 0.59 & 0.66 & 0.72 \\ 
  5-fold & 1.69 & 0.71 & 0.96 & 1.26 & 3.57 & 0.71 & 1.67 & 2.95 & 1.42 & 0.58 & 0.95 & 1.15 \\ 
  5-fold cross fit & 0.96 & 0.69 & 0.47 & 0.57 & 1.35 & 0.72 & 0.76 & 0.99 & 0.83 & 0.58 & 0.55 & 0.56 \\ 
  5-fold combined & 1.07 & 0.54 & 0.54 & 0.78 & 2.48 & 0.47 & 0.90 & 1.97 & 1.32 & 0.53 & 0.58 & 1.01 \\ 
  50:50 cross-fit median & \underline{0.89} & 0.59 & 0.59 & 0.60 & 1.36 & 0.57 & 0.92 & 1.10 & 0.92 & 0.57 & 0.64 & 0.66 \\ 
  double split cross-fit median & \underline{0.90}& 0.64 & 0.52 & 0.57 & 1.24 & 0.65 & 0.78 & 0.94 & 0.88 & 0.60 & 0.56 & 0.59 \\ 
  5-fold cross-fit median & 0.91 & 0.69 & 0.42 & \underline{0.52} & \underline{1.12} & 0.72 & 0.58 & 0.74 & \underline{0.77} & 0.58 & 0.49 & 0.50 \\ 
  5-fold combined median & 1.00 & 0.54 & 0.74 & 0.73 & 1.91 & 0.48 & 1.24 & 1.57 & 1.18 & 0.54 & 0.86 & 0.90 \\

\hline \hline \\
Scenarios & \multicolumn{4}{c}{J} & \multicolumn{4}{c}{K} & \multicolumn{4}{c}{L}   \\    \\ \hline

 & $\overline{MSE}$ & $\overline{|Bias|}$ & $\overline{SD}$  & \textit{Median MSE}  & $\overline{MSE}$ & $\overline{|Bias|}$ & $\overline{SD}$  & \textit{Median MSE} & $\overline{MSE}$ & $\overline{|Bias|}$ & $\overline{SD}$  & \textit{Median MSE}\\
  \hline
naive & 0.90 & 0.78 & 0.40 & 0.80 & \underline{0.70} & 0.60 & 0.41 & 0.47 & \underline{0.22} & 0.22 & 0.38 & 0.15 \\ 
  50:50 & 1.30 & 0.44 & 1.00 & 0.96 & 1.89 & 0.66 & 1.09 & 1.39 & 0.91 & 0.32 & 0.86 & 0.69 \\ 
  50:50 cross-fit & 0.87 & 0.47 & 0.71 & 0.67 & 1.33 & 0.64 & 0.80 & 0.88 & 0.62 & 0.32 & 0.67 & 0.46 \\
  double split & 1.51 & 0.47 & 1.08 & 1.19 & 2.25 & 0.71 & 1.20 & 1.75 & 1.07 & 0.33 & 0.95 & 0.82 \\ 
  double split cross-fit & 0.79 & 0.48 & 0.66 & 0.63 & 1.29 & 0.70 & 0.73 & 0.87 & 0.51 & 0.34 & 0.58 & 0.40 \\ 
  5-fold & 1.27 & 0.44 & 0.98 & 1.10 & 1.97 & 0.63 & 1.16 & 1.51 & 0.85 & 0.27 & 0.86 & 0.73 \\ 
  5-fold cross fit & 0.56 & 0.43 & 0.52 & 0.45 & 1.05 & 0.65 & 0.61 & 0.66 & 0.32 & 0.27 & 0.45 & 0.26 \\ 
  5-fold combined & 1.11 & 0.40 & 0.56 & 0.79 & 1.73 & 0.62 & 0.66 & 1.19 & 0.90 & 0.35 & 0.47 & 0.66 \\ 
  50:50 cross-fit median & 0.72 & 0.46 & 0.63 & 0.54 & 1.22 & 0.67 & 0.71 & 0.78 & 0.48 & 0.34 & 0.55 & 0.36 \\ 
  double split cross-fit median & 0.68 & 0.50 & 0.56 & 0.52 & 1.15 & 0.70 & 0.62 & 0.73 & 0.41 & 0.35 & 0.47 & 0.30 \\ 
  5-fold cross-fit median & \underline{0.52} & 0.43 & 0.48 & 0.40 & 0.99 & 0.66 & 0.56 & 0.60 & 0.27 & 0.28 & 0.39 & 0.21 \\ 
  5-fold combined median & 0.97 & 0.41 & 0.83 & 0.70 & 1.54 & 0.62 & 0.94 & 1.05 & 0.78 & 0.34 & 0.77 & 0.57 \\ 
\hline \hline \\

\end{tabular}
\begin{tablenotes}
      \small
      \item \textit{Notes:} Number of observations is 500. Averages over 300 repetitions. Lowest MSE values are underlined. If values are similar we underline the median. For the median estimators we take the median over 20 iterations of sample splitting.  
      \end{tablenotes}
  \end{threeparttable}
}
\end{table}

\begin{table}[ht]
\centering
\resizebox{\textwidth}{!}{%
    \begin{threeparttable}
\caption{Performance measures for estimators based on the R-learner without lasso and linear models.}
\label{tab:R-Results_without_LASSO_2000}

\begin{tabular}{c|rrrrrrrrrrrr}
\hline \hline \\

Scenarios & \multicolumn{4}{c}{A} & \multicolumn{4}{c}{B} & \multicolumn{4}{c}{C}   \\    \\ \hline

 & $\overline{MSE}$ & $\overline{|Bias|}$ & $\overline{SD}$  & \textit{Median MSE}  & $\overline{MSE}$ & $\overline{|Bias|}$ & $\overline{SD}$  & \textit{Median MSE} & $\overline{MSE}$ & $\overline{|Bias|}$ & $\overline{SD}$  & \textit{Median MSE}\\
  \hline
naive & 0.64 & 0.44 & 0.58 & 0.48 & 1.25 & 0.60 & 0.78 & 0.77 & 2.45 & 0.60 & 1.32 & 1.36 \\ 
  50:50 & 0.73 & 0.51 & 0.56 & 0.50 & 1.40 & 0.69 & 0.76 & 0.82 & 2.04 & 0.44 & 1.28 & 1.23 \\ 
  50:50 cross-fit & 0.60 & 0.51 & 0.44 & 0.37 & 1.15 & 0.69 & 0.57 & 0.59 & 1.25 & 0.43 & 0.93 & 0.70 \\ 
  double split & 0.82 & 0.52 & 0.62 & 0.57 & 1.60 & 0.74 & 0.83 & 0.98 & 2.05 & 0.44 & 1.29 & 1.28 \\ 
  double split cross-fit & 0.62 & 0.53 & 0.42 & 0.37 & 1.17 & 0.73 & 0.53 & 0.60 & 1.03 & 0.43 & 0.81 & 0.58 \\ 
  5-fold & 0.89 & 0.56 & 0.62 & 0.60 & 1.43 & 0.69 & 0.81 & 0.94 & 1.75 & 0.37 & 1.21 & 1.26 \\ 
  5-fold cross fit & 0.63 & 0.56 & 0.37 & 0.35 & 0.98 & 0.69 & 0.44 & 0.50 & 0.61 & 0.36 & 0.59 & 0.38 \\ 
  5-fold combined & 0.62 & 0.48 & 0.50 & 0.41 & 1.28 & 0.68 & 0.68 & 0.68 & 1.87 & 0.47 & 1.18 & 1.04 \\ 
  50:50 cross-fit median & \underline{0.58} & 0.51 & 0.41 & 0.35 & 1.10 & 0.69 & 0.53 & 0.55 & 1.09 & 0.43 & 0.84 & 0.58 \\ 
  double split cross-fit median & 0.60 & 0.53 & 0.39 & 0.35 & 1.12 & 0.73 & 0.48 & 0.55 & 0.92 & 0.43 & 0.73 & 0.49 \\ 
  5-fold cross-fit median & 0.62 & 0.56 & 0.35 & 0.34 & \underline{0.96} & 0.69 & 0.42 & 0.48 &\underline{0.55} & 0.36 & 0.53 & 0.32 \\ 
  5-fold combined median & 0.60 & 0.48 & 0.48 & 0.39 & 1.24 & 0.68 & 0.65 & 0.65 & 1.70 & 0.47 & 1.11 & 0.93 \\ 

\hline \hline \\
Scenarios & \multicolumn{4}{c}{D} & \multicolumn{4}{c}{E} & \multicolumn{4}{c}{F}   \\    \\ \hline

 & $\overline{MSE}$ & $\overline{|Bias|}$ & $\overline{SD}$  & \textit{Median MSE}  & $\overline{MSE}$ & $\overline{|Bias|}$ & $\overline{SD}$  & \textit{Median MSE} & $\overline{MSE}$ & $\overline{|Bias|}$ & $\overline{SD}$  & \textit{Median MSE}\\
  \hline
naive & 1.82 & 0.59 & 1.09 & 1.02 & 2.30 & 0.54 & 1.29 & 1.28 & 2.23 & 0.48 & 1.29 & 1.10 \\ 
  50:50 & 1.38 & 0.39 & 1.02 & 0.78 & 1.72 & 0.39 & 1.20 & 1.07 & 1.82 & 0.35 & 1.24 & 1.01 \\ 
  50:50 cross-fit & 0.88 & 0.4 & 0.74 & 0.44 & 1.02 & 0.39 & 0.86 & 0.63 & 1.03 & 0.35 & 0.87 & 0.54 \\ 
  double split & 1.46 & 0.41 & 1.06 & 0.86 & 1.74 & 0.42 & 1.19 & 1.15 & 1.78 & 0.34 & 1.24 & 1.05 \\ 
  double split cross-fit & 0.78 & 0.42 & 0.66 & 0.38 & 0.87 & 0.41 & 0.75 & 0.53 & 0.80 & 0.33 & 0.75 & 0.42 \\ 
  5-fold & 1.20 & 0.34 & 1.01 & 0.84 & 1.49 & 0.37 & 1.12 & 1.13 & 1.05 & 0.16 & 1.00 & 0.75 \\ 
  5-fold cross fit & 0.42 & 0.33 & 0.48 & 0.25 & 0.54 & 0.39 & 0.55 & 0.37 & 0.26 & 0.16 & 0.45 & 0.17 \\ 
  5-fold combined & 1.29 & 0.40 & 0.96 & 0.67 & 1.57 & 0.39 & 1.12 & 0.93 & 1.78 & 0.40 & 1.18 & 0.89 \\ 
  50:50 cross-fit median & 0.78 & 0.40 & 0.67 & 0.36 & 0.88 & 0.39 & 0.77 & 0.51 & 0.88 & 0.35 & 0.78 & 0.43 \\ 
  double split cross-fit median & 0.70 & 0.42 & 0.59 & 0.31 & 0.75 & 0.42 & 0.66 & 0.44 & 0.67 & 0.33 & 0.65 & 0.32 \\ 
  5-fold cross-fit median & \underline{0.37}& 0.33 & 0.43 & 0.22 & \underline{0.47} & 0.38 & 0.49 & 0.31 & \underline{0.20} & 0.17 & 0.37 & 0.12 \\ 
  5-fold combined median & 1.18 & 0.40 & 0.90 & 0.59 & 1.43 & 0.39 & 1.06 & 0.83 & 1.62 & 0.40 & 1.10 & 0.77 \\ 

\hline \hline \\

\end{tabular}
\begin{tablenotes}
      \small
      \item \textit{Notes:} Number of observations is 2000. Averages over 300 repetitions. Lowest MSE values are underlined. If values are similar we underline the median. For the median estimators we take the median over 20 iterations of sample splitting.  
    \end{tablenotes}
  \end{threeparttable}
}
\end{table}

\begin{table}[ht]
\centering
\resizebox{\textwidth}{!}{%
    \begin{threeparttable}
\caption{Performance measures for estimators based on the R-learner without lasso and linear models.}
\label{tab:R-Results_without_LASSO_500}

\begin{tabular}{c|rrrrrrrrrrrr}
\hline \hline \\

Scenarios & \multicolumn{4}{c}{G} & \multicolumn{4}{c}{H} & \multicolumn{4}{c}{I}   \\    \\ \hline

 & $\overline{MSE}$ & $\overline{|Bias|}$ & $\overline{SD}$  & \textit{Median MSE}  & $\overline{MSE}$ & $\overline{|Bias|}$ & $\overline{SD}$  & \textit{Median MSE} & $\overline{MSE}$ & $\overline{|Bias|}$ & $\overline{SD}$  & \textit{Median MSE}\\
  \hline
naive & 0.96 & 0.47 & 0.78 & 0.75 & 1.79 & 0.68 & 0.99 & 1.17 & 2.54 & 0.57 & 1.39 & 1.63 \\ 
  50:50 & 1.10 & 0.56 & 0.77 & 0.81 & 1.86 & 0.75 & 0.96 & 1.24 & 2.28 & 0.50 & 1.35 & 1.61 \\ 
  50:50 cross-fit & 0.84 & 0.56 & 0.58 & 0.57 & 1.39 & 0.74 & 0.68 & 0.81 & 1.33 & 0.50 & 0.94 & 0.90 \\ 
  double sp	lit & 1.29 & 0.61 & 0.84 & 0.95 & 2.08 & 0.80 & 1.02 & 1.43 & 2.34 & 0.51 & 1.37 & 1.75 \\ 
  double split cross-fit & 0.87 & 0.61 & 0.54 & 0.55 & 1.39 & 0.80 & 0.59 & 0.76 & 1.17 & 0.54 & 0.82 & 0.75 \\ 
  5-fold & 1.31 & 0.64 & 0.82 & 0.95 & 1.84 & 0.76 & 0.95 & 1.30 & 2.08 & 0.53 & 1.27 & 1.67 \\ 
  5-fold cross fit & 0.84 & 0.64 & 0.43 & 0.48 & 1.18 & 0.77 & 0.46 & 0.62 & 0.78 & 0.52 & 0.58 & 0.50 \\ 
  5-fold combined & 0.88 & 0.52 & 0.68 & 0.63 & 1.65 & 0.72 & 0.86 & 0.98 & 1.92 & 0.47 & 1.22 & 1.25 \\ 
  50:50 cross-fit median & \underline{0.79}& 0.56 & 0.53 & 0.51 & 1.28 & 0.74 & 0.60 & 0.70 & 1.14 & 0.50 & 0.83 & 0.72 \\ 
  double split cross-fit median & 0.81 & 0.61 & 0.48 & 0.50 & 1.26 & 0.79 & 0.48 & 0.64 & 1.01 & 0.54 & 0.70 & 0.60 \\ 
  5-fold cross-fit median & 0.81 & 0.64 & 0.41 & 0.45 & \underline{1.12}& 0.77 & 0.40 & 0.56 & \underline{0.70}& 0.52 & 0.51 & 0.42 \\ 
  5-fold combined median & 0.86 & 0.52 & 0.66 & 0.61 & 1.58 & 0.72 & 0.82 & 0.93 & 1.77 & 0.47 & 1.15 & 1.14 \\ 

\hline \hline \\
Scenarios & \multicolumn{4}{c}{J} & \multicolumn{4}{c}{K} & \multicolumn{4}{c}{L}   \\    \\ \hline

 & $\overline{MSE}$ & $\overline{|Bias|}$ & $\overline{SD}$  & \textit{Median MSE}  & $\overline{MSE}$ & $\overline{|Bias|}$ & $\overline{SD}$  & \textit{Median MSE} & $\overline{MSE}$ & $\overline{|Bias|}$ & $\overline{SD}$  & \textit{Median MSE}\\
  \hline
naive & 2.22 & 0.78 & 1.16 & 1.48 & 2.40 & 0.56 & 1.32 & 1.50 & 1.85 & 0.37 & 1.24 & 1.14 \\ 
  50:50 & 1.87 & 0.56 & 1.17 & 1.32 & 1.96 & 0.45 & 1.27 & 1.40 & 1.75 & 0.32 & 1.24 & 1.19 \\ 
  50:50 cross-fit & 1.15 & 0.55 & 0.81 & 0.78 & 1.11 & 0.46 & 0.85 & 0.77 & 0.93 & 0.31 & 0.85 & 0.60 \\ 
  double split & 2.01 & 0.57 & 1.22 & 1.53 & 2.20 & 0.52 & 1.31 & 1.59 & 1.76 & 0.31 & 1.25 & 1.27 \\ 
  double split cross-fit & 1.05 & 0.59 & 0.70 & 0.70 & 0.97 & 0.51 & 0.72 & 0.64 & 0.67 & 0.29 & 0.71 & 0.46 \\ 
  5-fold & 1.67 & 0.47 & 1.15 & 1.43 & 1.77 & 0.51 & 1.17 & 1.45 & 1.21 & 0.20 & 1.07 & 1.01 \\ 
  5-fold cross fit & 0.61 & 0.46 & 0.52 & 0.48 & 0.66 & 0.49 & 0.52 & 0.45 & 0.29 & 0.18 & 0.48 & 0.23 \\ 
  5-fold combined & 1.53 & 0.52 & 1.03 & 1.00 & 1.64 & 0.40 & 1.15 & 1.14 & 1.49 & 0.34 & 1.11 & 0.95 \\ 
  50:50 cross-fit median & 0.98 & 0.55 & 0.71 & 0.64 & 0.94 & 0.46 & 0.75 & 0.61 & 0.71 & 0.31 & 0.71 & 0.44 \\ 
  double split cross-fit median & 0.91 & 0.58 & 0.60 & 0.58 & 0.81 & 0.51 & 0.61 & 0.50 & 0.50 & 0.29 & 0.57 & 0.31 \\ 
  5-fold cross-fit median & \underline{0.55}& 0.46 & 0.46 & 0.42 & \underline{0.57}& 0.49 & 0.44 & 0.37 & \underline{0.19} & 0.18 & 0.37 & 0.14 \\ 
  5-fold combined median & 1.41 & 0.52 & 0.97 & 0.90 & 1.46 & 0.40& 1.07 & 1.01 & 1.33 & 0.33 & 1.04 & 0.81 \\ 
\hline \hline \\

\end{tabular}
\begin{tablenotes}
      \small
      \item \textit{Notes:} Number of observations is 500. Averages over 300 repetitions. Lowest MSE values are underlined. If values are similar we underline the median. For the median estimators we take the median over 20 iterations of sample splitting.  
      \end{tablenotes}
  \end{threeparttable}
}
\end{table}

\clearpage

\textbf{Pseudo-code for the estimation procedure of the double sample splitting with cross-fitting and median averaging.}

\vskip 0.5cm
\begin{algorithm}[H] \label{pseudo:1}

 \textbf{create hold out sample:} Split data in two samples: $S$ and $T$ with $S \cupdot T$ \\
    \For{b=1 to B}{
    \textbf{create folds:} Split data $S$ in $K$ folds \\
   \For{k in 1 to K}{
    \textbf{regress} $Y_{i}^d =   \mu_{d}(X_{i}) + U_{i}$, with $i \in S_{1}$ \\
      \hskip 1.0cm \textbf{predict} $\hat{Y}_{i}^{d} = \hat{\mu}_{d}(X_{i})$, with $i \in S_{3}$ \\
    \textbf{regress} $D_{i} = e(X_{i}) + V_i $, with $i \in S_{2}$ \\
        \hskip 1.0cm \textbf{predict} $\hat{D_{i}} = \hat{e}(X_{i})$, with $i \in S_{3}$ \\
 \textbf{calculate} $\hat{\psi}_{f}$, with $i \in S_{3}$ (see Table \ref{tab:learners} for estimators)\\
    \textbf{regress} $\hat{\psi}_{f} = l(X_i) + W $ with $i \in S_{3}$ \\
    \hskip 1.0cm \textbf{predict} $\hat{\tau}_{k}(X_i) = \hat{l}(X_i)$ with $i \in T$ \\
    \textbf{cross-fitting:} Assign $S_1$, $S_2$ and $S_3$ to different folds than before
      }
   \textbf{average:} Take mean over $K$ folds: $\tilde{\tau}_{i,b} = \frac{1}{K}\sum \hat{\tau}_{i,k}$      
      }
  \textbf{average:} Take median over $B$ repetitions: $\tilde{\tau}_{i,median} = median\{\tilde{\tau}_{i,b}\}$   
  
  \caption{Cross-fitting and averaging} 
\end{algorithm}
\noindent
{\footnotesize   $Y_{i}^d$ contains either  $d = D_i \in \{0,1\}$ for the two separate conditional mean functions or all observations (like in the R-learner). The example shows the procedure for $K = 3$ folds.}

\vskip 0.5cm

\clearpage

\textbf{Pseudo-code for the generation of correlated covariates}

\vspace{5mm}

\begin{algorithm}[H] \label{pseudo:Cov_Generation}
    
  \textbf{Generate} random positive definite covariance matrix $\Sigma$ based on a uniform distribution over the space $k \times k$ of the correlation matrix. \\
  \textbf{Scale covariance matrix} This equals the correlation matrix and can be seen as the covariance matrix of the standardized random variables $\Sigma = \frac{X}{\sigma(X)}$. \\
  \textbf{Generate} random normal distributed variables $X_{N \times k}$ with mean = 0 and variance = $\Sigma$.\\
    
    \caption{Generation of Covariates}
\end{algorithm}

\addtocontents{toc}{\protect\enlargethispage{1\normalbaselineskip}}

\section{Supplementary}

\subsection{Results based on sample splitting and averaging} \label{sub:plots_50}

\begin{figure}[ht]
\begin{center}
\includegraphics[width=0.8\textwidth]{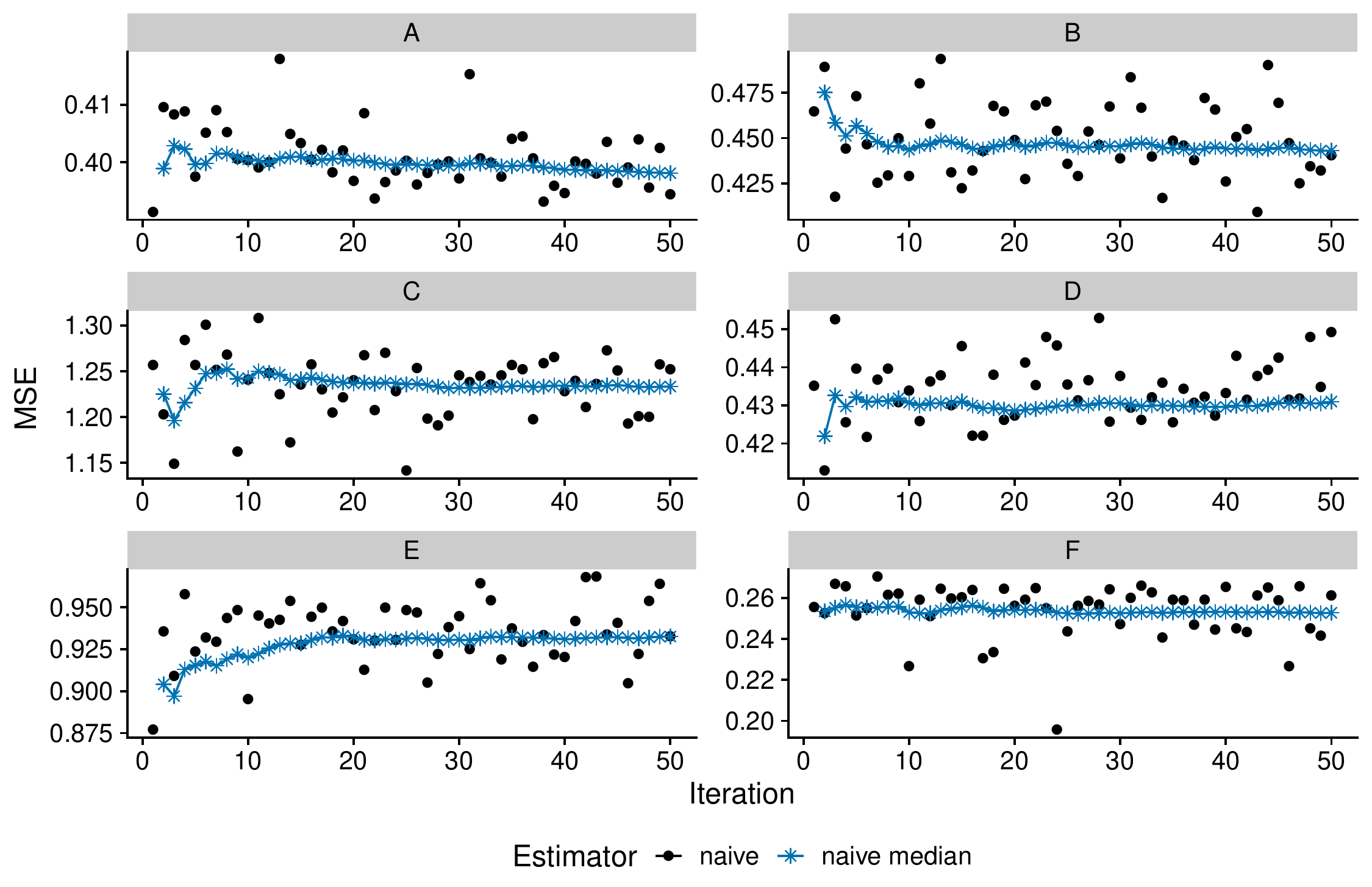}
\caption{MSE using the DR-learner without sample splitting (naive). We also consider the case where we take the median over multiple iterations. Number of observations in all settings is set to 2000.}
\label{dr_naive_2000}
\end{center}
\end{figure}

\begin{figure}[ht]
\begin{center}
\includegraphics[width=0.8\textwidth]{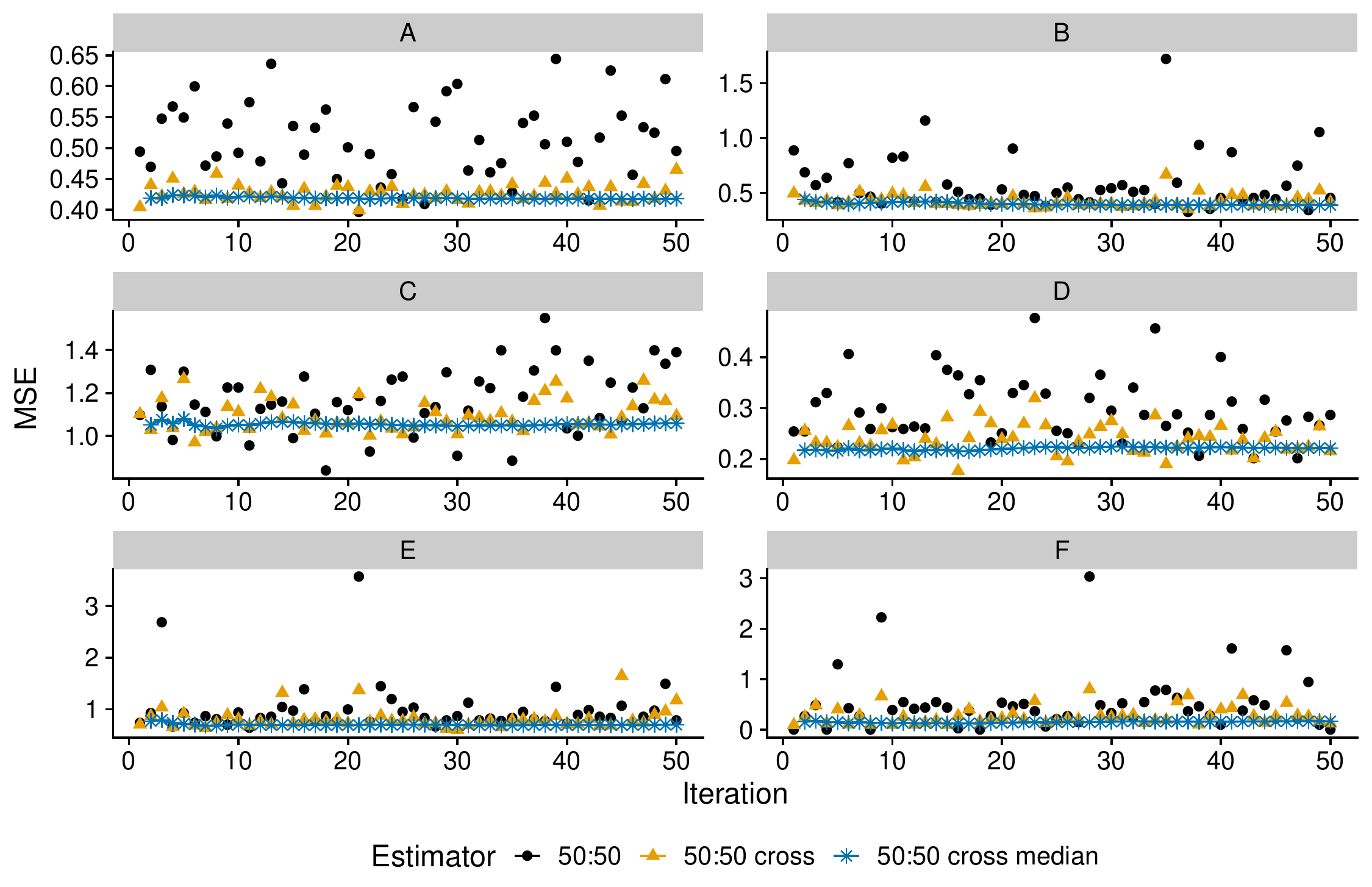}
\caption{MSE using the DR-learner on a 50:50 split, cross-fitting and taking the median. Number of observations in all settings is set to 2000.  }
\label{dr_50_50_2000}
\end{center}
\end{figure}

\begin{figure}[ht]
\begin{center}
\includegraphics[width=0.8\textwidth]{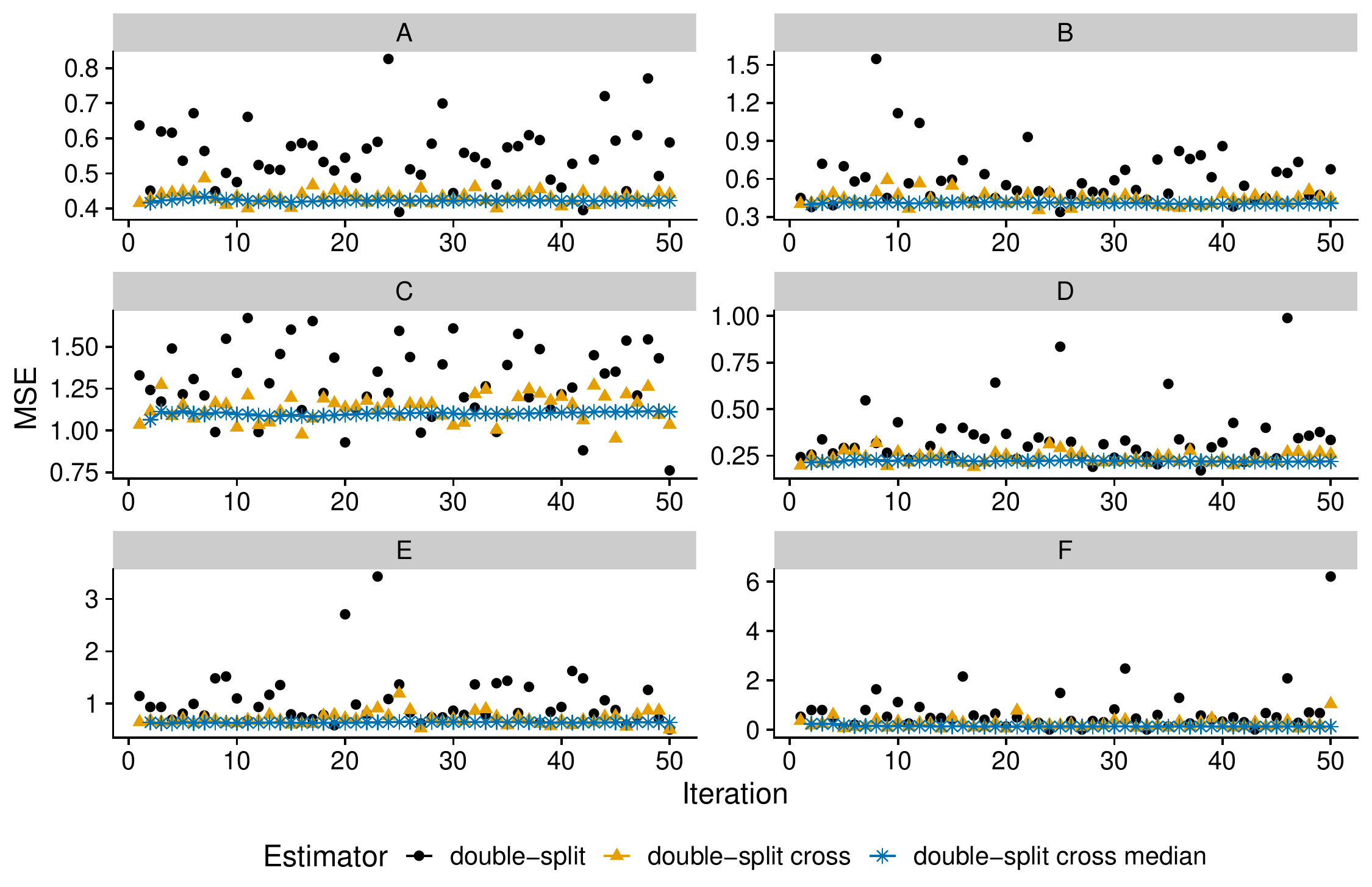}
\caption{MSE using the DR-learner with double sample splitting, cross-fitting and taking the median. Number of observations in all settings is set to 2000. }
\label{dr_double_2000}
\end{center}
\end{figure}

\begin{figure}[ht]
\begin{center}
\includegraphics[width=0.8\textwidth]{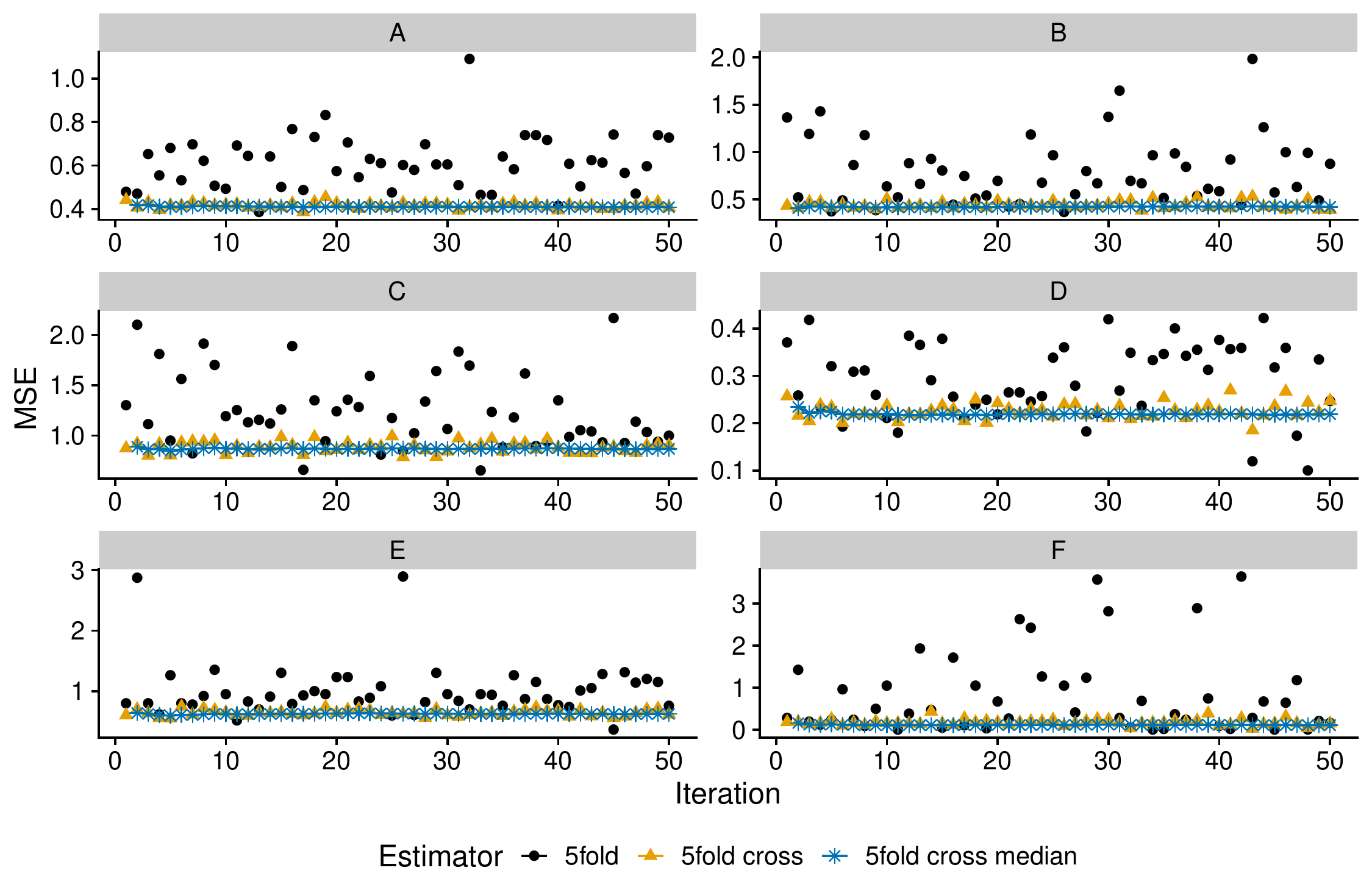}
\caption{MSE using the DR-learner on a 5-fold split, cross-fitting and taking the median. Number of observations in all settings is set to 2000. }
\label{dr_5fold_2000}
\end{center}
\end{figure}

The following plots are based on one sample dataset and give an intuition of sample splitting, cross-fitting and taking the median. 
For example, Figures \ref{dr_naive_2000}  to \ref{dr_5fold_2000} show the MSE (on the y-axis) for each estimator based on the DR-learner. We plot lines for the median results to see if the MSE becomes stable the more iterations we use (the x-axis). We show the same plots for all other learners that we consider in this paper.

Averaging by taking the median leads to robust estimates among all learners after at most 30 iterations. If we take the median over any 5-fold  cross-fit estimator this robustness can be achieved with even fewer iterations.  
We also see that given a specific sample split, extreme outliers arise. Especially the DR and the R-learner are prone to such outliers among all settings except for the balanced randomized control trial. With the T and X-learner the outlier effect is not that strong and only occurs here for settings that use 500 observations.

\begin{figure}[ht]
\begin{center}
\includegraphics[width=0.8\textwidth]{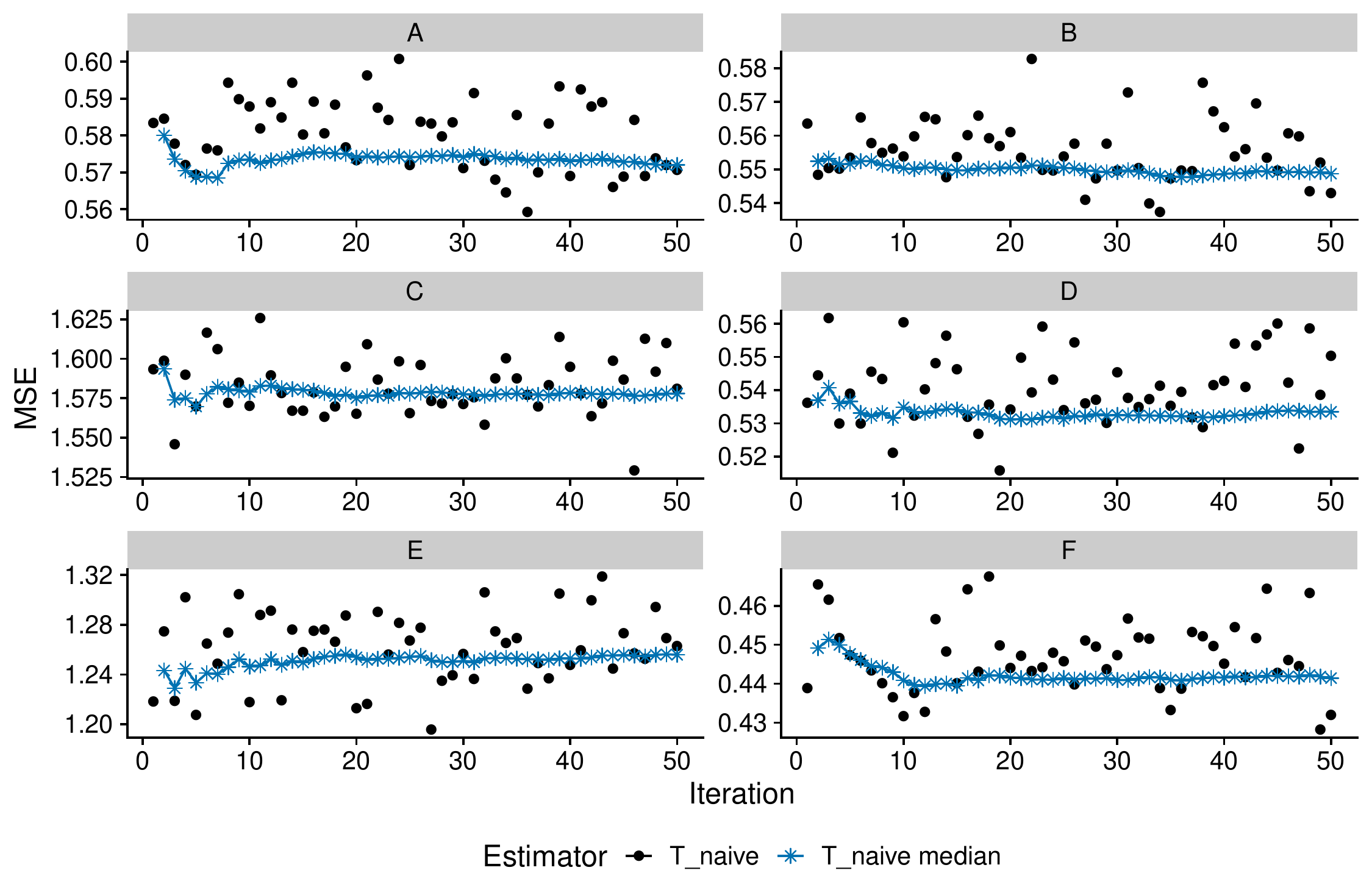}
\caption{MSE using the T-learner without sample splitting (naive). We also consider the case where we take the median over multiple iterations. Number of observations in all settings is set to 2000. }
\label{t_naive_2000}
\end{center}
\end{figure}

\begin{figure}[ht]
\begin{center}
\includegraphics[width=0.8\textwidth]{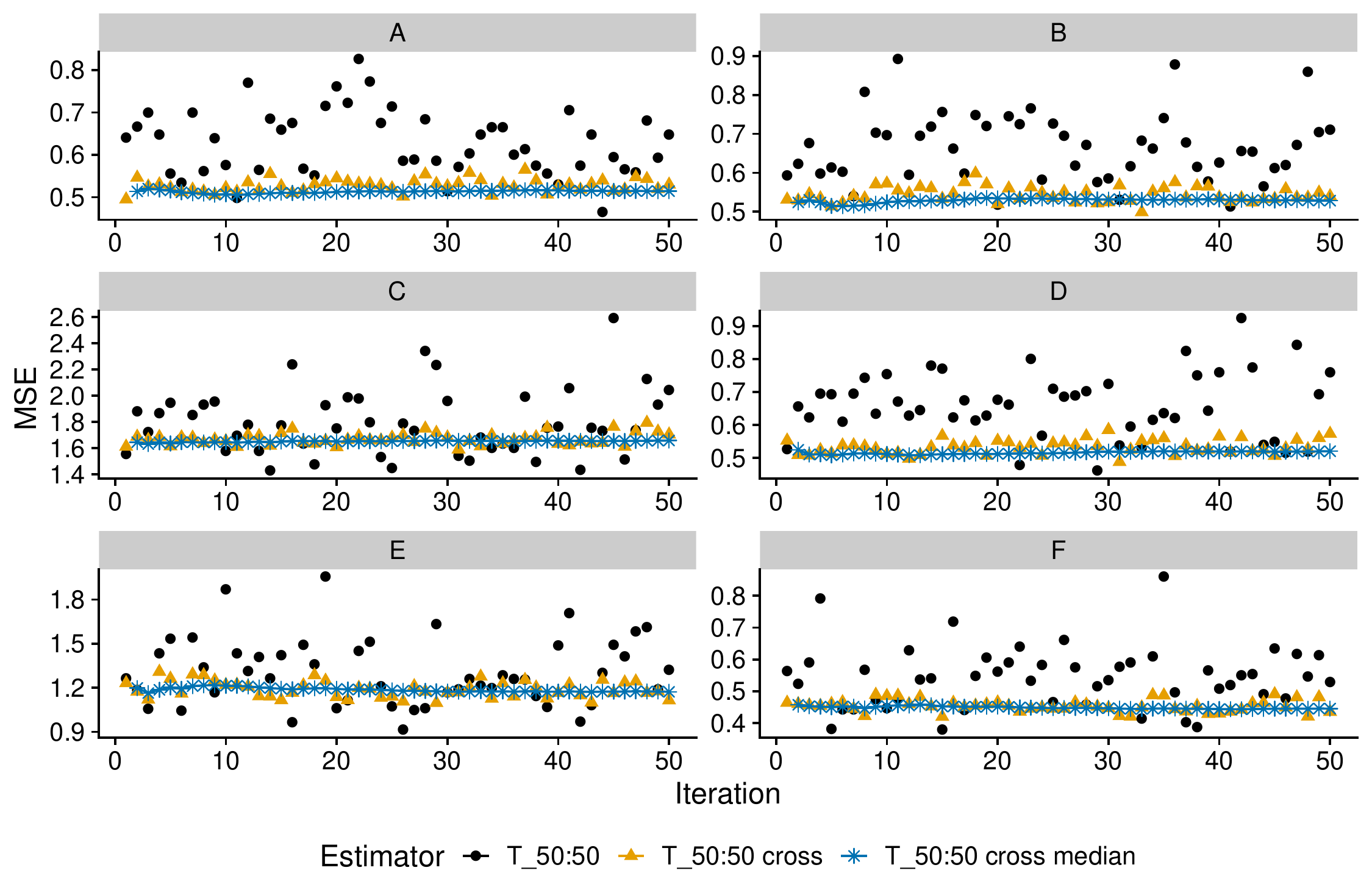}
\caption{MSE using the T-learner on a 50:50 split, cross-fitting and taking the median. Number of observations in all settings is set to 2000. }
\label{t_50_50_2000}
\end{center}
\end{figure}

\begin{figure}[ht]
\begin{center}
\includegraphics[width=0.8\textwidth]{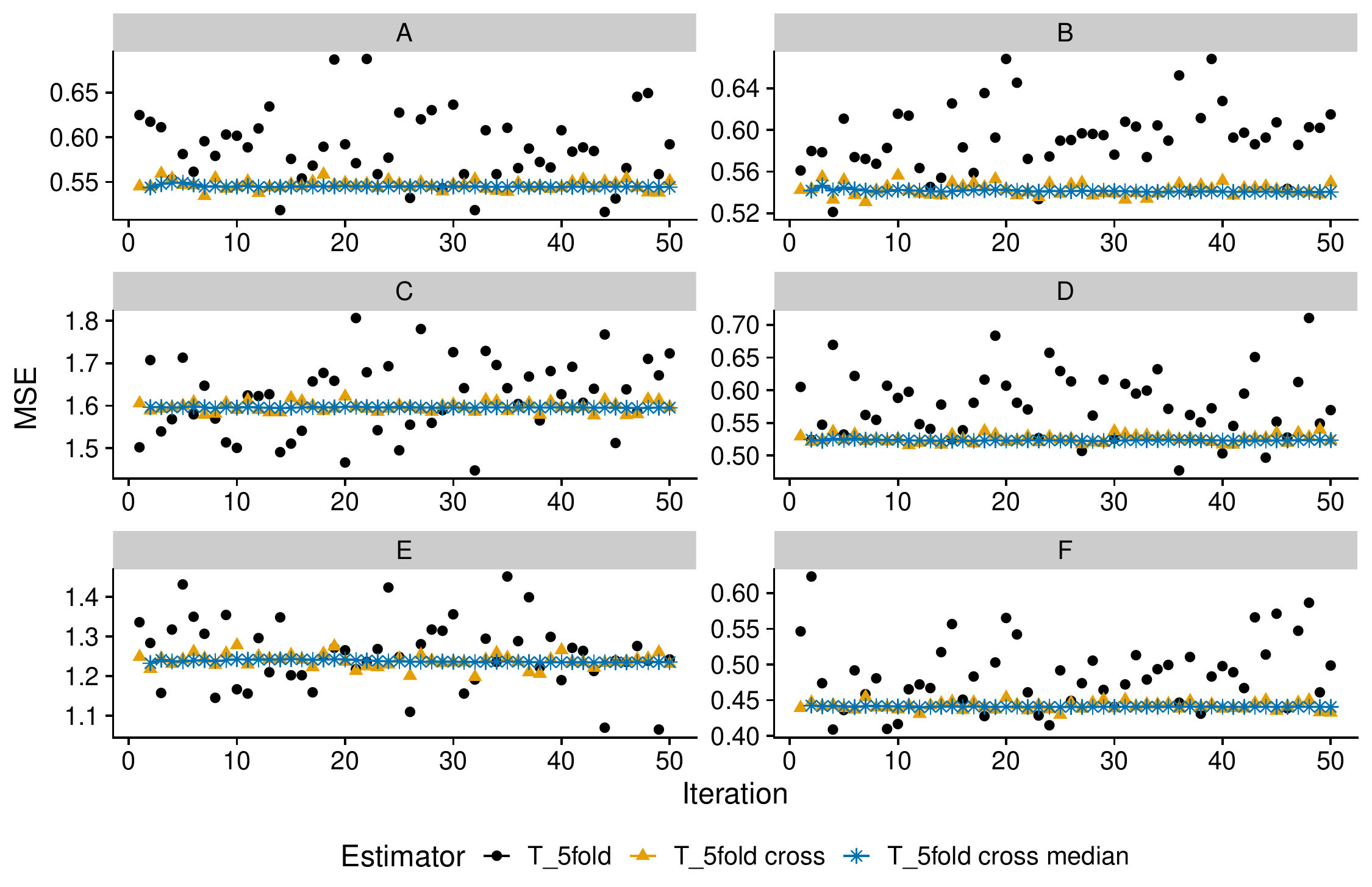}
\caption{MSE using the T-learner on a 5-fold split, cross-fitting and taking the median. Number of observations in all settings is set to 2000. }
\label{t_double_5fold}
\end{center}
\end{figure}


\begin{figure}[ht]
\begin{center}
\includegraphics[width=0.8\textwidth]{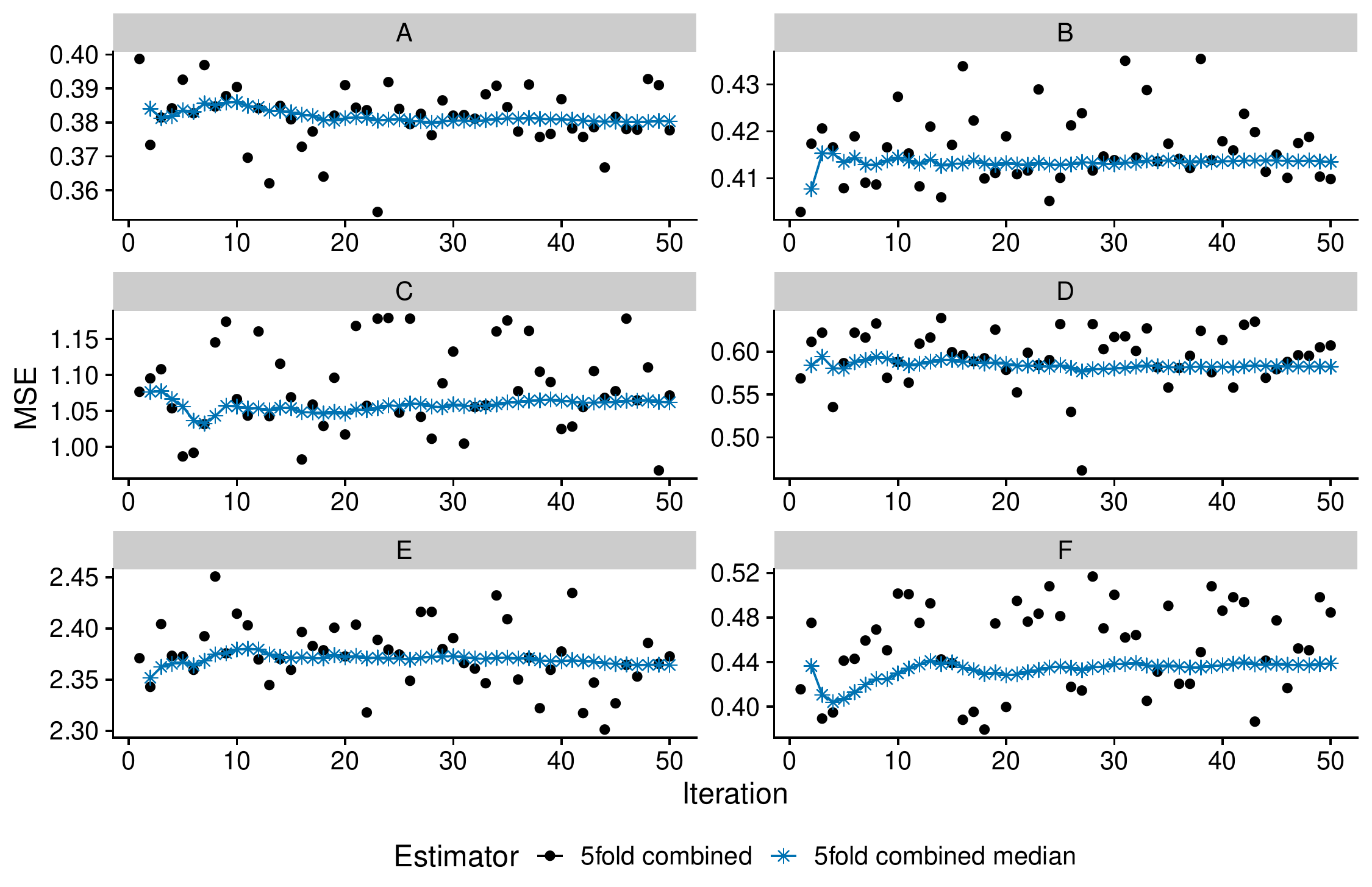}
\caption{MSE using the DR-learner on a 5-fold combined. Number of observations in all settings is set to 2000.  }
\label{dr_5fold_combined_2000}
\end{center}
\end{figure}

\begin{figure}[ht]
\begin{center}
\includegraphics[width=0.8\textwidth]{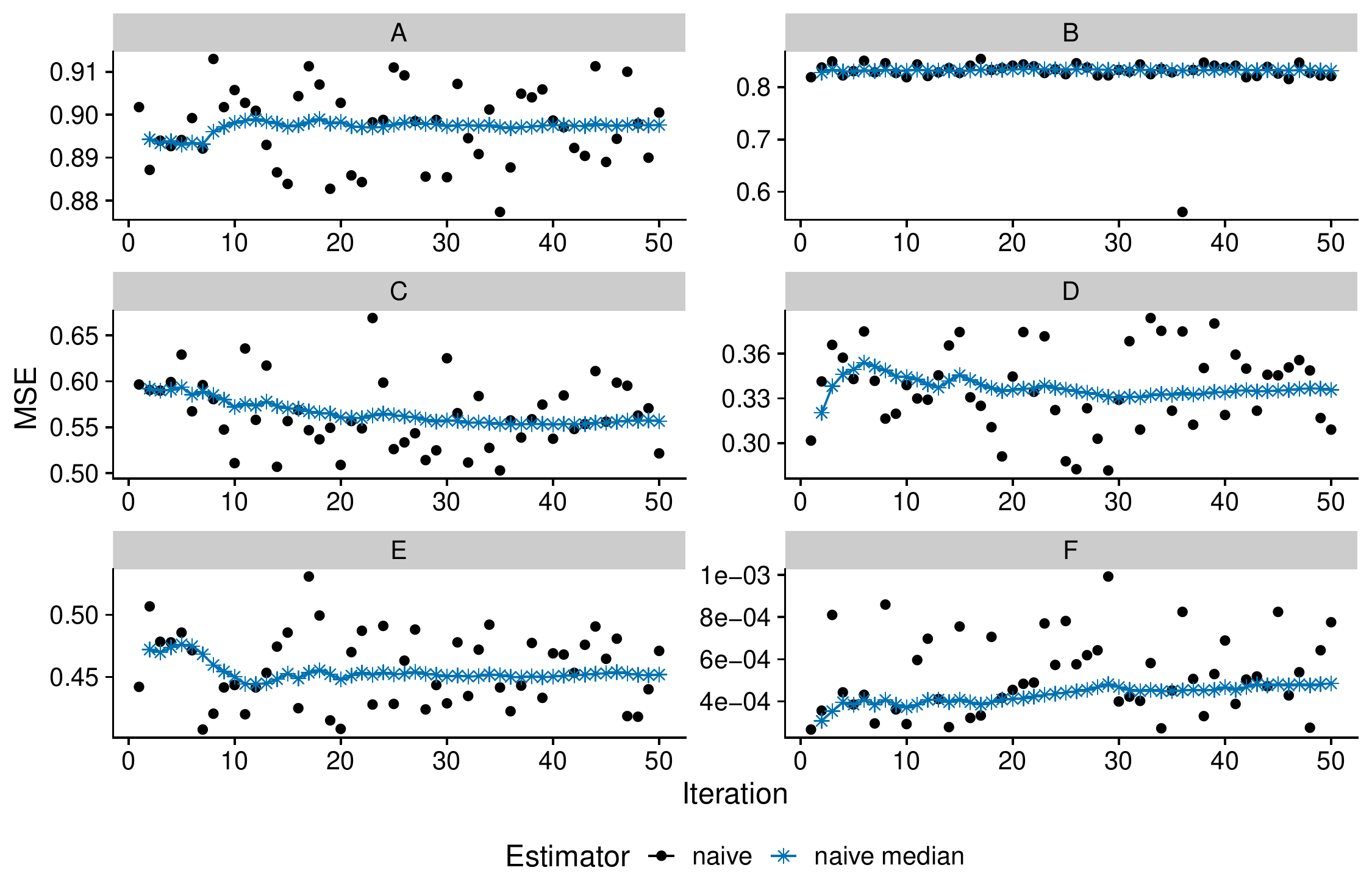}
\caption{MSE using the R-learner without sample splitting (naive). We also consider the case where we take the median over multiple iterations. Number of observations in all settings is set to 2000. }
\label{}
\end{center}
\end{figure}

\begin{figure}[ht]
\begin{center}
\includegraphics[width=0.8\textwidth]{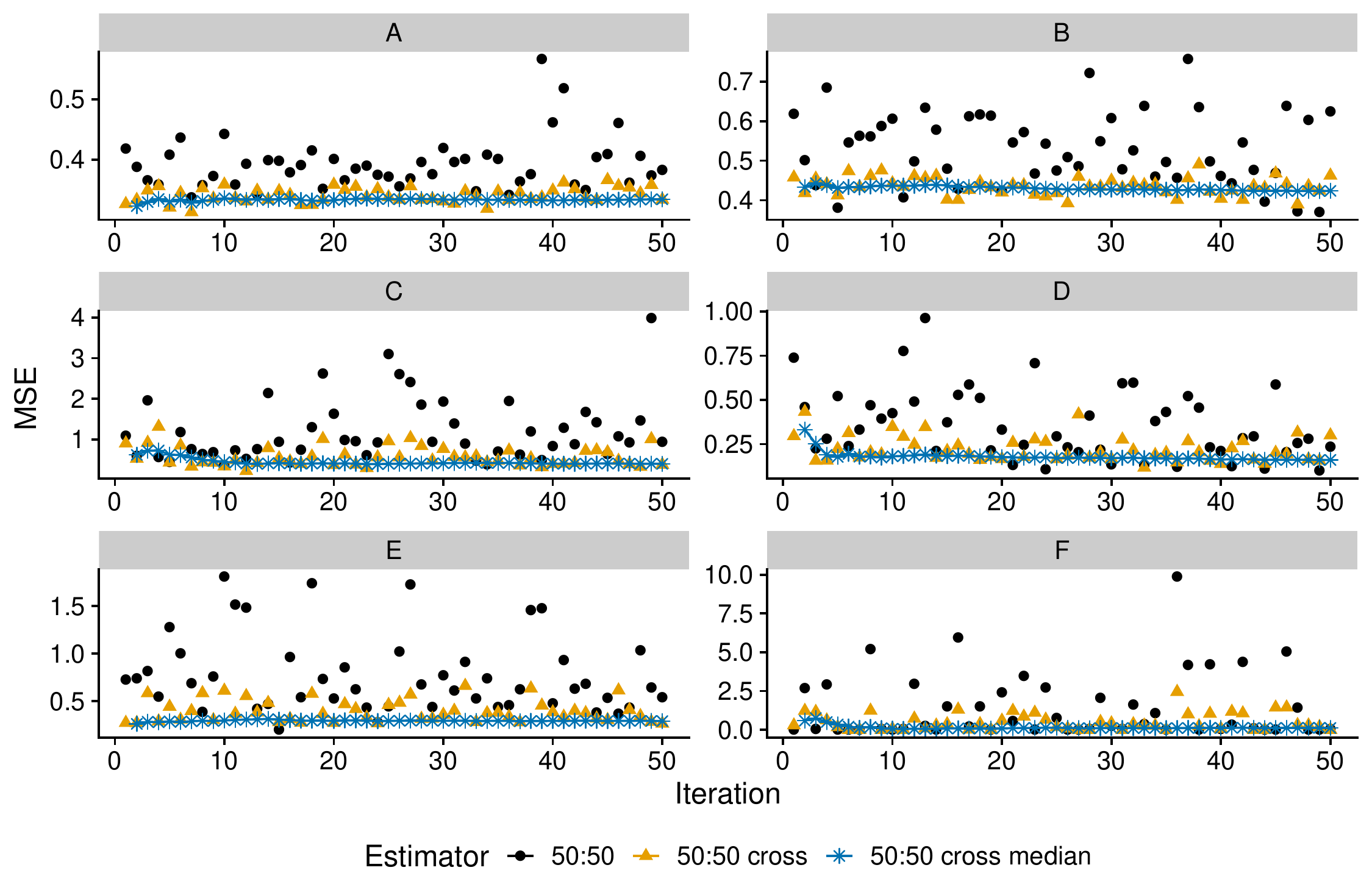}
\caption{MSE using the R-learner on a 50:50 split, cross-fitting and taking the median. Number of observations in all settings is set to 2000. }
\label{}
\end{center}
\end{figure}

\begin{figure}[ht]
\begin{center}
\includegraphics[width=0.8\textwidth]{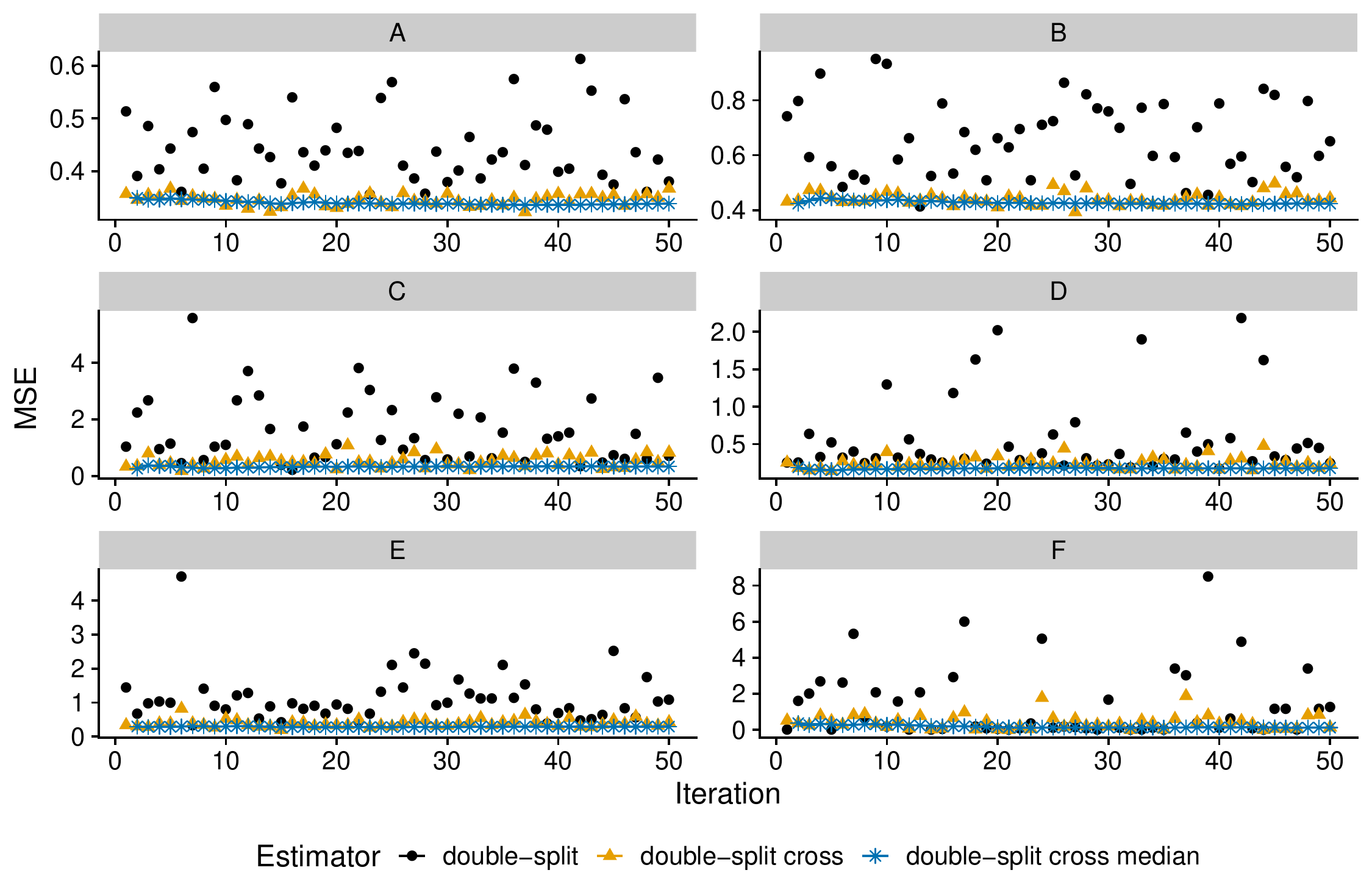}
\caption{MSE using the R-learner with double sample splitting, cross-fitting and taking the median. Number of observations in all settings is set to 2000. }
\label{}
\end{center}
\end{figure}

\begin{figure}[ht]
\begin{center}
\includegraphics[width=0.8\textwidth]{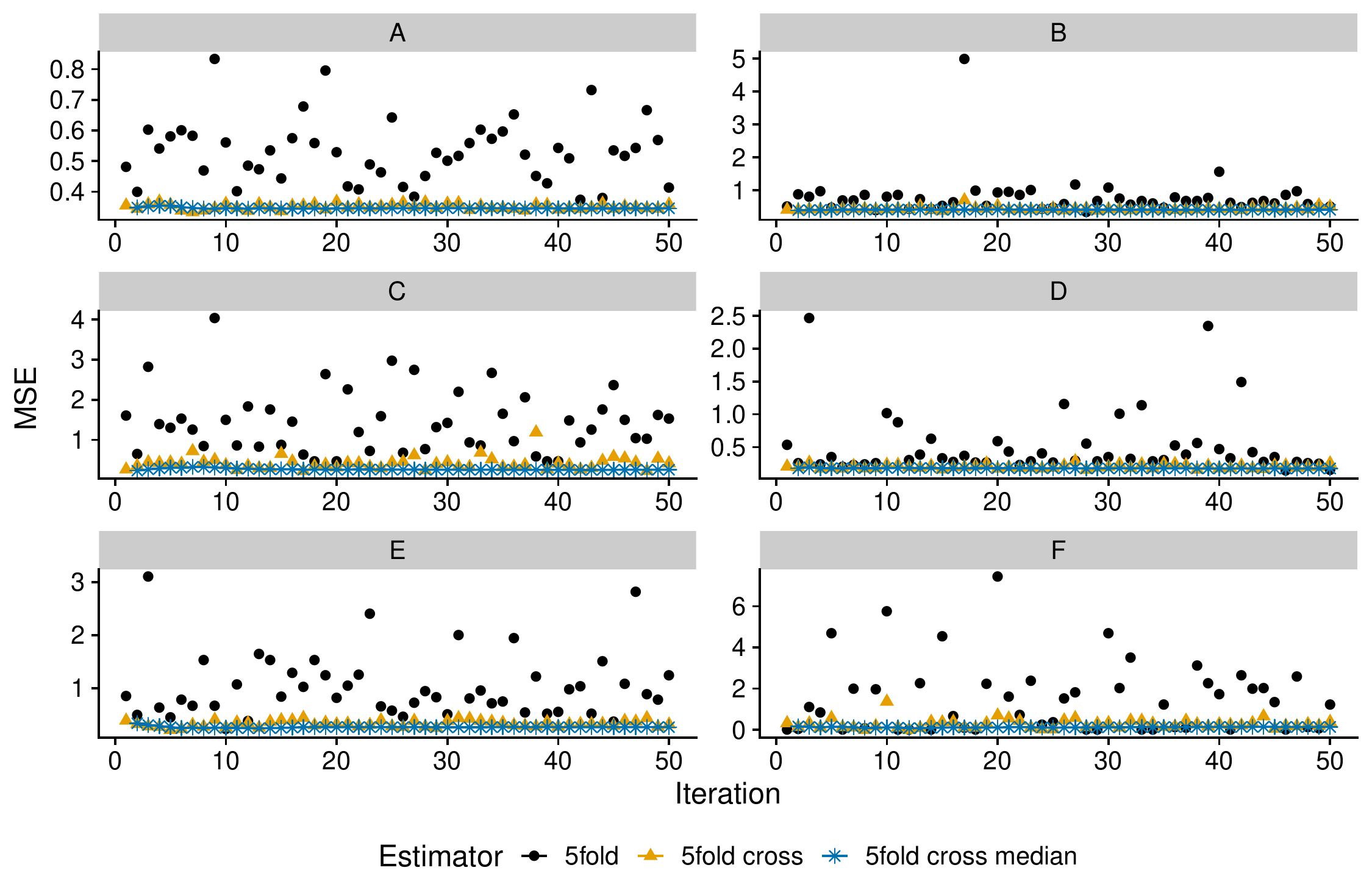}
\caption{MSE using the R-learner on a 5-fold split, cross-fitting and taking the median. Number of observations in all settings is set to 2000.}
\label{}
\end{center}
\end{figure}

\begin{figure}[ht]
\begin{center}
\includegraphics[width=0.8\textwidth]{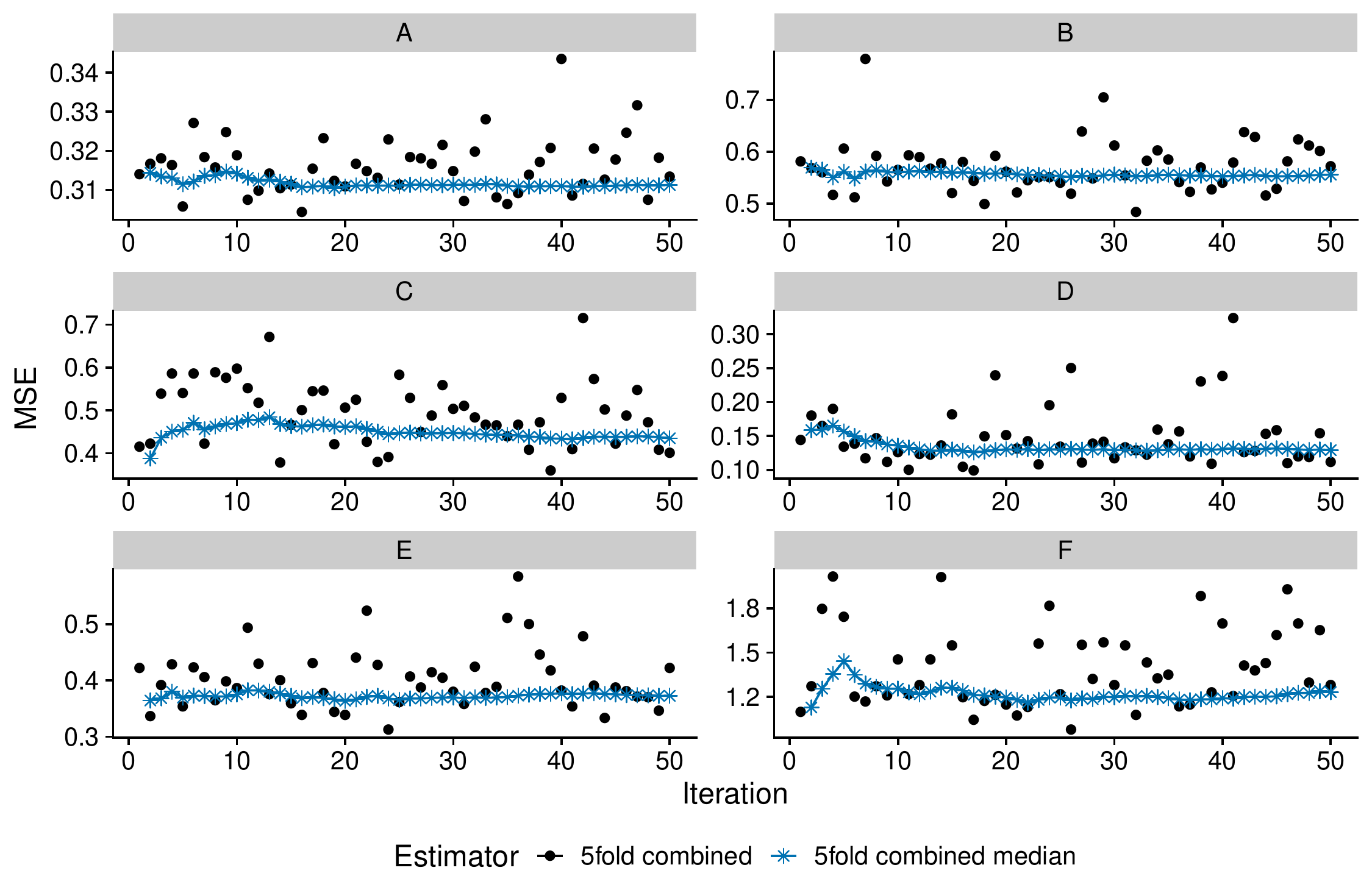}
\caption{MSE using the R-learner on a 5-fold combined. Number of observations in all settings is set to 2000. }
\label{r_5fold_combined_2000}
\end{center}
\end{figure}

\begin{figure}[ht]
\begin{center}
\includegraphics[width=0.8\textwidth]{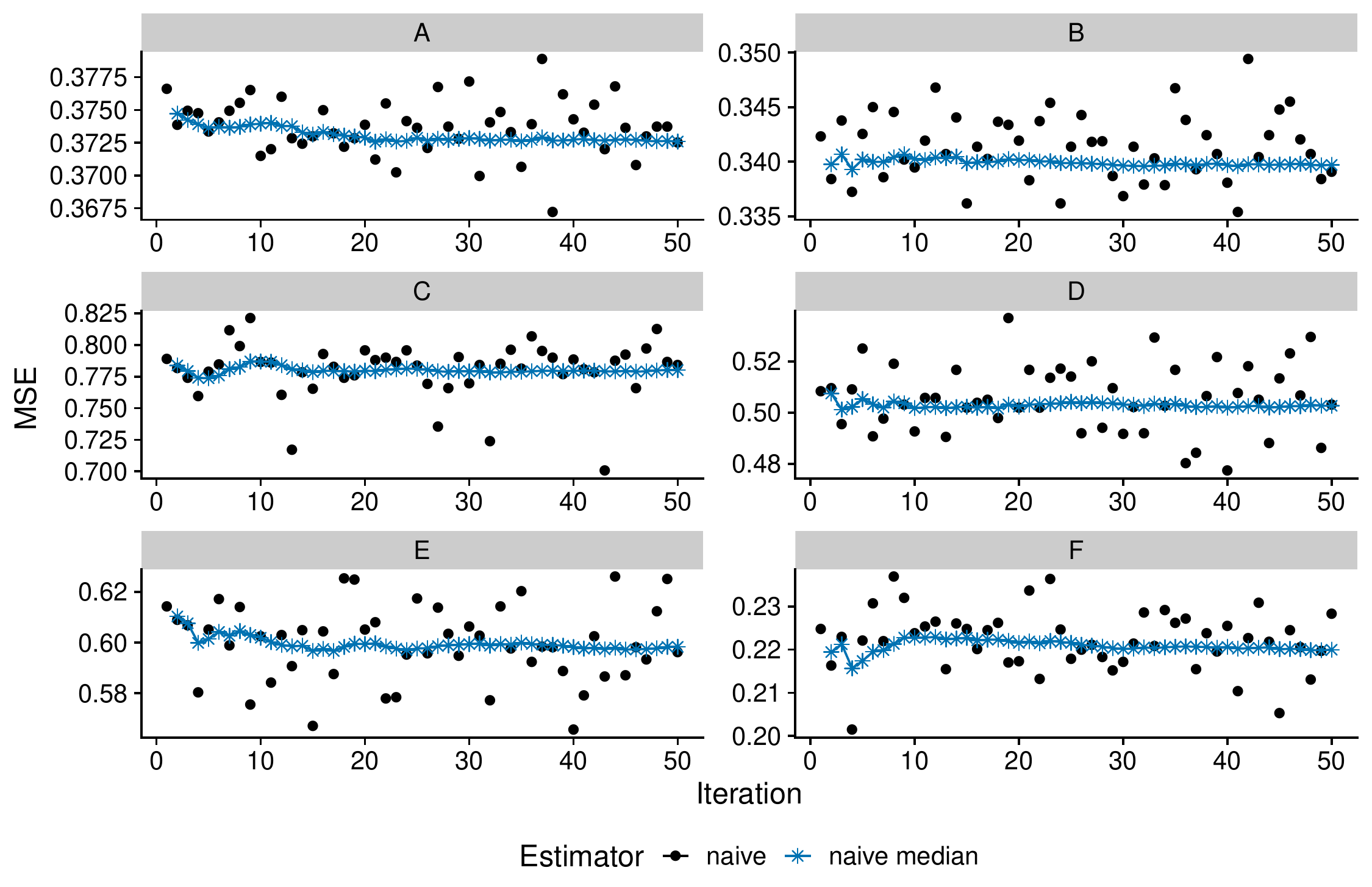}
\caption{MSE using the X-learner without sample splitting (naive). We also consider the case where we take the median over multiple iterations. Number of observations in all settings is set to 2000. }
\label{}
\end{center}
\end{figure}

\begin{figure}[ht]
\begin{center}
\includegraphics[width=0.8\textwidth]{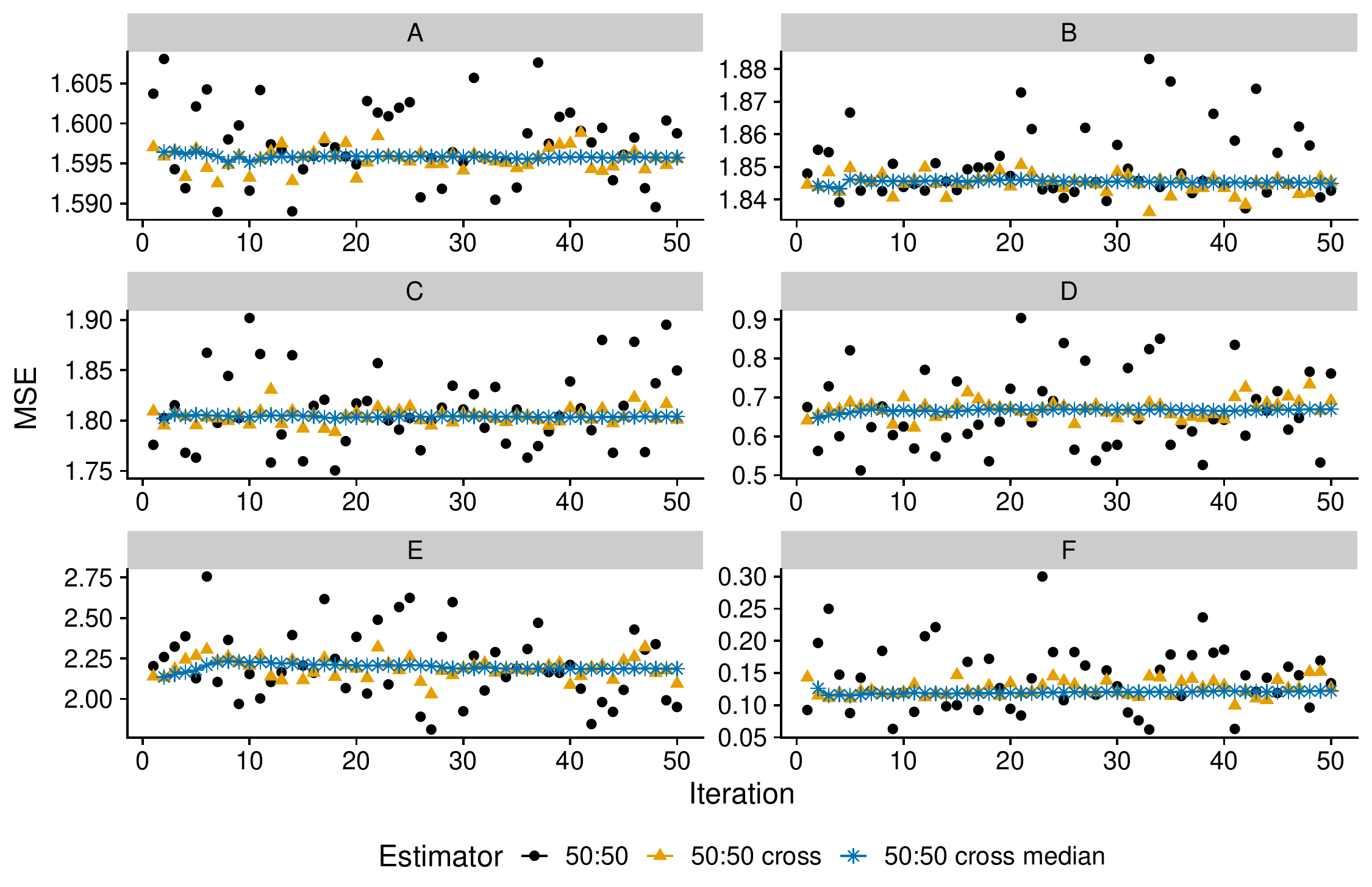}
\caption{MSE using the X-learner on a 50:50 split, cross-fitting and taking the median. Number of observations in all settings is set to 2000.  }
\label{}
\end{center}
\end{figure}

\begin{figure}[ht]
\begin{center}
\includegraphics[width=0.8\textwidth]{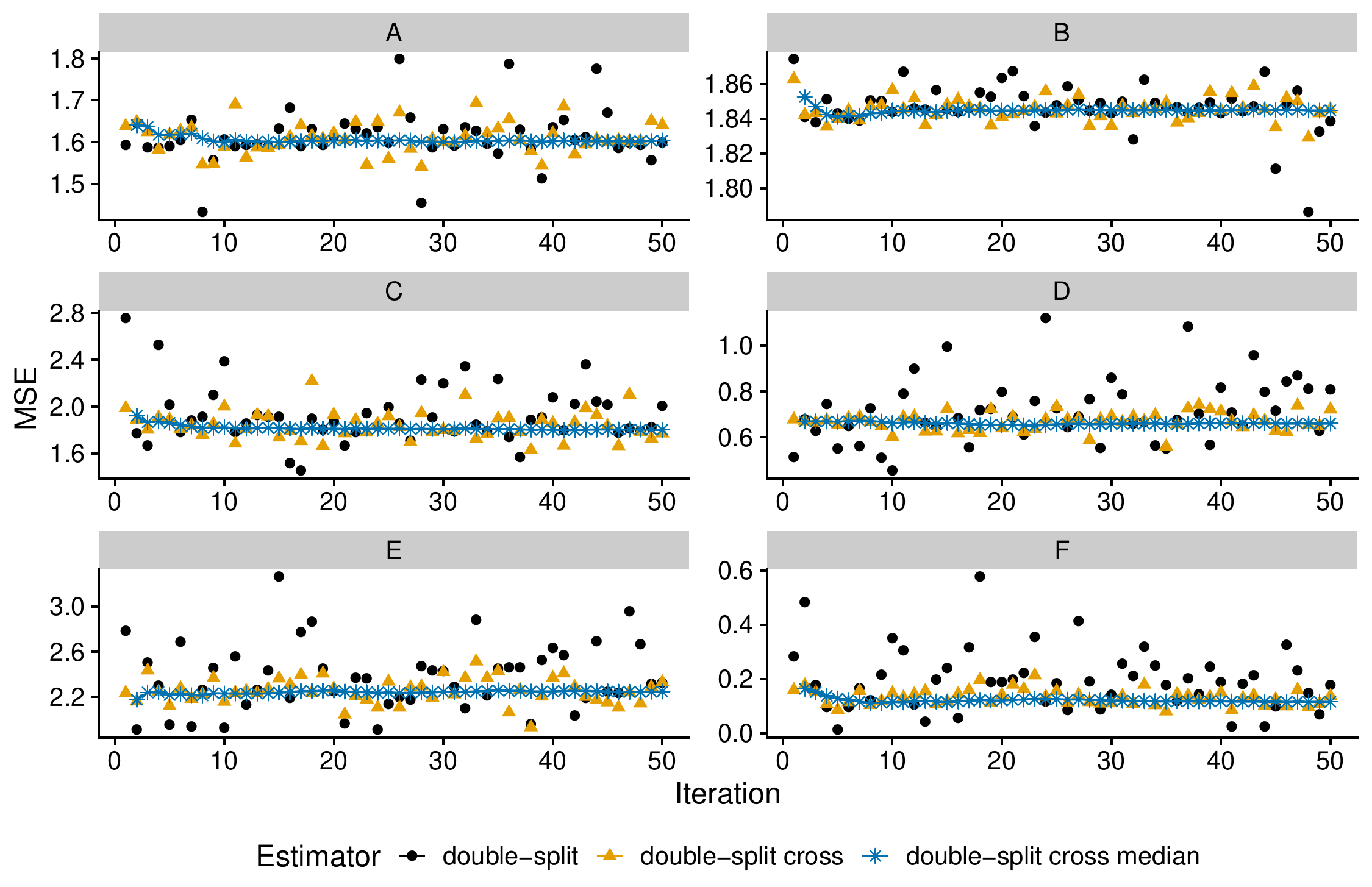}
\caption{MSE using the X-learner with double sample splitting, cross-fitting and taking the median. Number of observations in all settings is set to 2000.  }
\label{}
\end{center}
\end{figure}

\begin{figure}[ht]
\begin{center}
\includegraphics[width=0.8\textwidth]{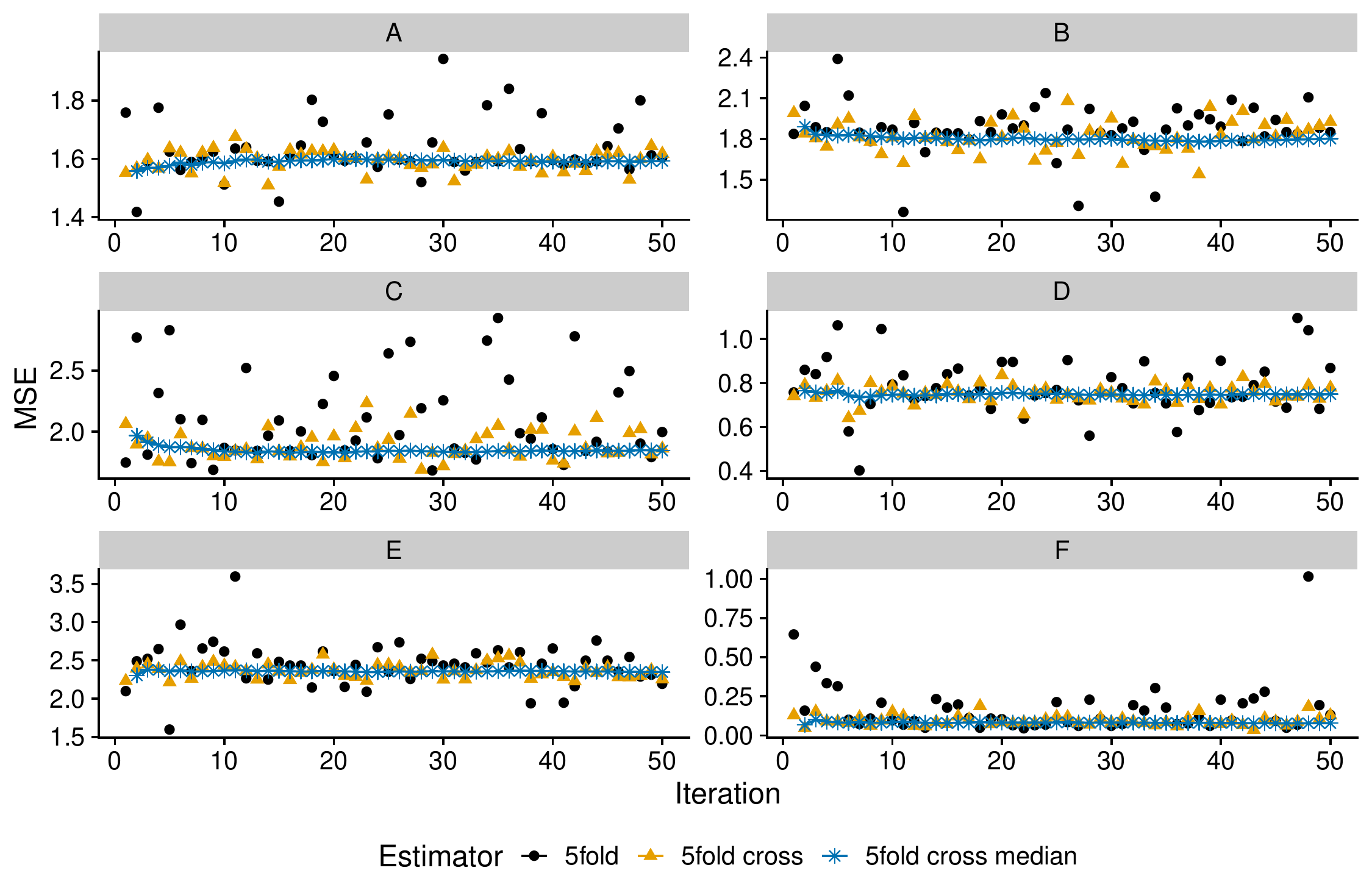}
\caption{MSE using the X-learner on a 5-fold split, cross-fitting and taking the median. Number of observations in all settings is set to 2000.  }
\label{}
\end{center}
\end{figure}


\begin{figure}[ht]
\begin{center}
\includegraphics[width=0.8\textwidth]{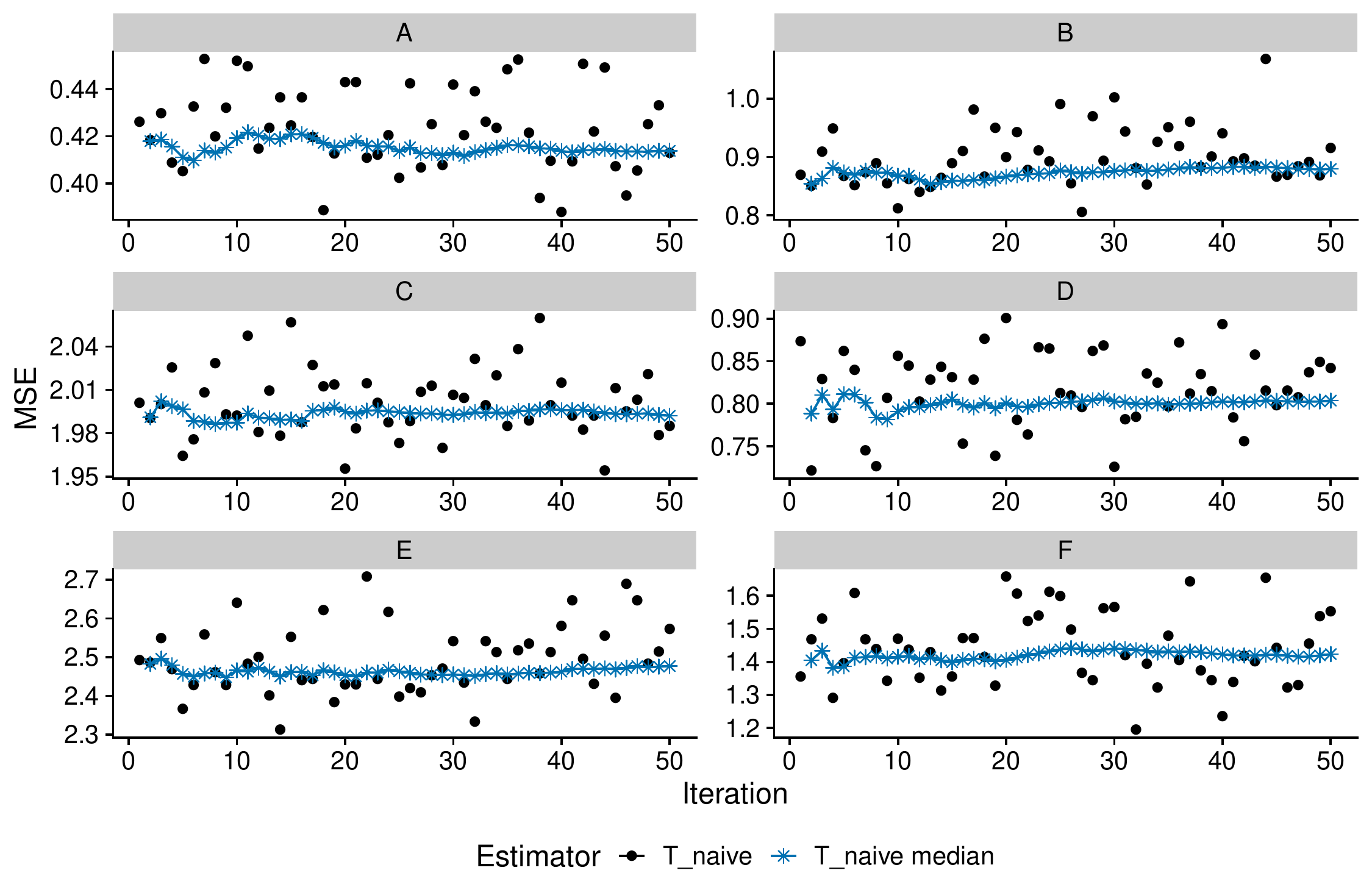}
\caption{MSE using the T-learner without sample splitting (naive). We also consider the case where we take the median over multiple iterations. Number of observations is 500. }
\label{t_naive_500}
\end{center}
\end{figure}

\begin{figure}[ht]
\begin{center}
\includegraphics[width=0.8\textwidth]{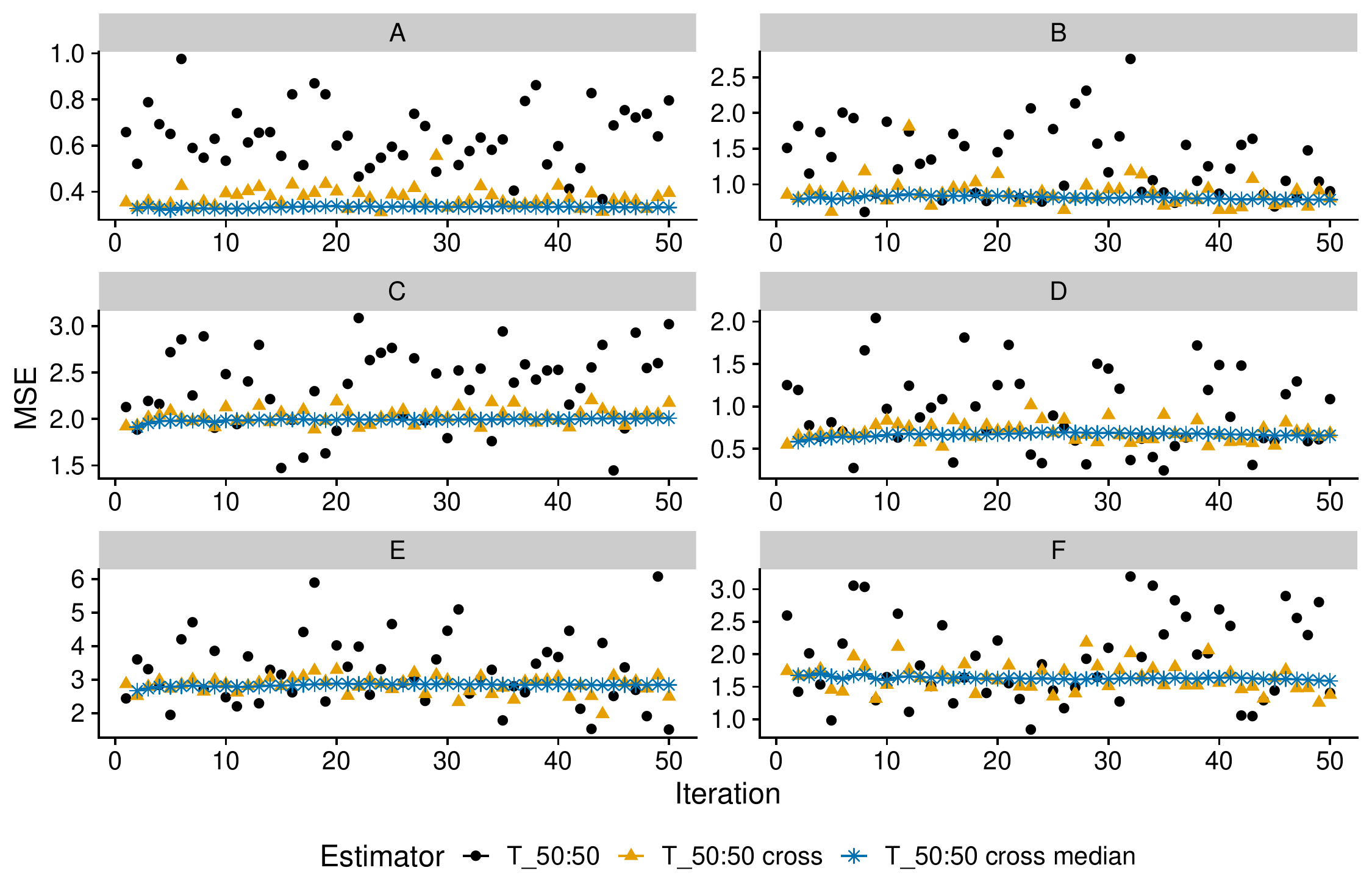}
\caption{MSE using the T-learner on a 50:50 split, cross-fitting and taking the median. Number of observations in all settings is set to 500. }
\label{t_50_50_500}
\end{center}
\end{figure}

\begin{figure}[ht]
\begin{center}
\includegraphics[width=0.8\textwidth]{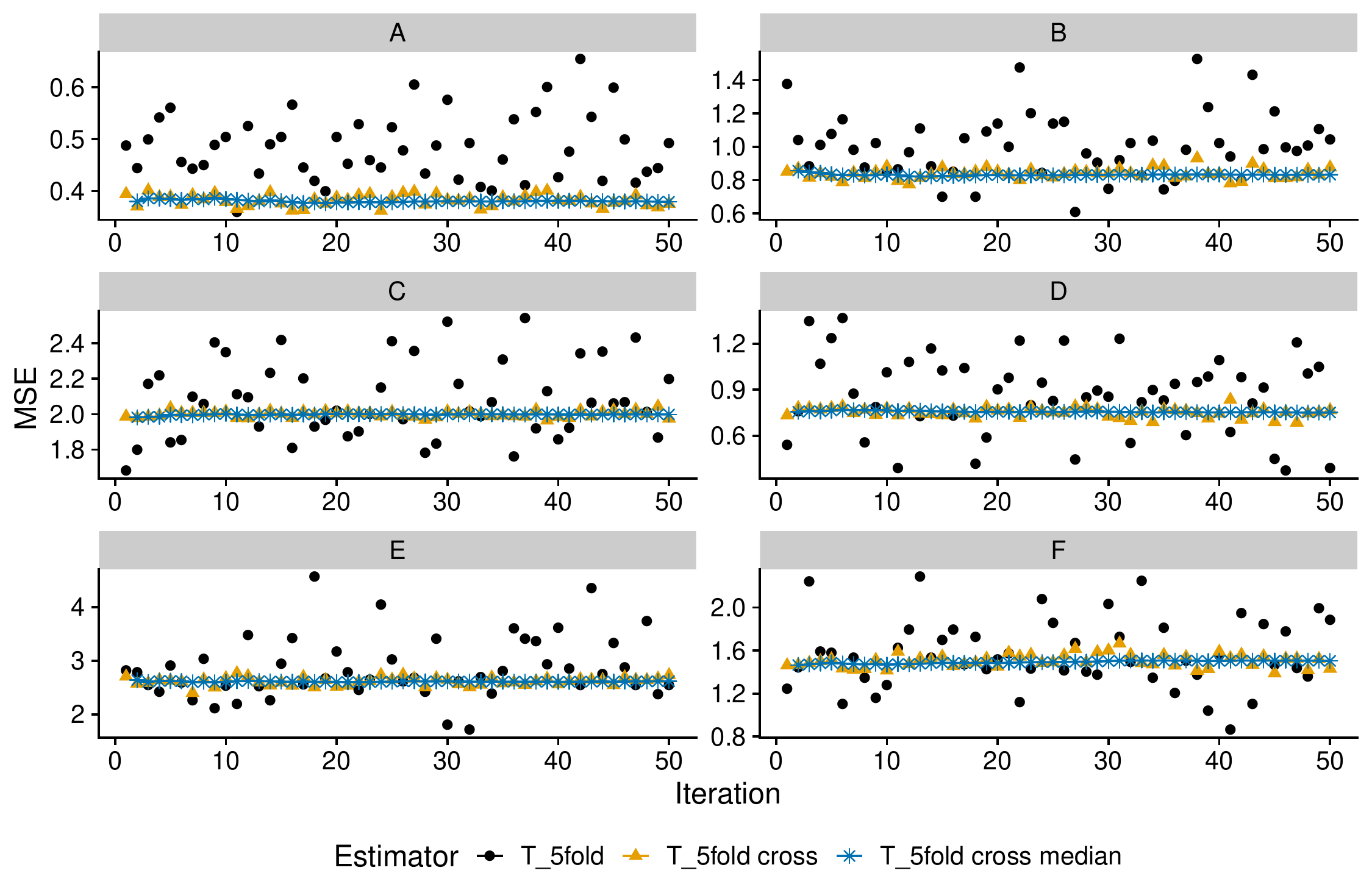}
\caption{MSE using the T-learner on a 5-fold split, cross-fitting and taking the median. Number of observations in all settings is set to 500.  }
\label{t_5fold}
\end{center}
\end{figure}


\begin{figure}[ht]
\begin{center}
\includegraphics[width=0.8\textwidth]{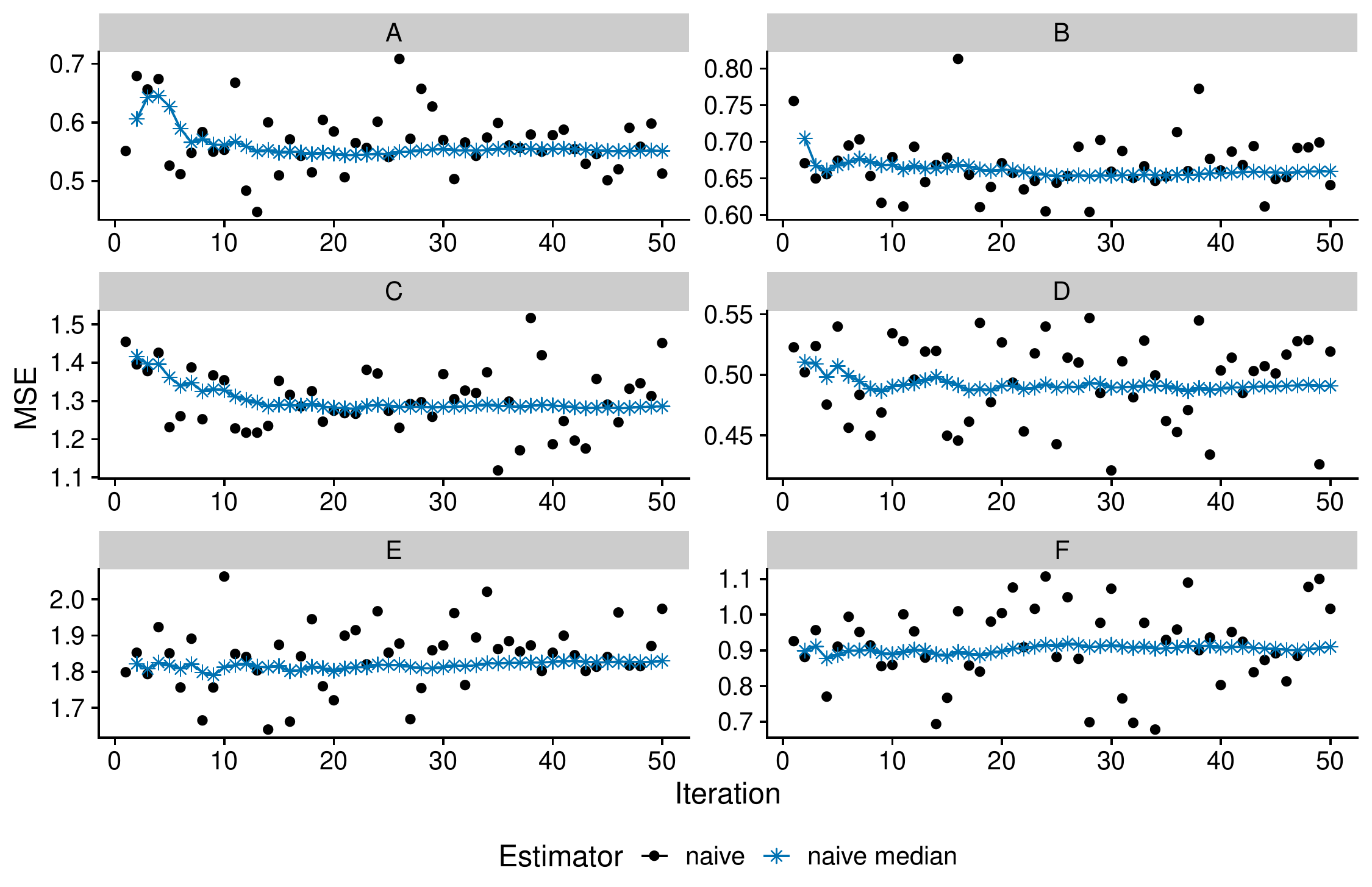}
\caption{MSE using the DR-learner without sample splitting (naive). We also consider the case where we take the median over multiple iterations. Number of observations is 500. }
\label{dr_naive_500}
\end{center}
\end{figure}

\begin{figure}[ht]
\begin{center}
\includegraphics[width=0.8\textwidth]{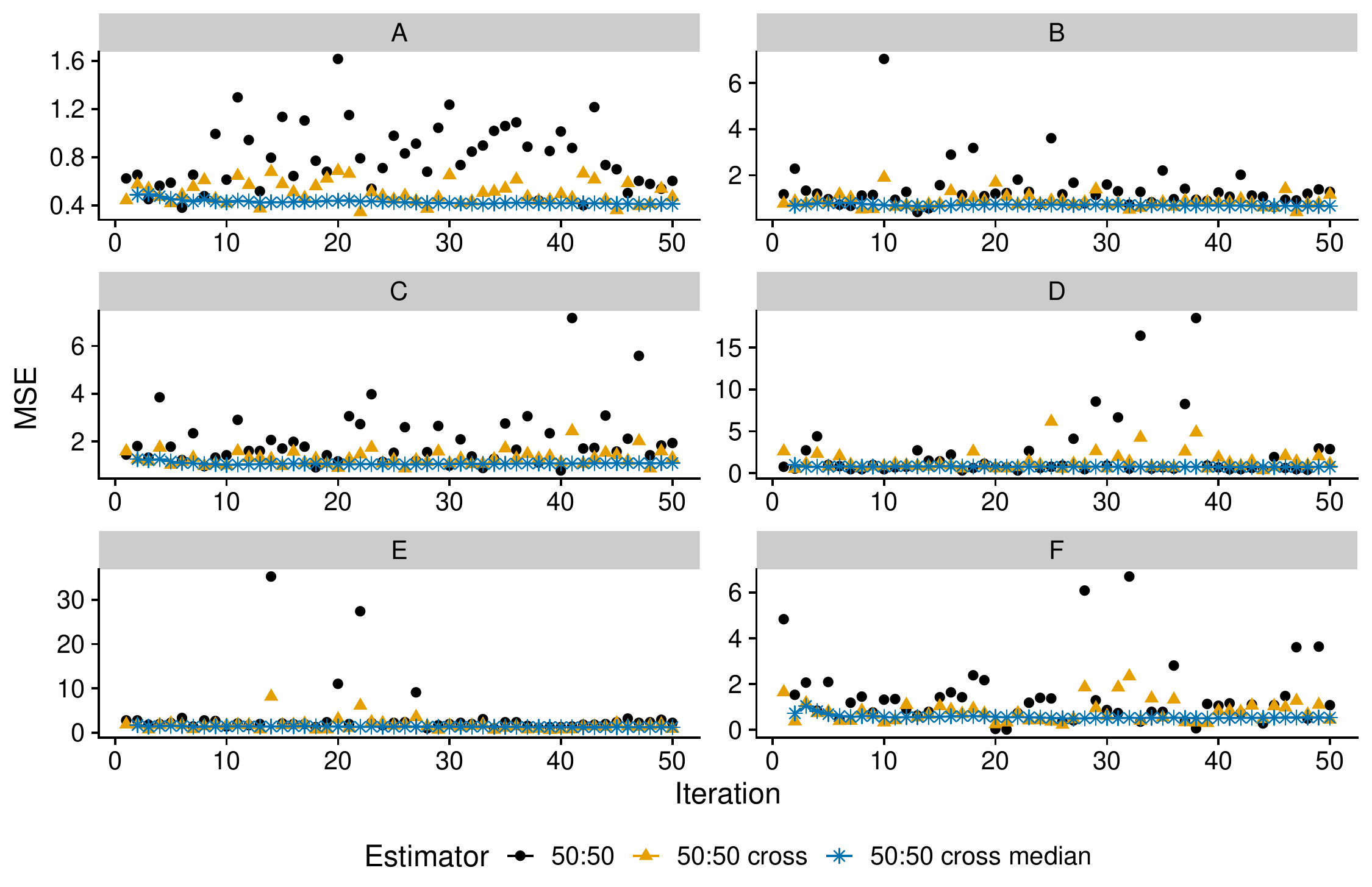}
\caption{MSE using the DR-learner on a 50:50 split, cross-fitting and taking the median. Number of observations in all settings is set to 500.  }
\label{dr_50_50_500}
\end{center}
\end{figure}

\begin{figure}[ht]
\begin{center}
\includegraphics[width=0.8\textwidth]{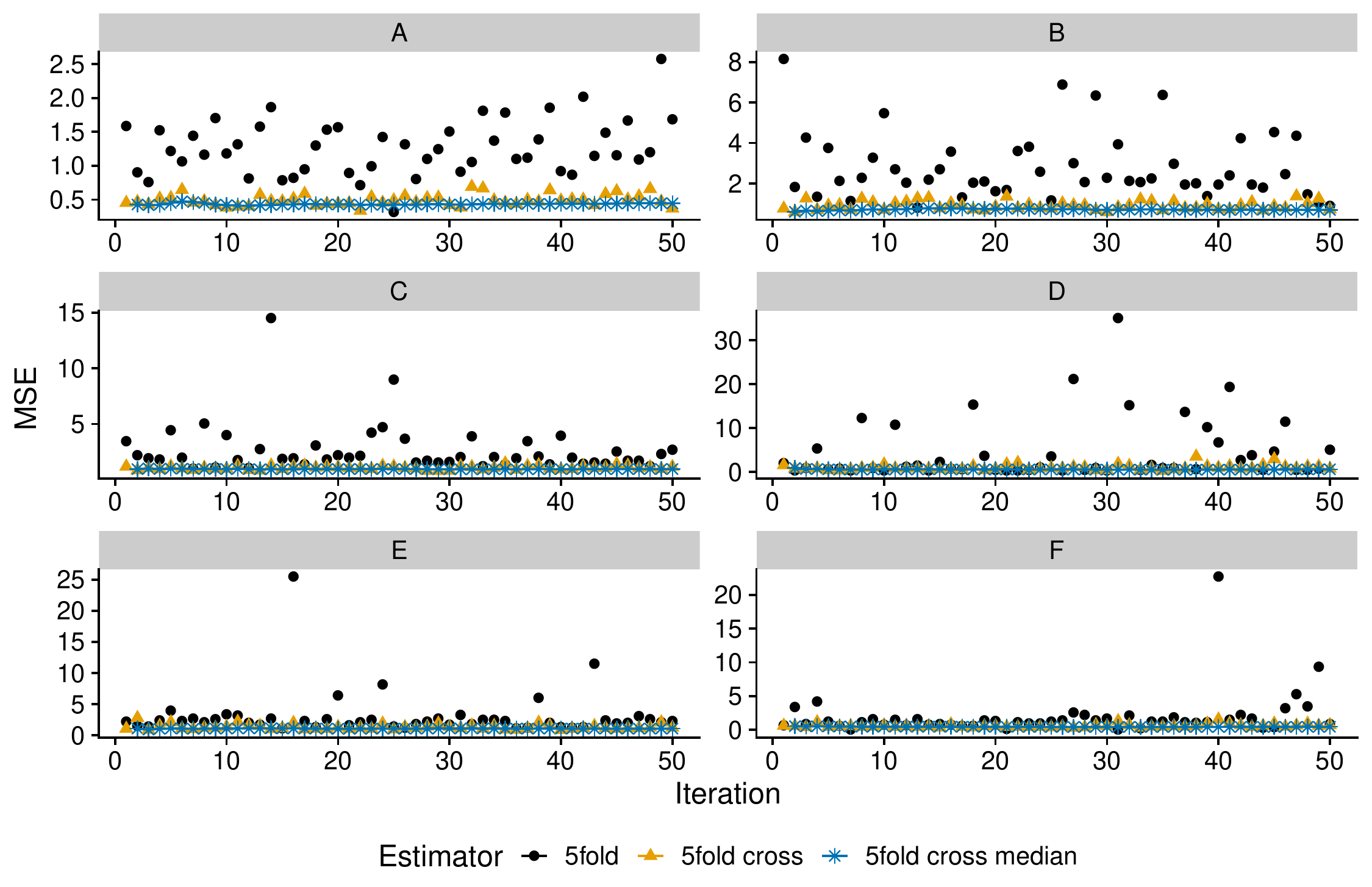}
\caption{MSE using the DR-learner on a 5-fold split, cross-fitting and taking the median. Number of observations in all settings is set to 500.  }
\label{dr_5fold_500}
\end{center}
\end{figure}

\begin{figure}[ht]
\begin{center}
\includegraphics[width=0.8\textwidth]{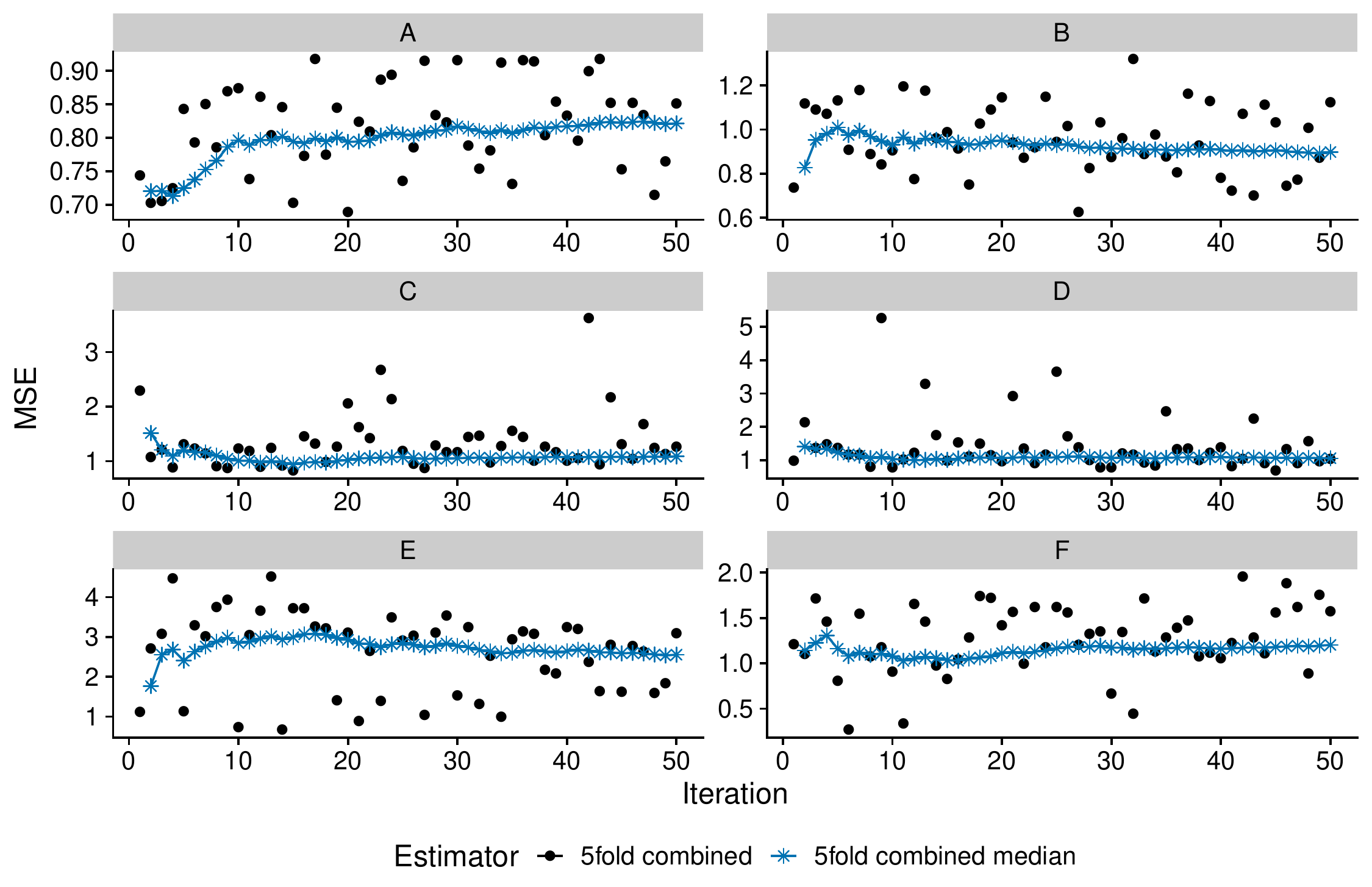}
\caption{MSE using the DR-learner on a 5-fold combined. Number of observations in all settings is set to 500.  }
\label{dr_5fold_combined_500}
\end{center}
\end{figure}

\begin{figure}[ht]
\begin{center}
\includegraphics[width=0.8\textwidth]{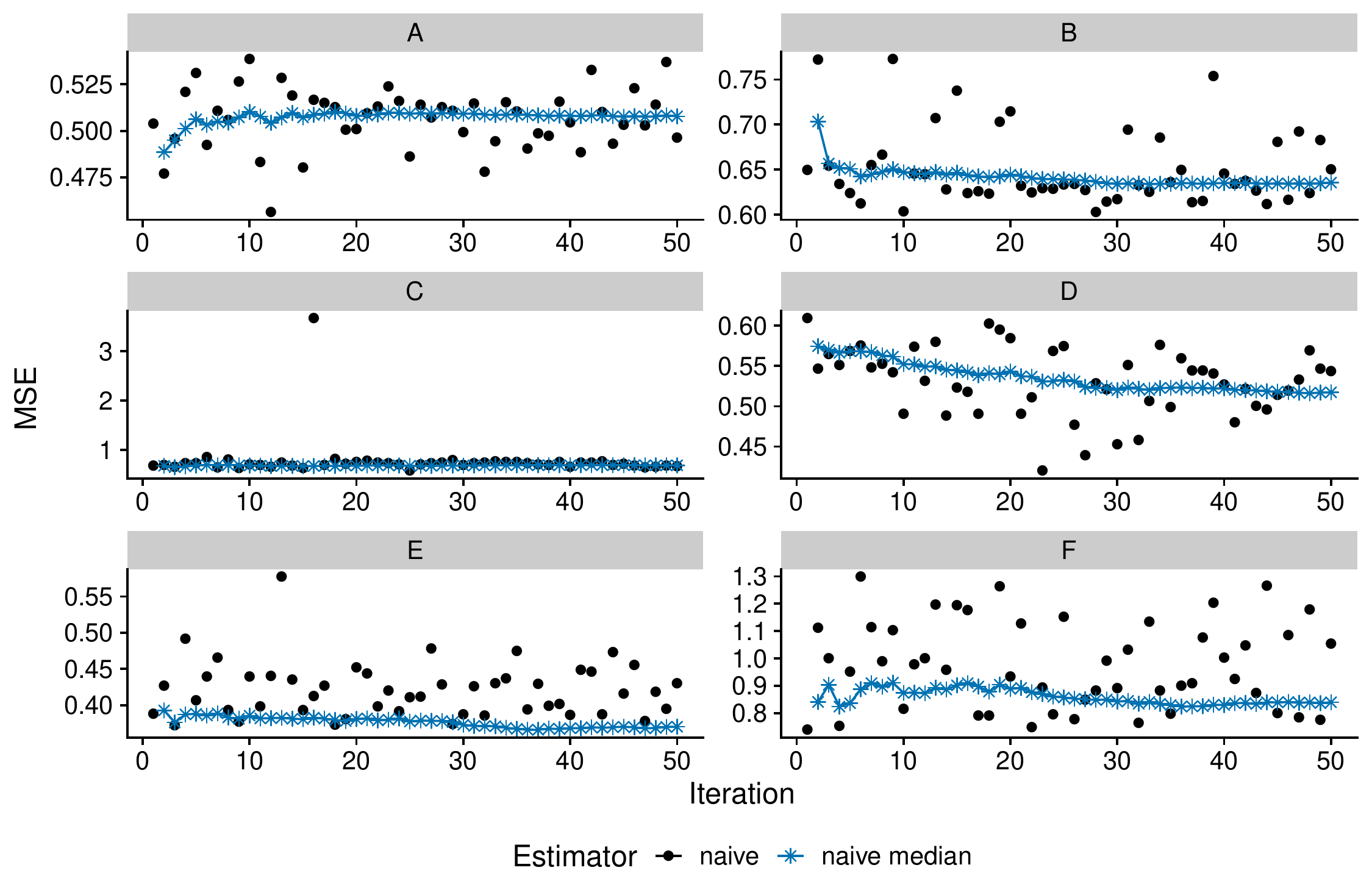}
\caption{MSE using the R-learner without sample splitting (naive). We also consider the case where we take the median over multiple iterations. Number of observations is 500. }
\label{}
\end{center}
\end{figure}

\begin{figure}[ht]
\begin{center}
\includegraphics[width=0.8\textwidth]{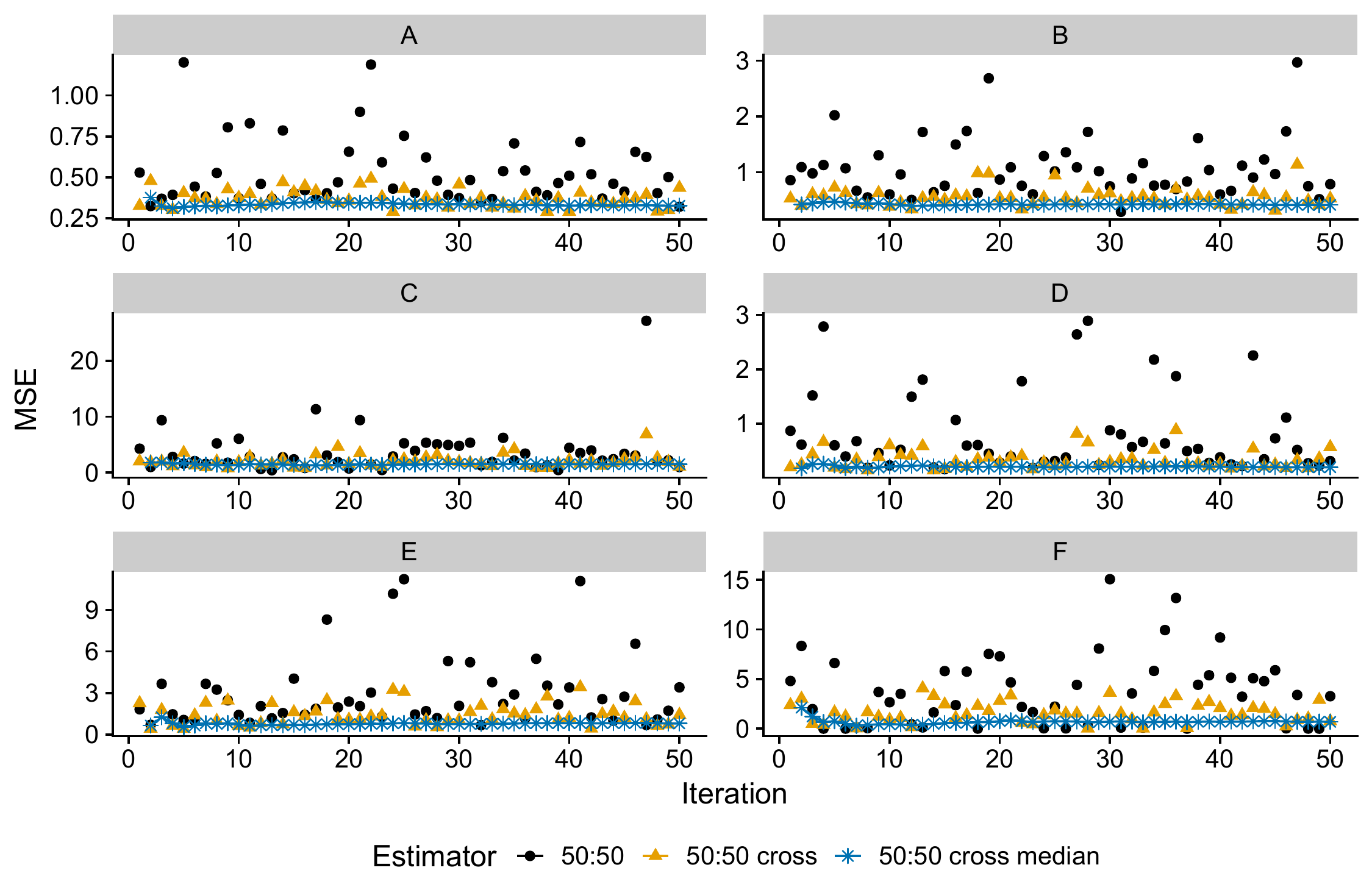}
\caption{MSE using the R-learner on a 50:50 split, cross-fitting and taking the median. Number of observations in all settings is set to 500.  }
\label{}
\end{center}
\end{figure}

\begin{figure}[ht]
\begin{center}
\includegraphics[width=0.8\textwidth]{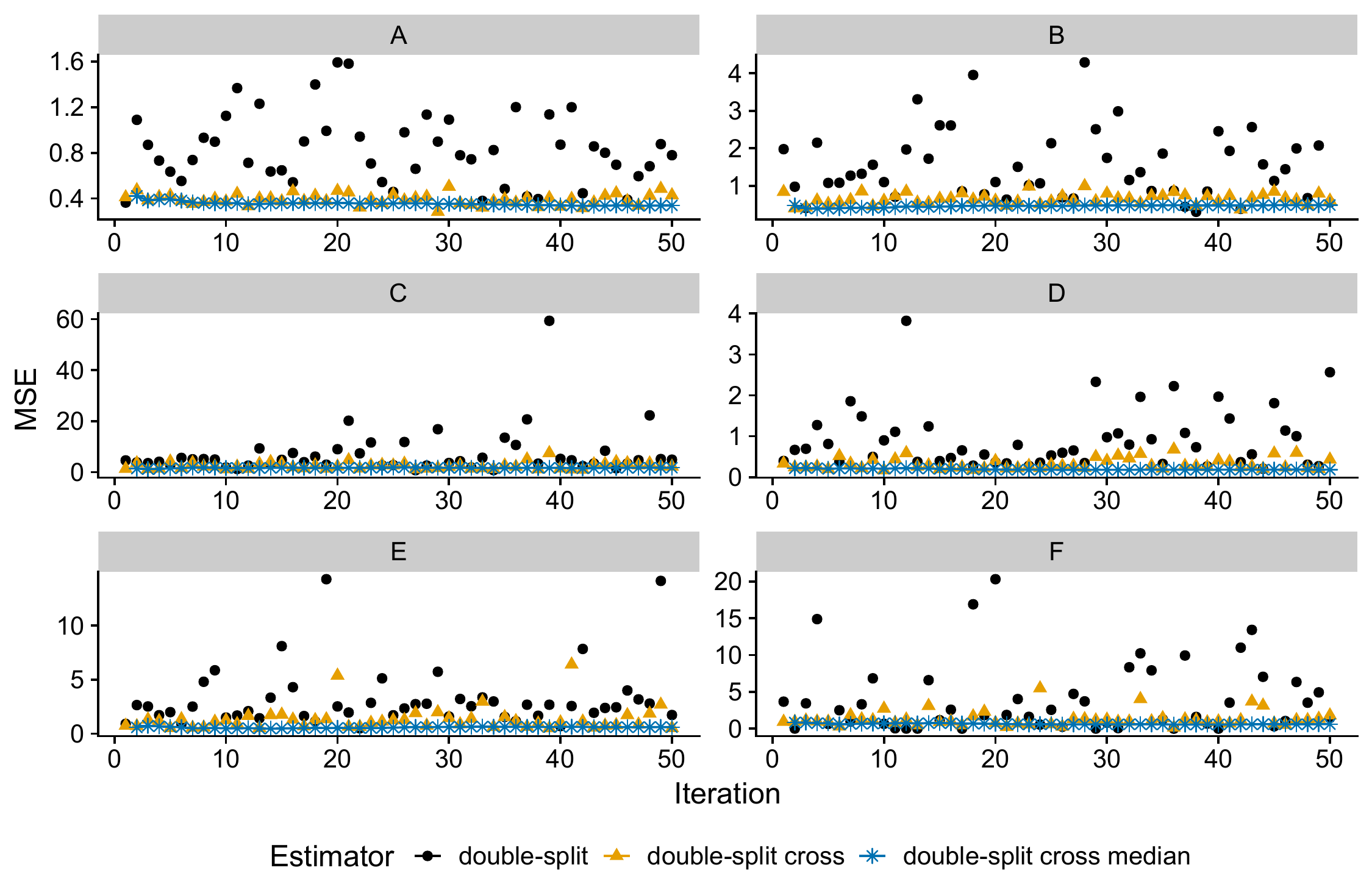}
\caption{MSE using the R-learner with double sample splitting, cross-fitting and taking the median. Number of observations in all settings is set to 500.  }
\label{}
\end{center}
\end{figure}

\begin{figure}[ht]
\begin{center}
\includegraphics[width=0.8\textwidth]{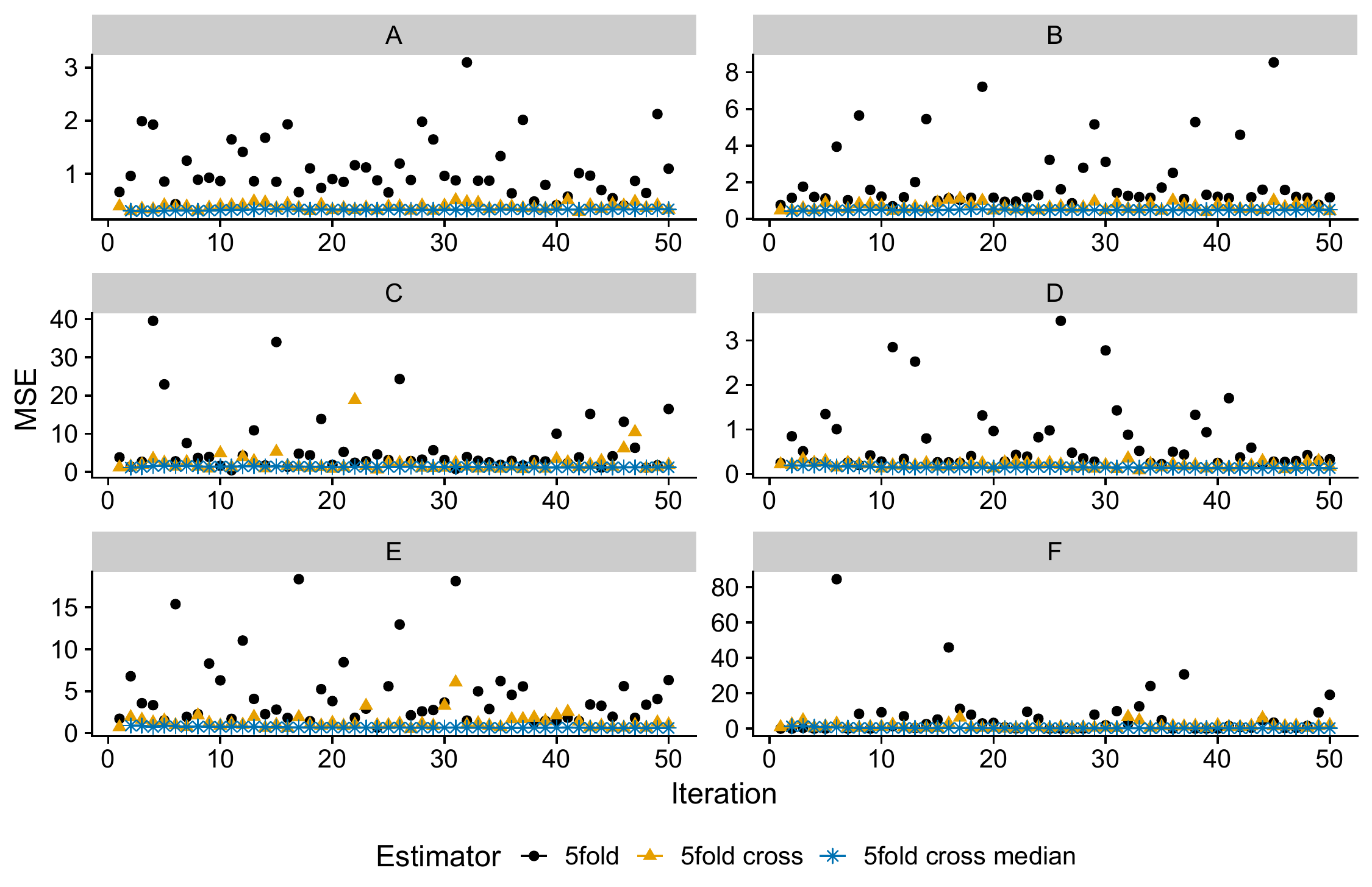}
\caption{MSE using the R-learner on a 5-fold split, cross-fitting and taking the median. Number of observations in all settings is set to 500.  }
\label{}
\end{center}
\end{figure}

\begin{figure}[ht]
\begin{center}
\includegraphics[width=0.8\textwidth]{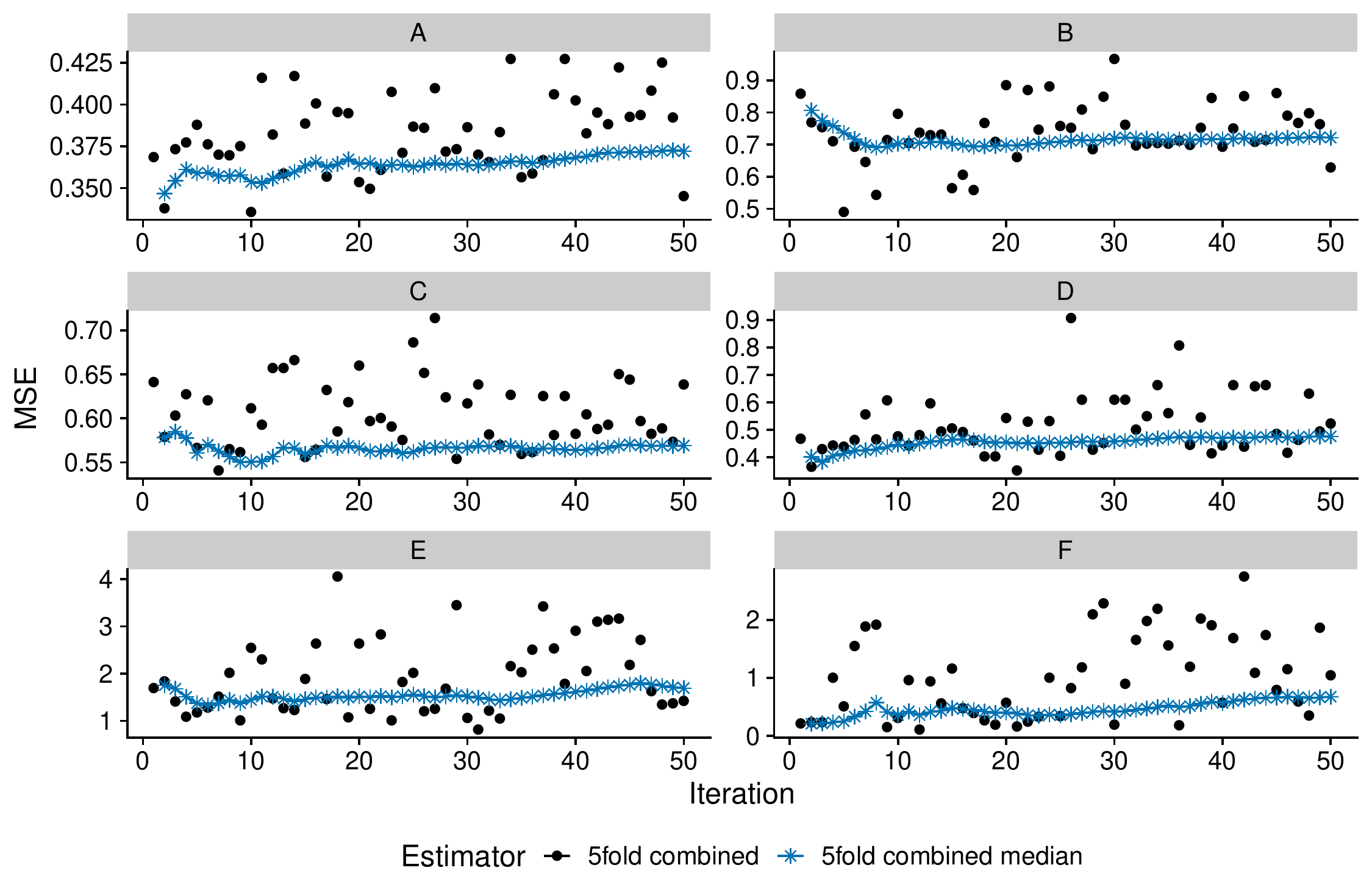}
\caption{MSE using the R-learner on a 5-fold combined. Number of observations in all settings is set to 500.  }
\label{r_5fold_combined_500}
\end{center}
\end{figure}

\begin{figure}[ht]
\begin{center}
\includegraphics[width=0.8\textwidth]{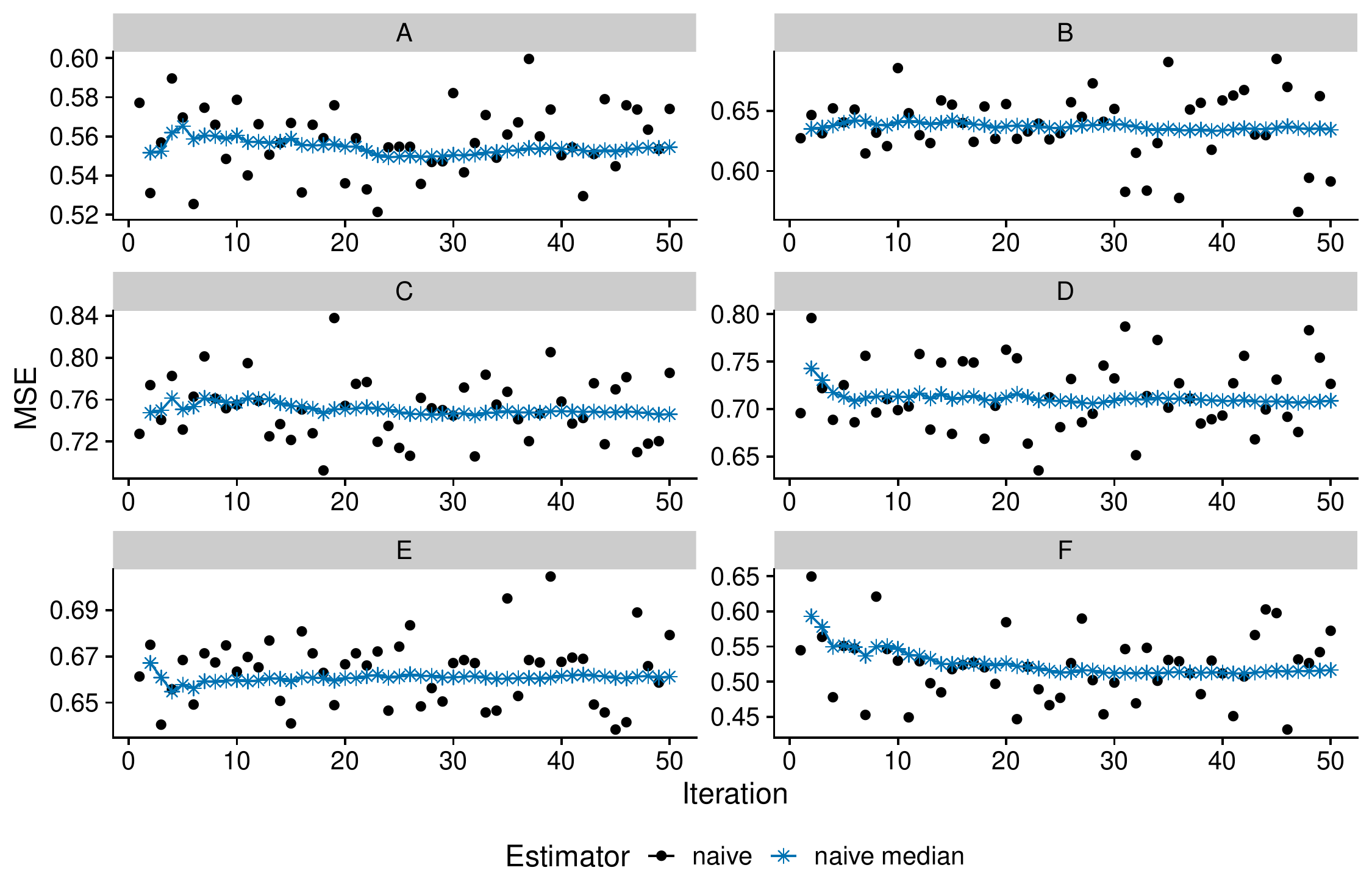}
\caption{MSE using the X-learner without sample splitting (naive). We also consider the case where we take the median over multiple iterations. Number of observations is 500. }
\label{}
\end{center}
\end{figure}

\begin{figure}[ht]
\begin{center}
\includegraphics[width=0.8\textwidth]{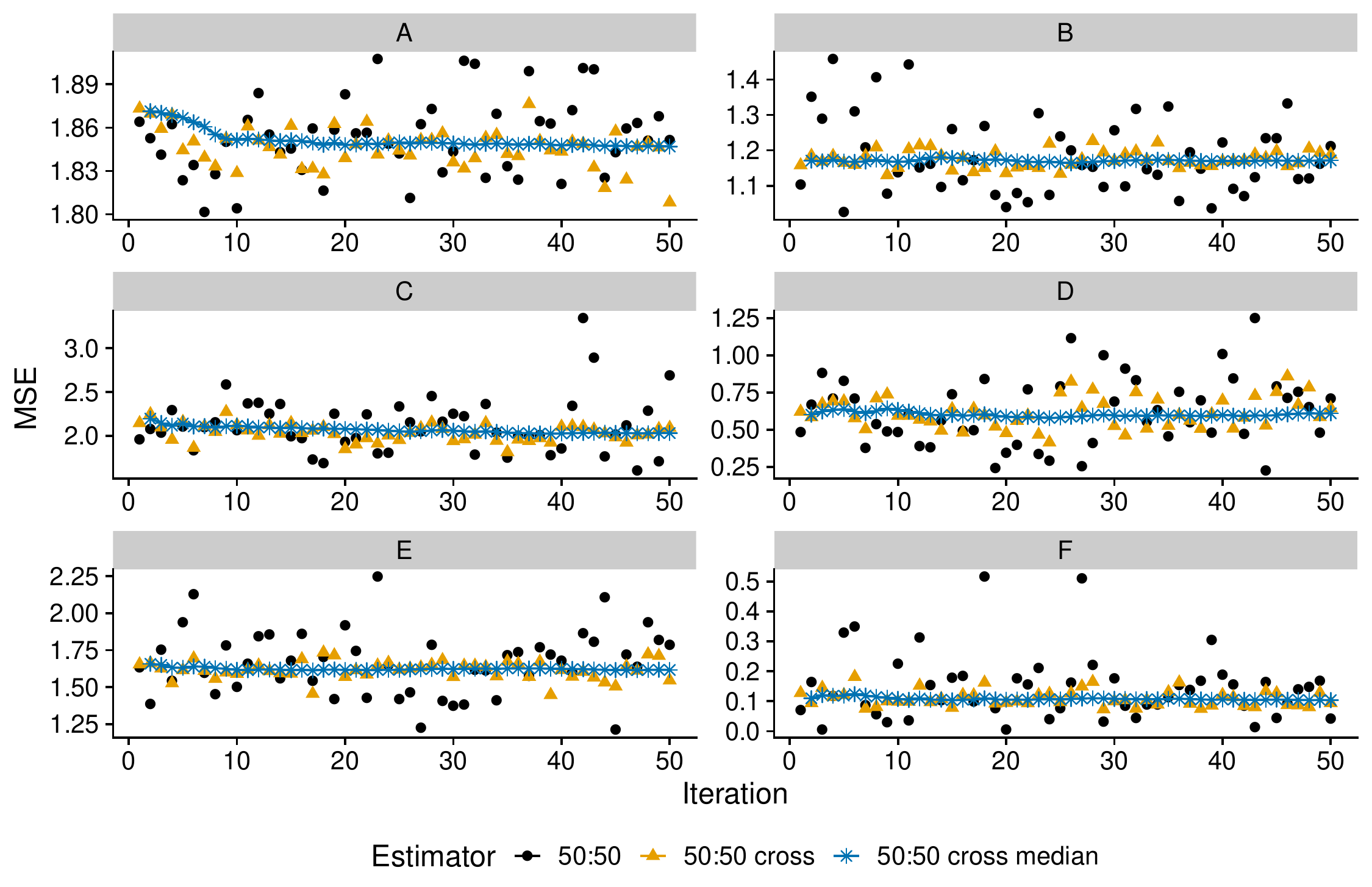}
\caption{MSE using the X-learner on a 50:50 split, cross-fitting and taking the median. Number of observations in all settings is set to 500.  }
\label{}
\end{center}
\end{figure}

\begin{figure}[ht]
\begin{center}
\includegraphics[width=0.8\textwidth]{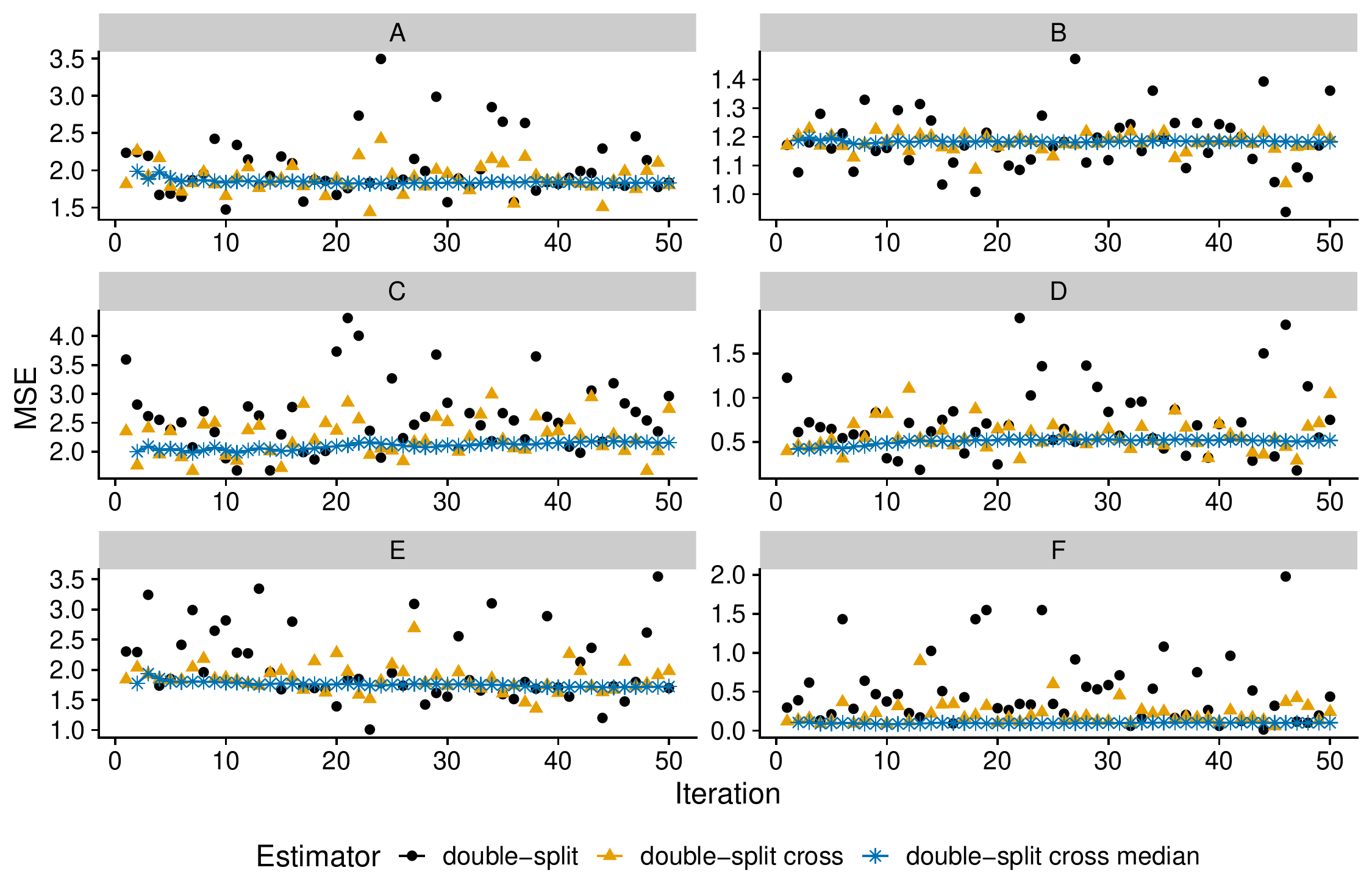}
\caption{MSE using the X-learner with double sample splitting, cross-fitting and taking the median. Number of observations in all settings is set to 500.  }
\label{}
\end{center}
\end{figure}

\begin{figure}[ht]
\begin{center}
\includegraphics[width=0.8\textwidth]{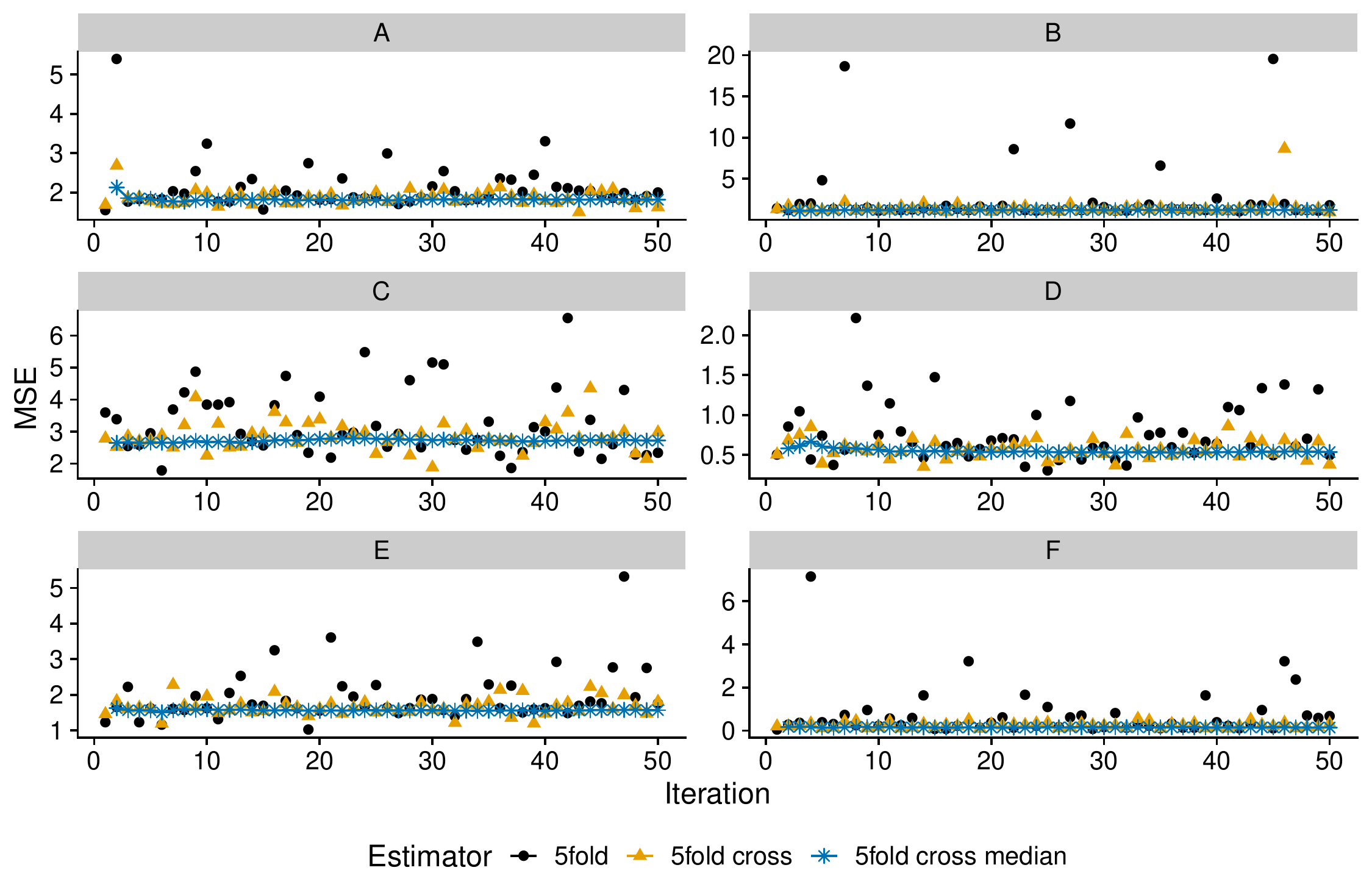}
\caption{MSE using the X-learner on a 5-fold split, cross-fitting and taking the median. Number of observations in all settings is set to 500.  }
\label{}
\end{center}
\end{figure}

\clearpage

\vspace{5mm}



\end{document}